\documentclass{aa}  

\usepackage{graphicx}
\usepackage{txfonts}
\usepackage{lipsum}
\usepackage{subcaption}         
\usepackage{lscape}             
\usepackage{placeins}           
\usepackage{wasysym}            
\usepackage{amsmath}	        
\usepackage{amssymb}	        
\usepackage{tikz}
\usepackage[LGR,T1]{fontenc}

\DeclareMathOperator{\sign}{sign}
\DeclareMathOperator{\lcm}{lcm}
\graphicspath{{./img/}}
                                
\usepackage[dvipsnames,table,xcdraw]{xcolor}
\usepackage{hyperref}
\hypersetup{colorlinks=true,citecolor=olive,linkcolor=olive,urlcolor=olive}

\newcommand{\Ham}{\mathcal{H}}
\newcommand{\Kam}{\mathcal{K}}
\newcommand{\Aam}{\mathcal{A}}
\newcommand{\Bam}{\mathcal{B}}
\newcommand{\Lam}{\mathcal{L}}

\newcommand{\drond}[2]{\frac{\partial #1}{\partial #2}}
\newcommand{\vect}[1]{\mathbf{\boldsymbol{#1}}}
\newcommand{\accolade}{\phantom{\frac{1}{1}\!\!\!\!\!}}

\newcommand{\LGR}{\fontencoding{LGR}\selectfont}
\newcommand{\Latin}{\fontencoding{\encodingdefault}\selectfont}
\newcommand{\sampi}{\text{\LGR \textsampi{} \Latin}\!}

\newcommand{\OO}{\mathcal{O}}
\newcommand{\G}{\mathcal{G}}

\begin{document}

   \title{
   	Dynamics of three-planet zeroth-order resonances with tides 
   }
   
   \subtitle{Shaping the architecture of close-in multiplanetary systems}
   
   \author{J\'er\'emy Couturier\inst{1,2}\fnmsep\thanks{Corresponding author. \email{jeremy.couturier@unige.ch}\\ Website: \url{https://jeremycouturier.com}
   	}
   	\and Adrien Leleu\inst{1}
   	\and Jean-Baptiste Delisle\inst{1}
   }
   
   \institute{Observatoire de Gen\`eve, Universit\'e de Gen\`eve, Chemin Pegasi, 51, 1290 Versoix, Switzerland
   	\and
   	LTE, Obs. de Paris, Univ. PSL, Sorbonne Univ., Univ. de Lille, LNE, CNRS, 61 Av. de l'Observatoire, 75014 Paris, France
   }   
   \date{Accepted November 25, 2026}

   
   \abstract{
   	Two-planet mean-motion resonances (2PR), where the period ratios between planets are close to main rational numbers, are natural outcomes in simulations of the formation of exoplanetary systems. However, observed pairs are rarely inside 2PR, and different physical processes, such as tidal evolution or post-disk instabilities, can be responsible for this distance. Since these processes affect three-planet resonances (3PR) differently than 2PR, a model of 3PR can be used to distinguish between these scenarios.
   	
   	In this study, we derived a zero-parameter one-degree-of-freedom pendulum-like analytical model of three-planet zeroth-order resonances (3P0R), with and without planetary tides. 
   	We checked the validity of the conservative model using a stability map built with the new software \texttt{Aptidal}, and we ran numerical integrations to verify the dissipative model. We showed analytically that 3P0R are always unstable to tides.
   	Numerically, we showed that planets in chains of 2PR undergoing tidal dissipation typically exit these 2PR while following 3P0R, then exit the 3P0R, before possibly getting captured in three-planet first-order resonance (3P1R) while staying close to 3P0R.
   	Triplets can be observed at any stage of this evolution, depending on the system's age and the tidal strength. 
   	We recovered the 3P0R state of known observed resonant chains, validated previous candidates, and found a new one.
   	
   	We used our model to classify the whole observed population.
   	We found a statistically significant excess of tidally affected triplets in 3P0R, consistent with a small fraction of the close-in systems forming in chains of 2PR and then tidally evolving without large-scale instabilities, although this is probably not a major mechanism of evolution.
   }

   
   \maketitle
   \nolinenumbers
   
\section{Introduction}\label{sec:intro}
   
Chains of mean-motion resonances are natural outcome in simulations of the formation of exoplanetary systems \citep{CresswellNelson2008}. However, these configurations are relatively rare in the observed population. To address this tension, planetary tides were investigated. Although they were shown to be unable to explain the whole observed period-ratio distribution \citep{Lee2013}, \cite{DelisleLaskar2014} showed evidence of tides in sculpting the period ratios of pairs of close-in planets (inner period at less than 5 days).
To explain the period ratio distribution of the whole Kepler population, \cite{Izidoro2017} proposed the breaking the chains model, in which most systems form in resonant chains but grow unstable afterwards \citep[\textit{e.g.},][]{HaddenWu2026,GoldbergPetit2026}, and emerging observational trends 
are consistent with this model \citep{Leleu2024,Dai2024,Leleu2026b}. While these trends relate to two-planet resonances (2PR), three-planet resonances (3PR) might also play a role in shaping exoplanetary systems \citep{Petit2021,Cerioni2022}.
   
   
In this work, mean-motion resonances are referred to as resonances. The general form of a $N$-planet resonance (NPR) is $k_1n_1+k_2n_2+\cdots+k_Nn_N=0$, where $n_j=2\pi/P_j$ denotes the mean motion of planet $j$, $P_j$ is its orbital period and the $k_j$ are integers. A $N$-planet \smash{$k^{\text{th}}$} order resonance (NPkR) is a NPR such that $k_1+k_2+\cdots+k_N=\pm k$.
\begin{figure}[h]
	\centering
	\includegraphics[width=\linewidth]{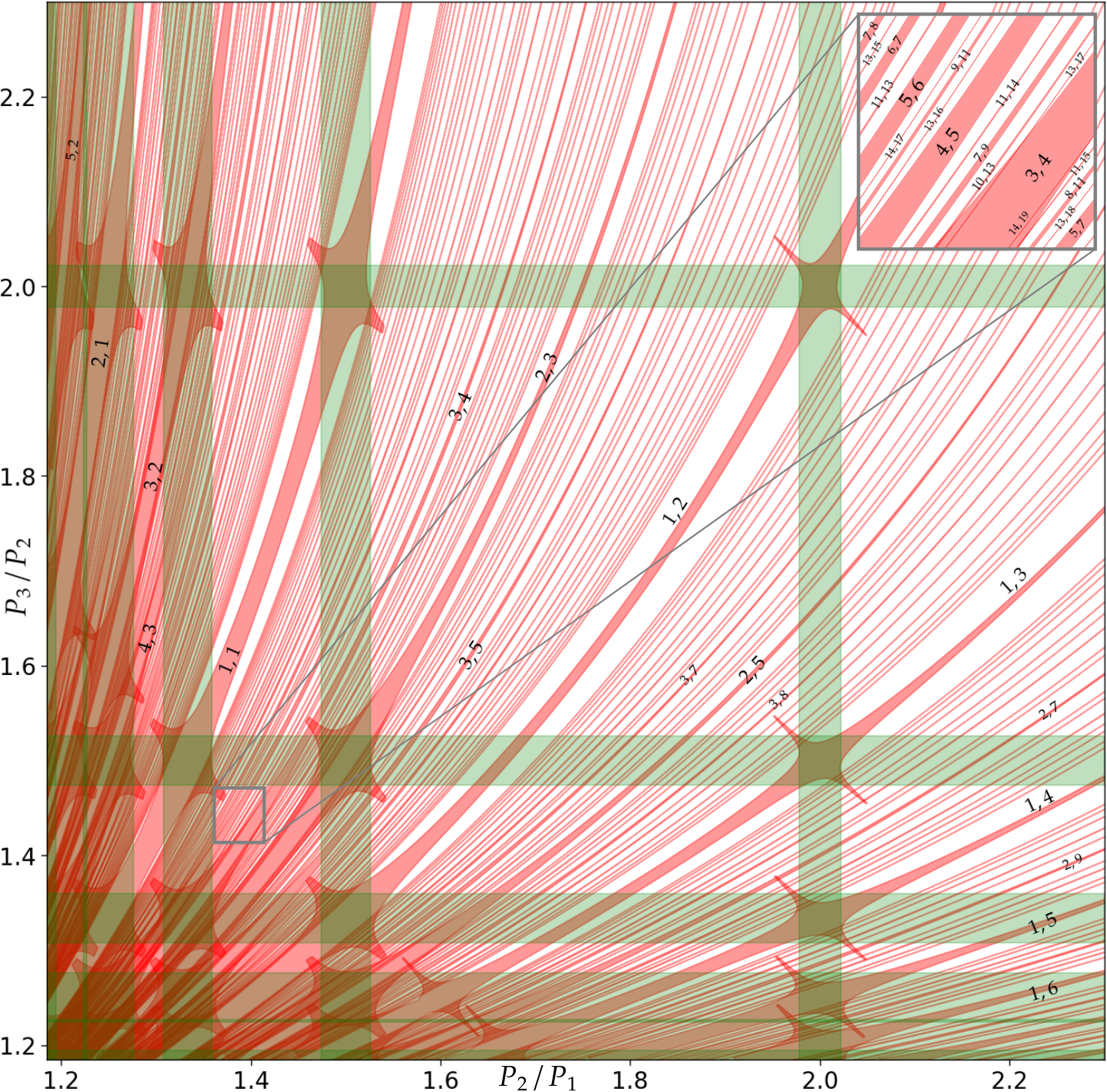}
	\caption{3P0R $pn_1-\left(p+q\right)n_2+qn_3=0$ for $p+q\leq33$ in the period ratio diagram, obtained from Eq. (\ref{eq:3P0R_width}), using \texttt{PerHam3pla} for the computation of $\Xi$. Red (\textit{resp.} green) shaded areas are within the separatrices of a 3P0R (\textit{resp.} 2P1R). For some 3P0R, $p,q$ are given. The masses are $m_1=m_2=m_3=2\times10^{-4}m_0$.}\label{fig:3P0R_width}
\end{figure} 
The perturbations due to NPkR scale in \smash{$\OO\left(\varepsilon^{N-1}e^k\right)$}, where $\varepsilon$ is the planet over star mass ratio and $e$ is the typical eccentricity. Therefore, besides 2PR, 3P0R are the strongest of resonances.
2P1R (in green in Fig. \ref{fig:3P0R_width}) are well understood \citep[\textit{e.g.},][]{Henrard1983,CoRoCo2022}, at least for moderate eccentricities and inclinations, expected for the bulk of the known multi-planetary systems. 
In multi-planetary systems, 2P1R dominate the dynamics as long as \smash{$k_jn_j-(k_j+1)n_{j+1}\lesssim\OO(\varepsilon^{2/3})$}, but 3P0R (in red in Fig. \ref{fig:3P0R_width}) become relevant when eccentricity damping (for example due to tides) increases the distance to 2PR while maintaining 3P0R commensurability \citep{Papaloizou2015,Charalambous_et_al_2023}. In Fig. \ref{fig:Period_ratio} we show a possible evolution where a capture in 3P1R follows an exit from 3P0R, but adiabatic systems can also get captured in 3P1R from non resonant period ratios \citep{Petit2021}. In the exoplanet population, Kepler-80, Trappist-1, K2-138 and TOI-178 are dominated by 3P0R dynamics \citep{Macdonald2016,Luger2017,Christiansen2018,Leleu2021}, while YZ-Ceti and Kepler-221 are strong candidates \citep{Stock2020,GoldbergBatygin2021}.
\begin{figure}[h]
	\includegraphics[width=\linewidth]{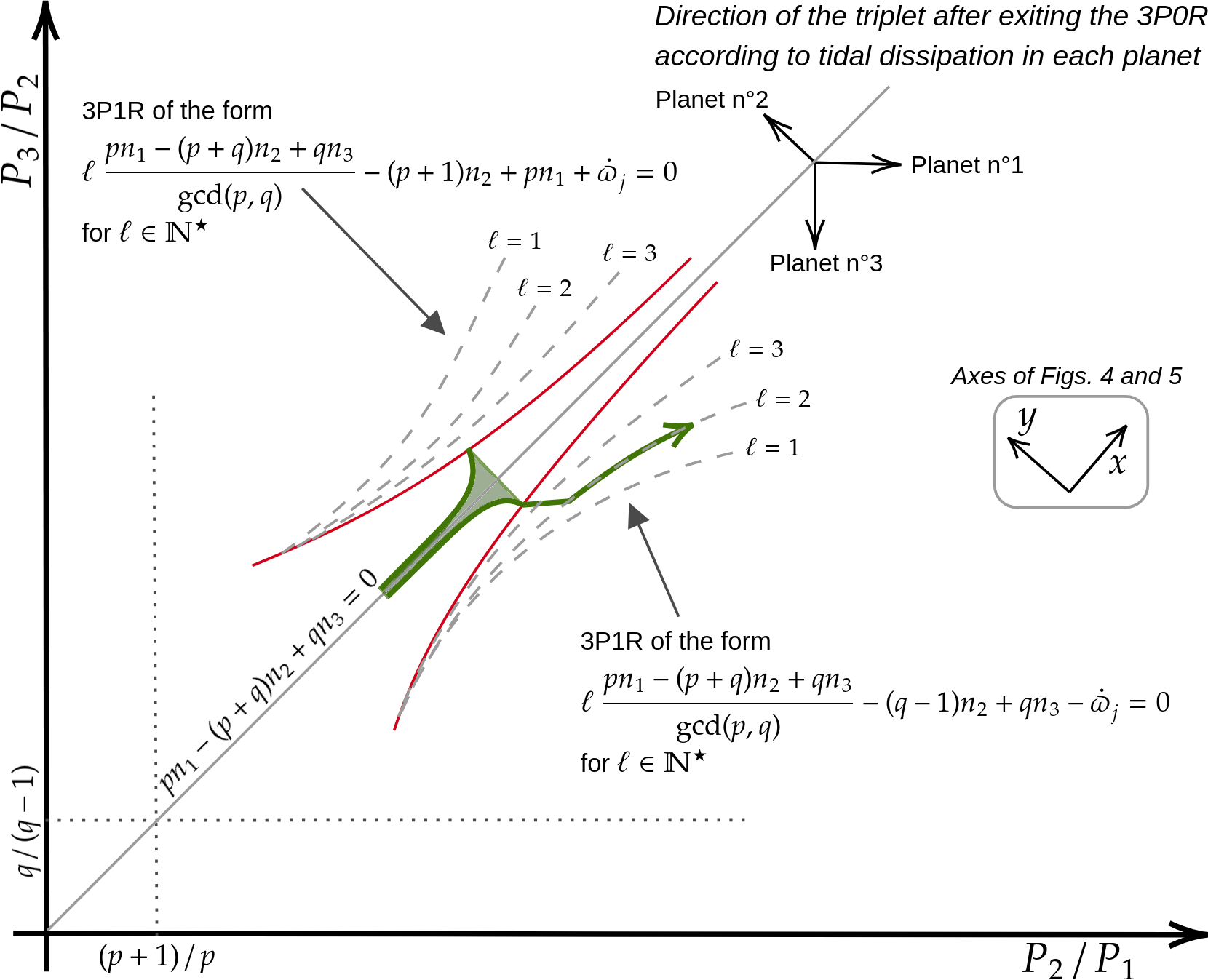}
	\caption{Schema of the exit from 3P0R and capture into 3P1R. The solid green curve shows the typical evolution of a triplet that started close to the 2P1R $pn_1-\left(p+1\right)n_2=0$ and $\left(q-1\right)n_2-qn_3=0$. The triplet first goes away from the 2P1R by tides, and the 3-planet angle libration amplitude increases exponentially until the separatrix is crossed. Then, the direction is determined by the dissipation in each planet (see arrows in top right), the 3P1R are crossed and capture is possible. Here, the triplet dissipates mostly in the inner planet, as is often the case.}\label{fig:Period_ratio}
\end{figure}
   
To understand resonant dynamics, one-degree-of-freedom (1-dof) models are useful as they highlight the combination of elliptic element and masses relevant to the dynamics, allowing comparisons between models and observations \citep{Leleu2026b}.  
\cite{Quillen2011} and \cite{Petit2020} derived 1-dof models of 3P0R, but neglected indirect terms and terms coming from non-consecutive planets. Therefore their expressions do not apply to the wide variety of period ratios in the population. In this work, 
we made our calculations with the open source python algebraic manipulator \href{https://gitlab.unige.ch/delisle/celeries}{\texttt{celeries}} \citep{Delisle2017}, where we implemented a class \texttt{PerHam3pla} able to compute the coefficients of any 3PkR at any order in eccentricity and inclination.
   
In Sect. \ref{sec:chaos}, we briefly describe our 1-dof model of 3P0R. 
We build a stability map of a 3P0R with the new software \href{https://github.com/Jeremycouturier/Aptidal}{\texttt{Aptidal}} to show the validity of our model in the conservative case. We use the model to interpret simulations of tidal evolution of chains initially in 2PR, classify triplets of observed multi-planetary systems and describe the overall distribution of observed triplets with respect to 3P0R. In Sect. \ref{sec:pendulum}, we derive the pendulum model, with and without tides, and we validate its tidal part with numerical simulations made by \texttt{Aptidal}. We conclude in Sect. \ref{sec:conclusion}. In Appendix \ref{sec:second_mass}, we average the planetary Hamiltonian to second-order in masses in order to make 3PR appear, while we present \href{https://github.com/Jeremycouturier/Aptidal}{\texttt{Aptidal}} in Appendix \ref{sec:Aptidal}, a symplectic planetary $N$-body integrator where tides are implemented pseudo-symplectically.

\section{Three-planet zeroth-order resonances}\label{sec:chaos}

In this section, we only explain the aspects of our model necessary to understand the results. Our model of 3P0R is thoroughly detailed in Sect. \ref{sec:pendulum}. We assume that the 3P0R $pn_1-\left(p+q\right)n_2+qn_3\sim0$ emanates from the 2P1R $pn_1-\left(p+1\right)n_2=0$ and $\left(q-1\right)n_2-qn_3=0$, but the system can be far from the 2P1R. The distance to 2P1R can be parameterized with $\Delta$, defined as
\begin{equation}\label{eq:rho}
	1+\Delta=\frac{p}{\left(p+1\right)}\frac{n_{1,0}}{n_{2,0}}=\frac{p}{\left(p+1\right)}\frac{P_{2,0}}{P_{1,0}},
\end{equation}
where $n_{j,0}$ and $P_{j,0}$ are nominal mean motions and periods defined in Sect. \ref{ssec:change_var}, close to their actual values, and such that \smash{$pn_{1,0}-\left(p+q\right)n_{2,0}+qn_{3,0}=0$}. However, a better parameterization is \citep[][Eq. (7) and Fig. 4]{Leleu2026b}
\begin{equation}\label{eq:D2P1R}
	\Delta_\text{2P1R}=\varepsilon^{-2/3}\left(\left(1+\Delta\right)^{1/3}-1\right)=\varepsilon^{-2/3}\OO(\Delta),
\end{equation}
where \smash{$\varepsilon=\left(m_1+m_2+m_3\right)/m_0$}. With this parameterization, the boundary of 2P1R is at \smash{$\Delta_\text{2P1R}=\OO(1)$}. The Hamiltonian of our 1-dof zero-parameter pendulum model of 3P0R reads (Eq. \ref{eq:0param})
\begin{equation}\label{eq:Ham_3P0R}
	\Ham(\Sigma;\sigma)=\frac{1}{2}\Sigma^2\pm\cos\sigma,
\end{equation}
where $\sigma=p\lambda_1-\left(p+q\right)\lambda_2+q\lambda_3$ is the 3-planet angle, $\Sigma$ is its conjugated momentum, that depends on $p$, $q$, the masses and the orbital periods, and the sign in front of $\cos\sigma$ matches that of $\Delta$.
\begin{figure}
	\begin{minipage}[c]{0.47\linewidth}
		\includegraphics[width=\linewidth]{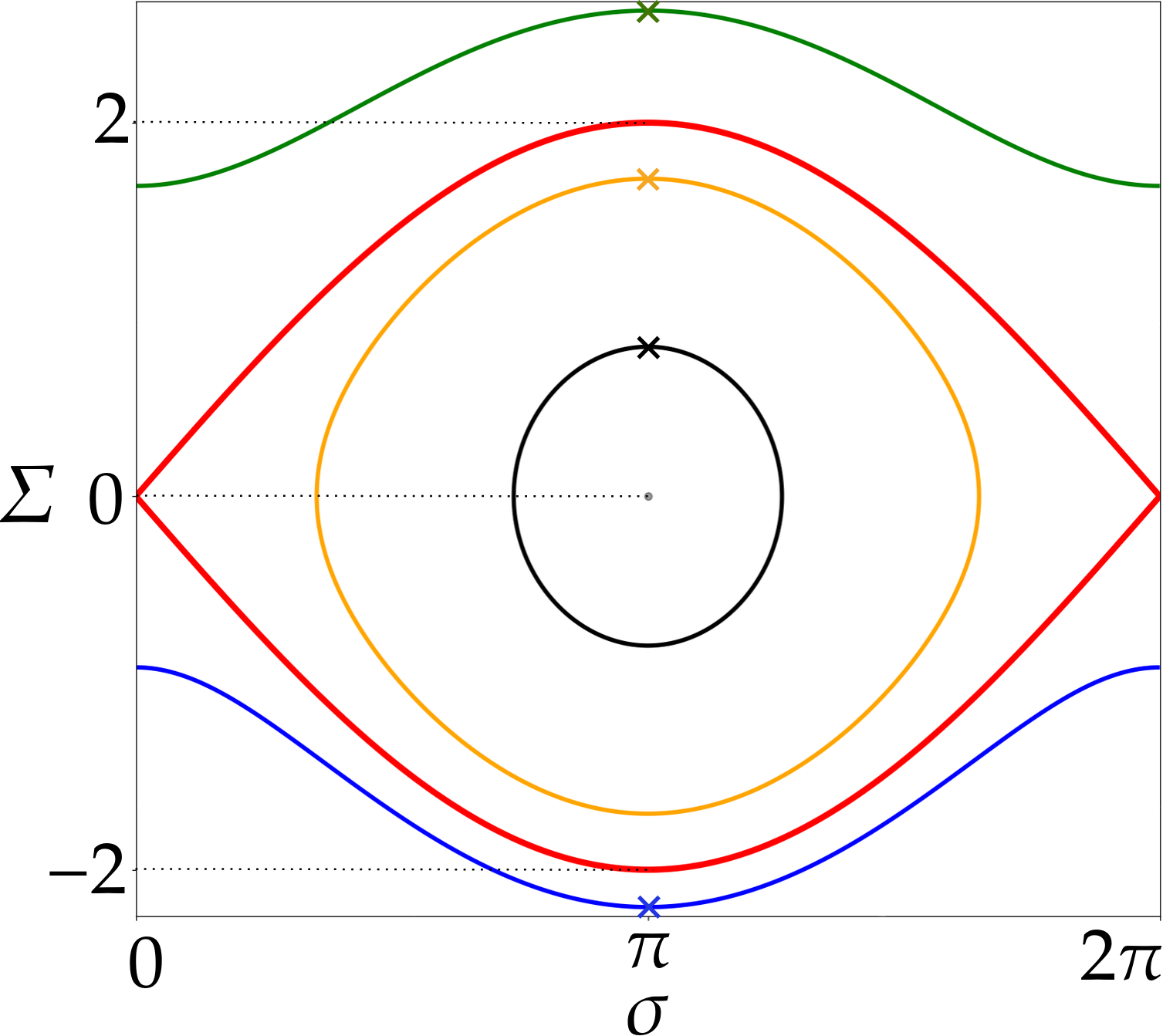}
	\end{minipage}\hfill
	\begin{minipage}[c]{0.49\linewidth}
		\caption{Phase space of the Hamiltonian of Eq. (\ref{eq:Ham_3P0R}) when $\Delta>0$. The black and orange orbits are in 3P0R (inside the red separatrix) whereas the green and blue orbits are not. When $\Delta<0$, the phase space is the same but the resonance center is at $\sigma=0$.}\label{fig:3P0RPhaseSpace}
	\end{minipage}
\end{figure}
The phase space of the pendulum modeling 3P0R is shown in Fig. \ref{fig:3P0RPhaseSpace}. For an orbit in this pendulum, we consider the extremum $\Sigma_\text{ex}$ reached by $\Sigma$, that is, we define
\begin{equation}\label{eq:D3P0R}
	\Delta_\text{3P0R}=\Sigma_\text{ex}=\pm\sqrt{2+2\Ham},
\end{equation}
where the $\pm$ sign is positive if the triplet is within the separatrix or above it and negative if the triplet is below the separatrix. This definition is shown with diagonal crosses in Fig. \ref{fig:3P0RPhaseSpace}.
Hereafter, we work with $\Sigma_\text{ex}$ and forfeit the information on $\sigma$, since $\Sigma_\text{ex}$ alone characterizes the orbit in Fig. \ref{fig:3P0RPhaseSpace}.
The separatrix delimits the boundary of the 3P0R and corresponds to \smash{$\Delta_\text{3P0R}=\pm2$}. As explained in Sect. \ref{ssec:change_var}, our pendulum model of 3P0R is valid only outside of 2P1R, when \smash{$\Delta_\text{2P1R}\geq\OO(1)$}.
\begin{figure*}[h]
	\centering
	\includegraphics[width=\linewidth]{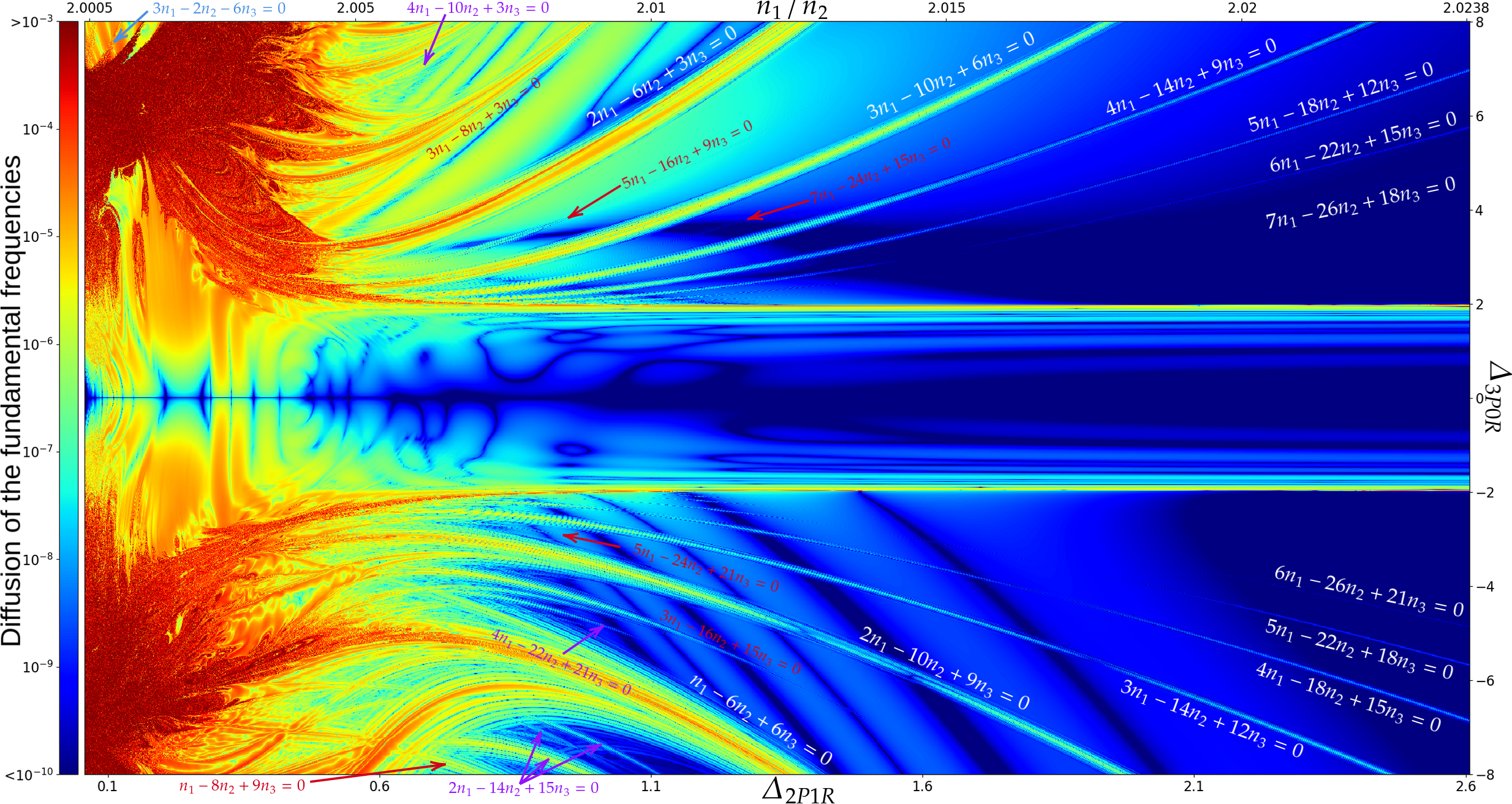}
	\caption{Stability map of the 3P0R $n_1-4n_2+3n_3=0$. On the $x$-axis are the libration centers of the 3P0R (Sect. \ref{ssec:libration}), parameterized by $n_1/n_2$ (top axis) or by $\Delta_\text{2P1R}$ (bottom axis). The separatrix of the 3P0R is at \smash{$\Delta_\text{3P0R}=\pm1.97$} close to the predicted \smash{$\Delta_\text{3P0R}=\pm2$} (Fig. \ref{fig:3P0RPhaseSpace}). The wide 3P0R $n_1-4n_2+3n_3=0$ is visible. 3P1R, 3P2R, 3P3R and 3P5R are indicated in white, red, purple and blue, respectively. See Sects. \ref{ssec:stability_map}, \ref{ssec:libration} and \ref{ssec:about_stability_map} for details. Similar figures for other $(p,q)$ and other masses are available \href{https://jeremycouturier.com/img/3P0R/}{here}\protect\footnotemark and in Appendix \ref{append:stability_map}.}\label{fig:stability_map}
\end{figure*}
\footnotetext{\url{https://jeremycouturier.com/img/3P0R/}}

\subsection{A stability map of the 3P0R $n_1-4n_2+3n_3\sim0$}\label{ssec:stability_map}

In order to visualize the 3P0R $n_1-4n_2+3n_3\sim0$ by the chaos that it generates in the phase space, we made Fig. \ref{fig:stability_map}. The $x$-axis and $y$-axis of this figure are \smash{$\Delta_\text{2P1R}$} and \smash{$\Delta_\text{3P0R}$}, respectively, defined by Eqs. (\ref{eq:D2P1R}) and (\ref{eq:D3P0R}). We also show $P_2/P_1$ on the top axis. Figure \ref{fig:stability_map} is further discussed and analyzed in Sect. \ref{ssec:about_stability_map}, where we explain the production of the figure with \texttt{Aptidal} and the definition of the colorbar. For now, we briefly introduce this figure which contains many 3PR structures. Along the $x$-axis of the figure, we follow the exact resonance centers (that is, the libration centers found by \texttt{Aptidal}, Sect. \ref{ssec:libration}) from deep in the 2P1R ($n_1/n_2=2.00042$) to far from the 2P1R ($n_1/n_2=2.02385$). Moving vertically on this figure, we modify $\Sigma$ while keeping constant the other initial conditions, and the semi-major axes are the only changing initial conditions when moving vertically in Fig. \ref{fig:stability_map}. In particular, the initial eccentricities are invariant on a given columns, and are the eccentricities of the libration center, going from \smash{$e_j\sim\left(5.9,4.9,4.8\right)\times10^{-2}$} on the leftmost column to \smash{$e_j\sim\left(1.4,1.5,1.5\right)\times10^{-3}$} on the rightmost column.

The color in Fig. \ref{fig:stability_map} is a measure of chaos by the diffusion of the fundamental frequencies. Blue regions are quasiperiodic while red regions are chaotic.
The 3P0R $n_1-4n_2+3n_3\sim0$ is visible due to mild chaos far from the 2P1R (right of the figure) at \smash{$\Delta_\text{3P0R}=\pm1.97$}, close to the value \smash{$\Delta_\text{3P0R}=\pm2$} expected from the model. This shows that our estimate of the resonance width in Eq. (\ref{eq:3P0R_width}), as well as our computation of the Hamiltonian coefficient $\Xi$ at second order in mass (Sect. \ref{ssec:Laplace}), are both correct.

Besides the 3P0R $n_1-4n_2+3n_3\sim0$, other 3PR structures are visible. In particular, the 3P1R schematized in Fig. \ref{fig:Period_ratio} are visible and labeled in Fig. \ref{fig:stability_map}. 3PkR up to $k=5$ are visible and labeled. Our stability map can be considered as a detailed and zoomed-in version of the stability map presented by \citet[][Fig. 2]{Petit2021}.

\subsection{Classification in the pendulum phase space}\label{ssec:classification}

\begin{figure*}[h]
	\centering
	\includegraphics[width=\linewidth]{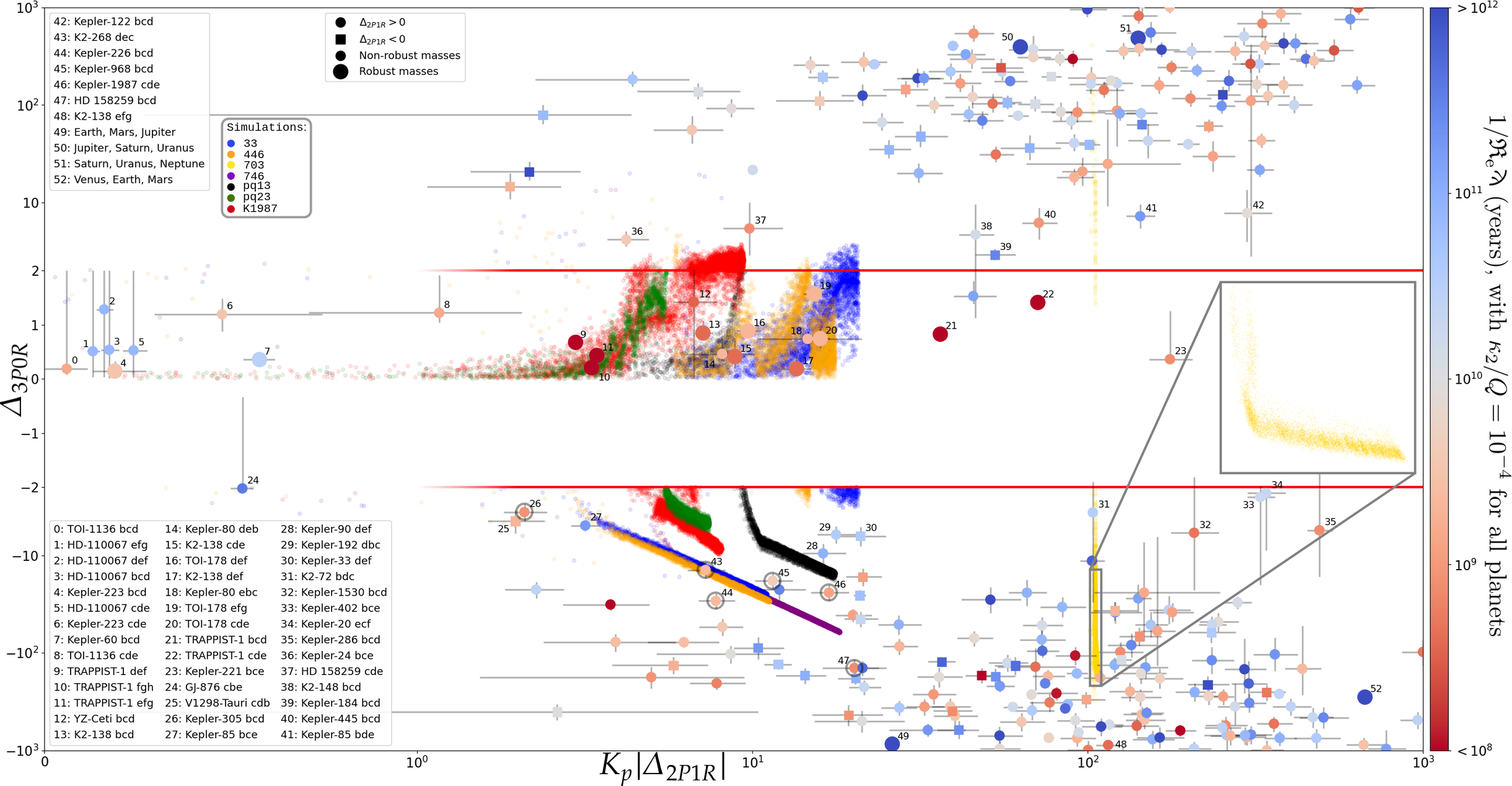}
	\caption{Classification of exoplanet triplets in the 3P0R phase space. The color is the e-folding time of 3-planet angle libration amplitude increase. The clouds of dots are simulations described in Appendix \ref{ssec:simulations}. Only the outer triplet of simulation \texttt{703} is plotted. The positions depend on the masses. The $x$ and $y$ axes are linear in \smash{$\left[0,1\right]$} and \smash{$\left[-2,2\right]$}, respectively, and logarithmic outside.}\label{fig:classification}
\end{figure*}
This section aims at classifying observed triplets of exoplanets in the phase space of 3P0R. Such classification is presented in Fig. \ref{fig:classification}, where, similarly to Fig. \ref{fig:stability_map}, the $x$-axis and $y$-axis of this figure are \smash{$\Delta_\text{2P1R}$} and \smash{$\Delta_\text{3P0R}$}, respectively. In order for the $x$-axis of our Fig. \ref{fig:classification} to match more closely the quantity $\Delta_{MMR}$ defined by \citet[][Figs 3 and 4, Eq. 7]{Leleu2026b} to quantify the distance to 2P1R, we use the quantity $K_p\Delta_{2P1R}$ in the $x$-axis of Fig. \ref{fig:classification}. The coefficient $K_p$ is given in Table \ref{tab:Kp} for small $p$ and defined in Appendix B of \cite{Leleu2026b}. In the case of similar masses, we have\footnote{If we decided to use $\varepsilon=\left(m_1+m_2\right)/m_0$ as in \cite{Leleu2026b}, we would have \smash{$\Delta_{MMR}=K_p\Delta_{2P1R}$} when $m_1=m_2$, but because we use $\varepsilon=\left(m_1+m_2+m_3\right)/m_0$, the two quantities are only close to each other.} \smash{$\Delta_{MMR}\approx K_p\Delta_{2P1R}$}. Since the pendulum model is only valid in the domain $\Delta_\text{2P1R}\gg\OO(1)$, the separatrix of the 3P0R, highlighted by a solid red line in Fig. \ref{fig:classification}, vanishes in the left part of the figure. For each triplet, we need to select $p$ and $q$ of the 3-planet resonance $pn_1-\left(p+q\right)n_2+qn_3=0$. We proceeded in the following way
\begin{itemize}
	\item If they exist, we use the smallest $p$ and $q$ such that $0.99<pn_1/\left(\left(p+1\right)n_2\right)<1.08$ and $0.99<\left(q-1\right)n_2/\left(qn_3\right)<1.08$.
	\item Otherwise, we choose \smash{$\left(p,q\right)$} that minimized \smash{$\left|\Delta_\text{3P0R}\right|$}.
\end{itemize}
Note that we do not need to truncate in $p$ and $q$ since our metric is rescaled by the width of the 3P0R, that goes to zero exponentially quickly when $p,q\to\infty$ (see Fig. \ref{fig:3P0R_width}). Therefore, \smash{$\left|\Delta_\text{3P0R}\right|$} is almost never minimized by $\max(p,q)>15$.
\begin{table}[h]
	\setlength{\tabcolsep}{3pt}
	\renewcommand{\arraystretch}{1.}
	\caption{Coefficients $K_p$ on the $x$-axis of Fig. \ref{fig:classification}}\label{tab:Kp}
	\vspace{-2mm}
	\centering
	\begin{tabular}{ccccccccc}
		\hline
		$p$&$1$&$2$&$3$&$4$&$5$&$6$&$7$\\
		\hline
		\hline
		$K_p$&$5.6405$&$3.5595$&$3.1781$&$2.9213$&$2.7316$&$2.5833$&$2.4628$\\
		\hline
	\end{tabular}
\end{table}

The conversion \smash{$(m_j, P_j, \lambda_j)\to(\Sigma,\sigma,\delta)$} from observed quantities to the model requires the masses, periods and mean longitude of each planet to be known. The eccentricities are assumed to be zero in the pendulum model. For transiting exoplanets, the periods are strongly constrained, and the mean longitudes $\lambda_j$ are obtained from the transit times. However, the masses are not constrained or not robust, and we estimated them from the mass-radius relations of \citet[][Fig. 1]{Parc2024}. In this case the errors bars on Fig. \ref{fig:classification} come from the errors provided by each system's authors on the radii, and on the bounds of the mass-radius relation. In general, the discrepancy between different authors is larger than the error of a single author. To account for this, we doubled all errors published, as to consider two standard deviations. Despite that, the actual uncertainty in Fig. \ref{fig:classification} can be larger than what is suggested by the error bars. Only when robust constraints on the masses exist, in the sense that robust posterior analyses of the system were published, did we use posterior samples \citep{Leleu2026a, Dai2023, Leleu2023, Hadden2017, Leleu2024TOI178, Agol2021}. We extracted them from the \href{https://dace.unige.ch/resonantStateObservationsCatalog/}{database} created during the Geneva resonant state workshop \citep{Leleu2026b}. Concerned systems are plotted with bigger points in Fig. \ref{fig:classification}. For these systems, the error bars show the $2.3$\% and $97.7$\% quantiles, and are meaningful.
For non-transiting exoplanets, we used the published masses directly. When the mean longitude $\lambda_j$ were not provided, we allowed the 3-planet angle to cover the entire range \smash{$\left[0,2\pi\right[$} for our error bars.

In Fig. \ref{fig:classification}, all triplets of consecutive exoplanets are plotted, as long as they are within the bounds of the figure. However, in order to keep the figure legible, not all triplets of non-consecutive planets are plotted. If a system had
\begin{itemize}
	\item at least one triplet of consecutive planets in 3P0R, that is, \smash{$\left|\Delta_\text{3P0R}\right|<2$}, we ignored non-consecutive triplets.
	\item at least one consecutive triplet in Fig. \ref{fig:classification}, that is, \smash{$\left|\Delta_\text{3P0R}\right|<10^3$}, we only plotted non-consecutive triplets with \smash{$\left|\Delta_\text{3P0R}\right|<10$}.
	\item no triplet of consecutive planets in Fig. \ref{fig:classification}, we considered all non-consecutive triplets.
\end{itemize}
The clouds of dots in Fig. \ref{fig:classification} correspond to numerical simulations with planetary tides, that we ran with \texttt{Aptidal}. Some of these simulations use initial conditions from the Bern model, while some others use arbitrary initial conditions near the 3P0R center and close to the 2P1R. See Appendix \ref{ssec:simulations} and Table \ref{tab:simulations} for details. Most of them follow the same typical evolution, described as follow:
\begin{itemize}
	\item In the early beginning, the system's triplets remain close to the 3P0R, but rapidly get away from the 2P1R as tides damp the eccentricities from a few $10^{-2}$ to typically less than $10^{-3}$ \citep[\textit{e.g.},][]{DelisleLaskar2014}.
	\item Once the eccentricities are $\lesssim10^{-3}$, the triplets keep moving towards the right of Fig. \ref{fig:classification} (away from the 2P1R) at a much slower pace. However, the libration amplitude of the 3-planet angle $\sigma=p\lambda_1-\left(p+q\right)\lambda_2+q\lambda_3$ increases until the triplets eventually cross the separatrix and exit the 3P0R.
	\item Once the separatrix is crossed, triplets keep getting further away from the 3P0R, and most systems have at least one triplet that gets captured in 3P1R, leading to characteristic thin straight lines in Fig. \ref{fig:classification}.
\end{itemize}
This typical evolution is schematically represented in Fig. \ref{fig:Period_ratio}. A notable fact is that the triplets always exit the 3P0R because of a separatrix crossing due to the increase of the libration amplitude of $\sigma$. In Sect. \ref{ssec:eigen}, we show that the eigenvalue $\sampi$ controlling the behavior of $\left(\sigma,\Sigma\right)$ in the vicinity of 3P0R center has a strictly positive real part, and we even give an analytical expression of $\mathfrak{R}\text{e}\,\sampi$ in Eq. (\ref{eq:Resampi}) as a function of the tidal parameters. Since $\mathfrak{R}\text{e}\,\sampi>0$, the 3-planet angle's libration amplitude increases exponentially with an e-folding time $1/\mathfrak{R}\text{e}\,\sampi$, and the triplets inexorably cross the separatrix of the 3P0R, given enough time.

This specificity of 3P0R is similar to what happens in the context of 2P0R (the co-orbital resonance $1:1$). Indeed, when tidal dissipation is involved, \cite{CoRoCo2021} computed the eigenvalue controlling the co-orbital angle $\lambda_1-\lambda_2$ in the vicinity of the equilateral equilibria, and showed that its real part is always positive (see $\sampi$ in their Eq. 60), leading to a systematic destruction of the co-orbital configuration.

While $1/\mathfrak{R}\text{e}\,\sampi$ approximates the e-folding time to within $1$\% for a lone triplet (Fig. \ref{fig:e-fold_3P0R}), additional planets can temporarily damp the $3$-planet angle, as we see in simulations \texttt{33} and \texttt{446} in Fig. \ref{fig:classification}. Whereas a triplet is out of 3P0R and in 3P1R, the other triplet undergoes some damping. This does not prevent it from eventually exiting the 3P0R, although on a timescale larger than predicted by $1/\mathfrak{R}\text{e}\,\sampi$. The same phenomenon, which we believe is due to secular resonances crossing, was previously described by \citet[][Figs. 9 and 10]{CoRoCo2022} for the 2P0R perturbed by a third planet. 

Although the triplets always cross the 3P0R separatrix because of planetary tides, the e-folding time $1/\mathfrak{R}\text{e}\,\sampi$ can be much longer than the system's age, and in Fig. \ref{fig:classification}, we plot observed triplets with a color indicating the e-folding time of the inner triplet, in years, assuming a typical \smash{$\kappa_2^{(j)}/Q_j=3\times10^{-4}$} for all planets. The inner triplet only is considered when computing $1/\mathfrak{R}\text{e}\,\sampi$, as it tends to drive the whole chain. We distinguish two populations. The red triplets, for which $1/\mathfrak{R}\text{e}\,\sampi\lesssim10^{10}$ years, are influenced by tides, whereas the blue triplets, for which $1/\mathfrak{R}\text{e}\,\sampi\gtrsim10^{10}$ years, are not. For the numerical simulations, $1/\mathfrak{R}\text{e}\,\sampi$ is given in Table \ref{tab:simulations} and is much smaller than the system's age (our simulations are all part of the red population).

In Fig. \ref{fig:classification}, several triplets are found to be in 3P0R resonance. Inside the resonance, red systems (subject to tides) tend to be on the right of blue systems (not subject to tides), because of eccentricity damping that brought red systems far from 2P1R. An exception is TOI-1136 bcd \citep{Dai2023}, that can be explained by the young age of this system ($\sim700$ Myr).

Another notable fact is the behavior of the simulations once the separatrix of the 3P0R is crossed. Our simulations show that a capture in 3P1R after exiting the 3P0R is a frequent scenario, although not certain. Additionally, all triplets seem to exit the 3P0R by the bottom, even though exiting on the top should a priori be as likely of a scenario.

\subsection{Exit of 3P0R and capture in 3P1R}\label{ssec:3P0R23P1R}

The fact that all triplets of our simulations exit the 3P0R by the bottom can be explained. Once a triplet crosses the separatrix of the 3P0R, the system becomes free of any MMR. It is mostly quasiperiodic (dark blue in Fig. \ref{fig:stability_map}), and the only remaining secular effects on semi-major axes are planetary tides, who preserve the total angular momentum of each star$-$planet pair. The total angular momentum of a star$-$planet pair is given by
\begin{equation}
	G_j=\Lambda_j-D_j=\beta_j\sqrt{\mu_ja_j\left(1-e_j^2\right)}.
\end{equation}
The angular momentum contained in the planetary rotation can be neglected, whereas the star's rotation is constant in the absence of stellar tides. Therefore, the conservation of $G_j$ is equivalent to the conservation of \smash{$a_j\left(1-e_j^2\right)$}. As tides keep damping the residual eccentricities, the semi-major axes $a_j$, and thus the orbital periods $P_j$, decrease in such a way that \smash{$a_j\left(1-e_j^2\right)$} is constant. Regarding the inner planet of the triplet, decreasing $P_1$ means going right in Fig. \ref{fig:Period_ratio}. Regarding the outer planet of the triplet, decreasing $P_3$ means going down in Fig. \ref{fig:Period_ratio}. As for the middle planet, decreasing $P_2$ means going top-left in Fig. \ref{fig:Period_ratio}. Both the inner and outer planet contribute in the triplet exiting the 3P0R by the bottom, while the middle planet contributes in the triplet exiting by the top (see arrows in the top right of Fig. \ref{fig:Period_ratio}). In order for the triplet to exit by the top, the middle planet needs to dissipate more than the inner and outer planet combined, which is unlikely since it orbits farther away from the star than the inner planet, and tides scale in \smash{$a_j^{-6.5}$}.
\begin{table}[h]
	\setlength{\tabcolsep}{1.5pt}
	\renewcommand{\arraystretch}{0.65}
	\caption{Triplets of our numerical simulations captured in 3P1R.}\label{tab:3P1R_simulations}
	\vspace{-2mm}
	\centering
	\begin{tabular}{cccc|cccc}
		\hline
		Name&Color&Triplet&$k_1,k_2,k_3$&Name&Color&Triplet&$k_1,k_2,k_3$\\
		\hline
		\hline
		\texttt{pq13}&Black&Inner&$2,-10,9$&\texttt{446}&Orange&Inner&$1,-4,4$\\
		\texttt{746}&Purple&Inner&$1,-4,4$&\texttt{703}&Yellow&Outer&$3,-7,5$\\
		\texttt{33}&Blue&Inner&$1,-4,4$&\texttt{K1987}&Red&Inner&$4,-12,9$\\
		\hline
	\end{tabular}
	\tablefoot{The inner triplet of \texttt{K1987} was captured in 3P1R around $t=5.7\,10^9P_1$ but exited the 3P1R around $t=9.5\,10^9P_1$. In Kepler-1987, the outer triplet is (possibly) in the 3P1R $\left(4,-12,9\right)$. The outer triplet of \texttt{703} was captured in 3P1R without following the evolution of Fig. \ref{fig:Period_ratio}.}
\end{table}

The typical evolution proposed in Fig. \ref{fig:Period_ratio} was followed by some triplets of our numerical simulations (see Fig. \ref{fig:classification}). In Table \ref{tab:3P1R_simulations}, we give all triplets of our simulations that were captured in 3P1R after exiting the 3P0R. In order to assess if some existing triplets followed the typical evolution proposed in Fig. \ref{fig:Period_ratio}, we give in Table \ref{tab:3P1R} triplets from the exoplanet population with the smallest distance to 3P1R. While our pendulum model provides us with an adapted metric to measure the distance to 3P0R, we do not have such metric to measure the distance to 3P1R. A relevant metric derived from the work of \cite{Petit2021} and using our module \texttt{PerHam3pla} in \texttt{celeries} (see Sect. \ref{ssec:Laplace}) could be designed, but it would depend on the unconstrained quantities $e_j$ and $\varpi_j$, and we leave it to a future work. Instead, we use the simple metric \smash{$\Delta_\text{3P1R}=\min\left|k_1n_1+k_2n_2+k_3n_3\right|/n_2$} for $1\leq k_1\leq5$, $1\leq k_3\leq27$, $k_1+k_2+k_3=\pm1$ and $k_1k_2k_3\neq0$. A red triplet is included in Table \ref{tab:3P1R} if $\Delta_\text{3P1R}<0.01$ and if it is in the zone where triplets of our simulations who followed the evolution of Fig. \ref{fig:Period_ratio} end up (bottom of Fig. \ref{fig:classification} with \smash{$2\lesssim K_p\Delta_\text{2P1R}\lesssim20$}).
\begin{table}[h]
	\setlength{\tabcolsep}{2.5pt}
	\renewcommand{\arraystretch}{0.65}
	\caption{Triplets of the exoplanet population possibly in 3P1R.}\label{tab:3P1R}
	\vspace{-2mm}
	\centering
	\begin{tabular}{ccc|ccc}
		\hline
		Name&N$^\circ$&$k_1,k_2,k_3$&Name&N$^\circ$&$k_1,k_2,k_3$\\
		\hline
		\hline
		Kepler-305 bcd&26&$1,-3,3$&K2-268 dec&43&$2,-6,5$\\
		Kepler-968 bcd&45&$1,-6,6$&Kepler-1987 cde&46&$4,-12,9$\\
		Kepler-226 bcd&44&$1,-4,4$&HD-158259 bcd&47&$3,-10,8$\\
		\hline
	\end{tabular}
	\tablefoot{These six triplets are circled and numbered in Fig. \ref{fig:classification}.} 
\end{table}

The outer triplet of K2-138 \citep[efg, number $48$ in Fig. \ref{fig:classification}, \text{e.g.},][]{Cerioni2023,Leleu2026a} is possibly in the 3P1R $2n_1-4n_2+3n_3=0$. Unlike triplets of Table \ref{tab:3P1R}, it did not follow the typical evolution proposed in Fig. \ref{fig:Period_ratio}, but it may have evolved similarly to the outer triplet of simulation \texttt{703}, visible in yellow in Fig. \ref{fig:classification}.



\subsection{Trends in the exoplanet population}\label{ssec:trends}

\begin{figure}[h]
	\centering
	\includegraphics[width=\linewidth]{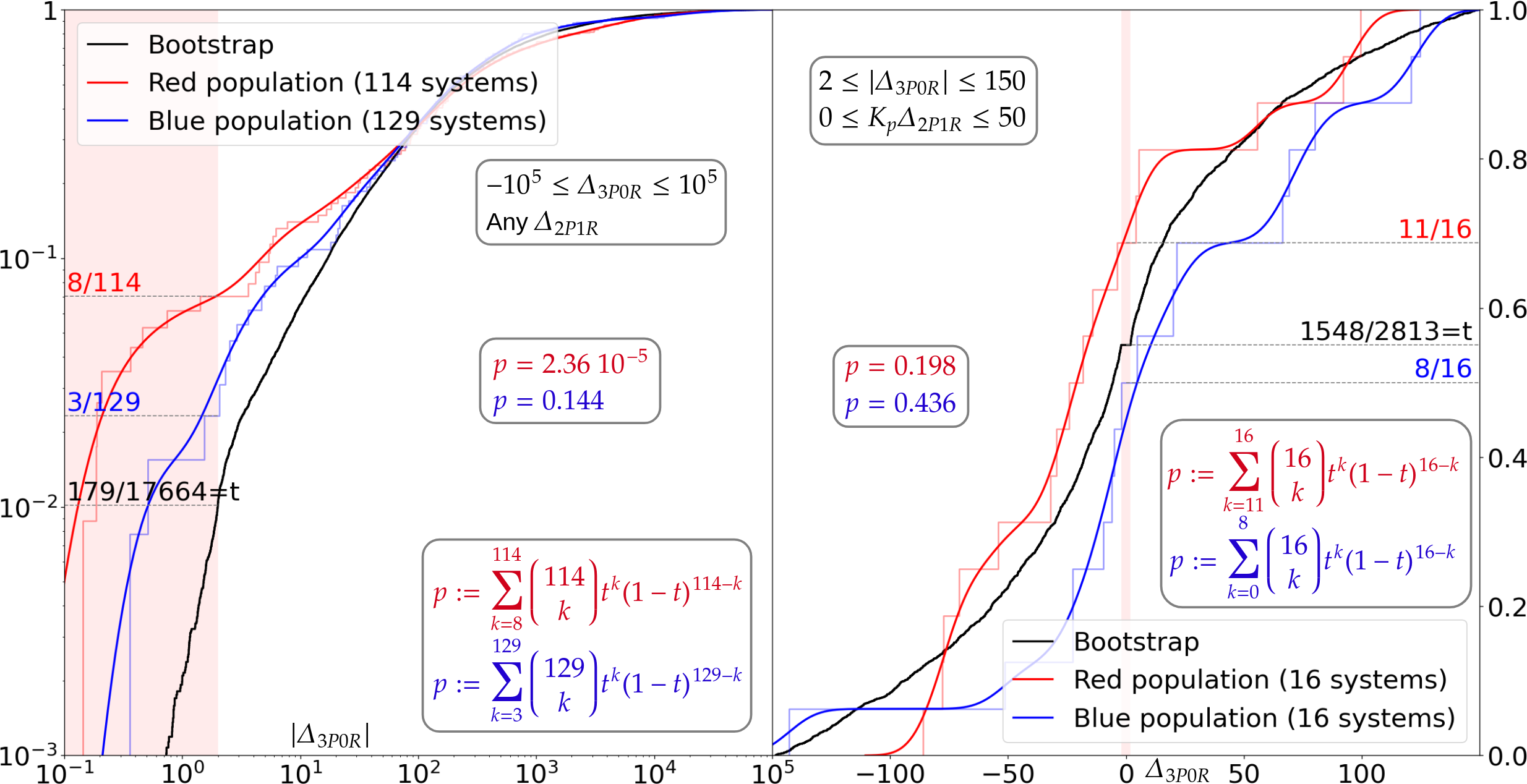}
	\caption{Cumulative distribution of the red ($1/\mathfrak{R}\text{e}\,\sampi\leq10^{10}$ yrs) and blue ($1/\mathfrak{R}\text{e}\,\sampi>10^{10}$ yrs) subpopulations and a bootstrap. Only one triplet per system is considered (smallest \smash{$\Delta_{3P0R}$}).}\label{fig:cumulative}
\end{figure}

In order to identify trends in the population of Fig. \ref{fig:classification}, we plot in Fig. \ref{fig:cumulative} cumulative distributions of the metric \smash{$\Delta_\text{3P0R}$} for the red (tidally affected) and blue (not tidally affected) subpopulations. We compare these two subpopulations with a bootstrapped population of 17664 triplets made by choosing $m_j/m_0$, $\lambda_j$, $P_2/P_1$ and $P_3/P_2$ at random within our global population. When choosing $P_2/P_1$ (\textit{resp}. $P_3/P_2$), the selection was within the inner pair (\textit{resp}. outer pair) of the global population triplets. In order to avoid giving unfair weight to large planetary systems, we only select from each system the triplet with the smallest distance \smash{$|\Delta_{3P0R}|$} to 3P0R. Therefore, only one triplet per system appears in Fig. \ref{fig:cumulative}. We also discard any triplet such that $n_j/n_{j+1}>7.5$.

In the left panel of Fig. \ref{fig:cumulative}, we plot the cumulative distribution of \smash{$|\Delta_{3P0R}|$} for both subpopulations. It shows an overabundance of red systems close to or in 3P0R \citep[\textit{e.g.},][]{GoldbergBatygin2021}, mostly driven by the resonant chains visible inside the 3P0R separatrices in Fig. \ref{fig:classification}. This overabundance is consistent with systems that were initially close to 2P1R, and tidally evolved without large-scale instabilities to their current position. These systems are still in 3P0R because their 3$-$planet angle did not yet cross the 3P0R separatrix.

In order to assert our claim that some systems of the exoplanet population could have followed the typical evolution schematically represented in Fig. \ref{fig:Period_ratio}, we plot in the right panel of Fig. \ref{fig:cumulative} the cumulative distribution of \smash{$\Delta_{3P0R}$} for both the red and blue subpopulation. We limit the statistic to systems outside of 3P0R (\smash{$|\Delta_{3P0R}|>2$}), and we truncate both subpopulations to \smash{$|\Delta_{3P0R}|<150$} and \smash{$0<K_p\Delta_{2P1R}<50$}, since systems who were initially deep in $2P1R$ and followed the evolution of Fig. \ref{fig:Period_ratio} are expected to be within these bounds. The tidally affected subpopulation (red) shows an overabundance of $~69$\% systems below the 3P0R, partly driven by the systems of Table \ref{tab:3P1R}, circled in Fig. \ref{fig:classification}. This overabundance is consistent with a proportion of the red population having followed the evolution proposed in Fig. \ref{fig:Period_ratio}.

In both panels, we computed with a binomial test the probability $p$ that both subpopulations are drawn from the same law as the bootstrap. If that was the case, observing 8 systems in 3P0R in the left population would have a probability \smash{$p=2.36\,10^{-5}$}, confirming that the overabundance of red system in 3P0R in the left panel is statistically significant. However, because of the small sample size, the overabundance of red systems below 3P0R in the right panel is not statistically significant. While it is clear\footnote{Chains such as Trappist-1, K2-138 or TOI-178, along with our numerical simulations, show that the evolution of Fig. \ref{fig:Period_ratio} is real, but the systems of Table \ref{tab:3P1R} do not prove it because we cannot verify that these systems are indeed in 3P1R.} that a non-zero proportion of the exoplanet population was protected from resonance breaking by tides and rather followed (or is following) the path of Fig. \ref{fig:Period_ratio}, Fig. \ref{fig:cumulative} indicates that this proportion is small. In the left panel, 8 red systems are in 3P0R, whereas only $\sim1.2$ system is expected from the bootstrap. This excess of $\sim7$ systems represents 6\% of the red population. While it misses systems that already exited 3P0R, the right panel shows that $\lesssim2$ systems are concerned. The evolution of Fig. \ref{fig:Period_ratio} therefore concerns 5 to 10\% of the tidally affected systems, and is not a major evolution mechanism. This is consistent with \cite{Leleu2026b}, who found that if the bulk of close-in systems formed in chains of resonance, most broke before tides played a role. This contrasts with our numerical simulations which mostly follow the evolution of Fig. \ref{fig:Period_ratio}. On one hand, some of them start nearly on the libration centers (\textit{e.g.}, \texttt{pq13} and \texttt{pq23}) and on the other hand, their tides are stronger than physical to limit the runtime (Sect. \ref{ssec:simulations}), which can damp the eccentricities before instabilities can develop.

\section{Pendulum model of 3P0R}\label{sec:pendulum}

In this section, we elaborate a zero-parameter 1-dof pendulum-like model of 3P0R and we introduce tidal dissipation in the model. We consider a planar $3$-planet system with a star of mass $m_0$ orbited by three planets of masses $m_j$ and radii $R_j$, for $1\leq j\leq3$. The planets are ordered by increasing distance to the star, and we denote \smash{$\iota=\sqrt{-1}$}.
The notations already introduced in previous sections are reused. We consider in this section that the system is far from all 2PR, but close
to the 3P0R $pn_1-\left(p + q\right)n_2 + qn_3 = 0$. The 3-planet angle is $\varphi=p\lambda_1-\left(p+q\right)\lambda_2+q\lambda_3$ and the Laplace angle is $\varphi'=\varphi/\gcd(p,q)$.

\subsection{Change of variables adapted to the resonance}\label{ssec:change_var}

For a planar system, we make use of the canonical Poincar\'e variables \smash{$\left(\Lambda_j, D_j; \lambda_j, -\varpi_j\right)$}, where $\lambda_j$ is the mean longitude of planet $j$ and $\varpi_j$ is its longitude of the periapsis. Refer to Eq. (\ref{eq:Poincare}) of Sect. \ref{sec:second_mass} for the definition of \smash{$\Lambda_j$} and \smash{$D_j$}. We define nominal values for the semi-major axis $a_j$ and \smash{$\Lambda_j$} with
\begin{equation}\label{eq:nominalLbd}
	\Lambda_{j,0}=\beta_j\sqrt{\mu_ja_{j,0}}\;\;\;\text{with}\;\;\;\mu_j=n_{j,0}^2a_{j,0}^3,
\end{equation}
where the nominal mean motions $n_{j,0}$ verify
\begin{equation}\label{eq:n_j0}
	pn_{1,0}-\left(p+q\right)n_{2,0}+qn_{3,0}=0.
\end{equation}
Averaging over the $2$-planet angles $p\lambda_1-\left(p+1\right)\lambda_2$ and $\left(q-1\right)\lambda_2-q\lambda_3$ is a good approximation only if the super frequencies $pn_1-\left(p+1\right)n_2$ and $\left(q-1\right)n_2-qn_3$ are fast with respect to the libration frequency of $\varphi$ (Eq. (\ref{eq:Imsampi}) in Sect. \ref{sssec:eigenvalues_pendulum}). When $\Delta\to0$, the libration frequency of $\varphi$ is \smash{$\OO(\varepsilon\Delta^{-1/2})$}, and the super frequencies are $\OO(\Delta)$. Therefore, averaging over the 2-planet angles is possible only if \smash{$\Delta\gg\OO\left(\varepsilon^{2/3}\right)$}, or equivalently, \smash{$\Delta_\text{2P1R}\gg\OO(1)$}. In that case, an appropriate transformation is \citep{Petit2020}
\begin{equation}\label{eq:change_var_far_angle}
	\begin{pmatrix}\varphi\\\varphi_2\\\varphi_3\end{pmatrix}=
	\begin{pmatrix}
		p&-\left(p+q\right)&q\\
		0&1&-1\\
		0&0&1
	\end{pmatrix}
	\begin{pmatrix}\lambda_1\\\lambda_2\\\lambda_3\end{pmatrix},
\end{equation}
which is made canonical via
\begin{equation}\label{eq:change_var_far_action}
	\begin{pmatrix}\Delta\Phi\\\Delta\Gamma\\\Delta\Upsilon\end{pmatrix}\!=\!
	\begin{pmatrix}
		1/p&\!\!\!\!0&\!\!\!\!0\\
		\frac{p+q}{p}&\!\!\!\!1&\!\!\!\!0\\
		1&\!\!\!\!1&\!\!\!\!1
	\end{pmatrix}
	\begin{pmatrix}\Delta\Lambda_1\\\Delta\Lambda_2\\\Delta\Lambda_3\end{pmatrix}\Leftrightarrow
	\begin{pmatrix}\Delta\Lambda_1\\\Delta\Lambda_2\\\Delta\Lambda_3\end{pmatrix}\!=\!
	\begin{pmatrix}
		\!p&\!\!\!\!\!\!0&\!\!\!\!0\\
		\!-\left(p\!+\!q\right)&\!\!\!\!\!\!1&\!\!\!\!0\\
		\!q&\!\!\!\!\!\!\!-1&\!\!\!\!1
	\end{pmatrix}
	\begin{pmatrix}\Delta\Phi\\\Delta\Gamma\\\Delta\Upsilon\end{pmatrix},
\end{equation}
where we wrote $\Lambda_j=\Lambda_{j,0}+\Delta\Lambda_j$. In these variables, the Hamiltonian of Eq. (\ref{eq:Ham_1dof}) does not depend on $\varphi_2$ and $\varphi_3$, and $\Delta\Gamma$ and $\Delta\Upsilon$ are constant. We choose the nominal mean motions $n_{j,0}$ in Eq. (\ref{eq:n_j0}) so that $\Delta\Gamma=\Delta\Upsilon=0$.

\subsection{Reduction to one degree of freedom}\label{sssec:to_1dof}

For our 3P0R model, we consider that the system is far from all 2P1R ($\Delta\gg\varepsilon^{2/3}$). Once averaged over all fast multi-planet angles and expanded to second order in $\varepsilon$, the Hamiltonian reads
\begin{equation*}
	\Ham=\Ham_K(\Delta\Lambda_j)+\varepsilon \Ham_{1}(\Delta\Lambda_j,D_j;-\varpi_j)+\varepsilon^2 \Ham_2(\Delta\Lambda_j,D_j;\lambda_j,-\varpi_j).
\end{equation*}
We assume that the \smash{$\Lambda_j$} remain close to their nominal values \smash{$\Lambda_{j,0}$} defined in Eq. (\ref{eq:nominalLbd}) and we expand the Hamiltonian in \smash{$\Delta\Lambda_j$}. In a general fashion, the expansion is pushed to second order for $\Ham_K$ but truncated to order zero for the perturbation. The perturbation \smash{$\varepsilon \Ham_{1}$} (\textit{resp}. \smash{$\varepsilon^2\Ham_2$}) is expanded to degree $1$ (\textit{resp}. $0$) in eccentricity. The Hamiltonian is now
\begin{equation}
	\Ham=\Ham_K(\Delta\Lambda_j)+\varepsilon \Ham_{1}(\Lambda_{j,0})+\varepsilon^2 \Ham_2(\Lambda_{j,0};\lambda_j)+\OO(\varepsilon^3).
\end{equation}
The secular term \smash{$\varepsilon \Ham_{1}$} is constant and removed. Using Eq. (\ref{eq:Keplerian}), the Keplerian Hamiltonian is
\begin{equation}
	\Ham_K=\sum_{1\leq j\leq3}n_{j,0}\Delta\Lambda_j-\frac{3}{2}\sum_{1\leq j\leq3}\frac{n_{j,0}}{\Lambda_{j,0}}\Delta\Lambda_j^2+\OO\left(\Delta\Lambda_j^3\right).
\end{equation}
Injecting the change of variables of Eqs. (\ref{eq:change_var_far_angle}) and (\ref{eq:change_var_far_action}), and using $\Delta\Gamma=\Delta\Upsilon=0$, it takes the form
\begin{equation}\label{eq:K}
	\Ham_K=-\frac{3}{2}K\Delta\Phi^2,\;\;\;K=\frac{n_{1,0}}{\Lambda_{1,0}}p^2\!+\!\frac{n_{2,0}}{\Lambda_{2,0}}\left(p\!+\!q\right)^2\!+\!\frac{n_{3,0}}{\Lambda_{3,0}}q^2.
\end{equation}
Using the $3$-planet d'Alembert rule (Eqs. \ref{eq:dAlembert3pla}), the perturbation at second order in mass and order zero in eccentricity reads
\begin{equation}\label{eq:H2_complete}
	\varepsilon^2\Ham_2=-\frac{m_1m_2}{m_0^2}n_{3,0}\Lambda_{3,0}\sum_{k\in\mathbb{N}}\Xi_k\cos k\varphi',
\end{equation}
where we recall that $\varphi'=\varphi/\gcd(p,q)$.  The term $k=0$ is constant and can be discarded, but infinitely many terms remain in the Hamiltonian. Because of the exponential decay of the Laplace coefficients, $\Xi_k$ goes to zero exponentially quickly when $k$ goes to infinity.
\begin{table}[h]
	\setlength{\tabcolsep}{4.2pt}
	\renewcommand{\arraystretch}{0.5}
	\caption{Exact values of $\Xi_1$ to $\Xi_5$ on one locus of the 3P0R.}\label{tab:Xi_gcd}
	\vspace{-2mm}
	\centering
	\begin{tabular}{cccccccc}
		\hline
		Chain&$p$&$q$&$\Xi_1/2$&$\Xi_2/2$&$\Xi_3/2$&$\Xi_4/2$&$\Xi_5/2$\\
		\hline
		\hline
		1:2 -- 2:3&1&3&\textcolor{red}{$19.874$}&$0.2740$&$0.0321$&$0.0054$&$0.0010$\\
		2:3 -- 1:2&2&2&$-0.253$&\textcolor{red}{$42.946$}&$0.8555$&$0.1833$&$0.0534$\\
		2:3 -- 2:3&2&3&\textcolor{red}{$74.556$}&$0.3617$&$0.0390$&$0.0059$&$0.0010$\\
		3:4 -- 2:3&3&3&$-8.316$&$-59.90$&\textcolor{red}{$70.429$}&$3.7389$&$1.0943$\\
		3:4 -- 3:4&3&4&\textcolor{red}{$99.726$}&$0.6333$&$0.0686$&$0.0104$&$0.0018$\\
		4:5 -- 1:2&4&2&$-1.752$&\textcolor{red}{$36.366$}&$1.0120$&$0.2152$&$0.0620$\\
		\hline
	\end{tabular}
	\tablefoot{The locus is $\Delta\!=\!10^{-2}$ and $pn_1\!-\!\left(p\!+\!q\right)n_2\!+\!qn_3\!=\!0$. $\Xi_{\gcd(p,q)}$ is red.}
\end{table}
In Table \ref{tab:Xi_gcd}, we show $\Xi_k$ ($1\leq k\leq 5$) for some chains, on one locus of the 3P0R. For most 3P0R, \smash{$\Xi_{\gcd(p,q)}$} is much larger that the other coefficients and we simplify \smash{$\varepsilon^2 \Ham_2$} to
\begin{equation}\label{eq:H2}
	\varepsilon^2 \Ham_2=-\frac{m_1m_2}{m_0^2}n_{3,0}\Lambda_{3,0}\Xi\cos\varphi,
\end{equation}
where $\Xi=\Xi_{\gcd(p,q)}$ and $\varphi=p\lambda_1-\left(p+q\right)\lambda_2+q\lambda_3$. When $\Xi_{\gcd(p,q)}$ is not the largest coefficient, we retain the largest harmonic in absolute value in Eq. (\ref{eq:H2_complete}). Sometimes, several harmonics can have the same order of magnitude and could all be considered while still having a 1-dof model, but the Hamiltonian (\ref{eq:0param}) would be more complicated with dependency on a parameter. For simplicity, we only keep the largest harmonic in Eq. (\ref{eq:H2_complete}). We introduced a minus sign in Eq. (\ref{eq:H2_complete}) so that $\Xi$ has the same sign as $\Delta$ (Eq. \ref{eq:rho}). Note that $\Xi$ in Table \ref{tab:Xi_gcd},
is highly dependent on $\Delta$ because of denominators $\left(pn_1-\left(p+1\right)n_2\right)^{-1}$ and $\left(\left(q-1\right)n_2-qn_3\right)^{-1}$ that both diverge to $\infty$ when $\Delta\to0$.
For $\Delta\neq0.01$, $\Xi$ in Eq. (\ref{eq:H2}) can be obtained exactly in python using \texttt{Xi=-2*h3.PerHam3pla(degree=0).angle((p,-p-q,q))}\footnote{See this \href{https://github.com/adleleu/resonantstate/blob/main/src/resonantstate/ell_to_3P0R.py}{notebook} for the conversion elliptic elements \smash{$\to\;\left(\Sigma, \sigma, \delta, \Xi\right)$}}.
 
\href{https://www.astro.unige.ch/~delisle/celeries/doc/_autosummary/celeries.perham3pla.PerHam3pla.html#celeries.perham3pla.PerHam3pla}{\texttt{PerHam3pla}} is a module that we implemented in the python algebraic manipulator \href{https://gitlab.unige.ch/delisle/celeries}{\texttt{celeries}} able to analytically expand the Hamiltonian to second order in mass at any order in eccentricity and inclination. A comprehensive framework of this expansion, that makes 3PR appear, is provided in Appendix \ref{sec:second_mass}, and the implementation details of \texttt{PerHam3pla} are given in Sect. \ref{ssec:Laplace}. We obtain the 1-dof Hamiltonian
\begin{equation}\label{eq:Ham_1dof}
	\Ham(\Delta\Phi;\varphi)=-\frac{3}{2}K\Delta\Phi^2-\frac{m_1m_2}{m_0^2}n_{3,0}\Lambda_{3,0}\Xi\cos\varphi,
\end{equation}
which is a simple pendulum, as was noticed by \citet{Quillen2011} and \citet{Petit2020}.

\subsection{Eigenvalues and resonance width}\label{sssec:eigenvalues_pendulum}

The two fixed points of this pendulum are \smash{$\left(\Delta\Phi,\varphi\right)=\left(0,\varphi_\text{eq}\right)$}, where \smash{$\varphi_\text{eq}\in\left\lbrace0,\pi\right\rbrace$}. When $\Delta>0$ (that is, $\Xi>0$), \smash{$\varphi_\text{hyp}=0$} is hyperbolic, with eigenvalues $\pm\lambda$, and \smash{$\varphi_\text{ell}=\pi$} is elliptic, with eigenvalues $\pm\iota\lambda$, where
\begin{equation}\label{eq:Imsampi}
	\lambda=\mathfrak{I}\text{m}\,\sampi=\!\sqrt{3\frac{m_1m_2}{m_0^2}Kn_{3,0}\Lambda_{3,0}\Xi}=\OO(\varepsilon\Xi^{1/2})\!=\!\OO(\varepsilon\Delta^{-1/2}),
\end{equation}
is the libration frequency in the vicinity of the elliptic fixed point. When $\Delta<0$, the elliptic fixed point corresponds to \smash{$\varphi_\text{ell}=0$} instead, and \smash{$\varphi_\text{hyp}=\pi$} is hyperbolic. The half-width \smash{$\delta\Phi$} of the resonance is obtained from \smash{$\Ham(0,\varphi_\text{hyp})=\Ham(\delta\Phi,\varphi_\text{ell})$}. It reads\footnote{Beware that Eq. (51) of \citet{Petit2020} has a factor $\sqrt{2}$ too much.}
\begin{equation}\label{eq:3P0R_width}
	\delta\Phi=2\sqrt{\frac{m_1m_2}{m_0^2}\frac{n_{3,0}\Lambda_{3,0}}{3K}\Xi}.
\end{equation}
For each 3P0R in Fig. \ref{fig:3P0R_width}, we obtain the mean motions $n_j$ at the resonance center with $pn_1-\left(p+q\right)n_2+qn_3=0$, we convert to semi-major axes $a_j$ and then to $\Delta\Lambda_j$ and finally to the coordinates \smash{$\left(\Delta\Phi,\Delta\Gamma,\Delta\Upsilon\right)$} using Eq. (\ref{eq:change_var_far_action}). The upper and lower separatrices of the 3P0R are obtained by converting \smash{$\left(\Delta\Phi\pm\delta\Phi,\Delta\Gamma,\Delta\Upsilon\right)$} back to the mean motions $n_j$. In Fig. \ref{fig:3P0R_width}, all points within the separatrices of a 3P0R are then red-shaded. In order to identify points within the separatrices of a 2P1R, we use the intersection of the 2P1R separatrix with $e_r=0$ in the top-right panel of Fig. 4 of \cite{Leleu2026b}. We green-shade the corresponding points.

\subsection{Reduction to no parameter}

It is possible to use techniques described by \citet{Henrard1983} in order to remove all parameters in the Hamiltonian of (Eq. \ref{eq:Ham_1dof}). We perform the normalization
\begin{equation}\label{eq:Renorm_0param}
	\Kam=-\Ham/\omega,\;\;\Sigma=\Delta\Phi/B,\;\;\sigma=\varphi,\;\;\tau=-\Omega t,
\end{equation}
where $t$ is the time and $B$, $\omega$ and $\Omega$ are constants. We choose $B$ and $\omega$ in such a way that the Hamiltonian is (Eq. \ref{eq:Ham_3P0R})
\begin{equation}\label{eq:0param}
	\Kam(\Sigma;\sigma)=\frac{1}{2}\Sigma^2\pm\cos\sigma,
\end{equation}
where the $\pm$ sign matches the sign of $\Delta$ and $\Xi$. This gives \smash{$\omega=\pm m_1m_2n_{3,0}\Lambda_{3,0}\Xi/m_0^2$} and \smash{$B=\sqrt{\omega/(3K)}$}. Choosing $\Omega=\omega/B$, we recover the usual form of the Hamilton equations \smash{$d\Sigma/d\tau=-\partial_\sigma\Kam$} and \smash{$d\sigma/d\tau=\partial_\Sigma\Kam$}.

\subsection{Fixed points and libration centers}\label{ssec:libration}

\begin{figure*}[h]
	\centering
	\includegraphics[width=\linewidth]{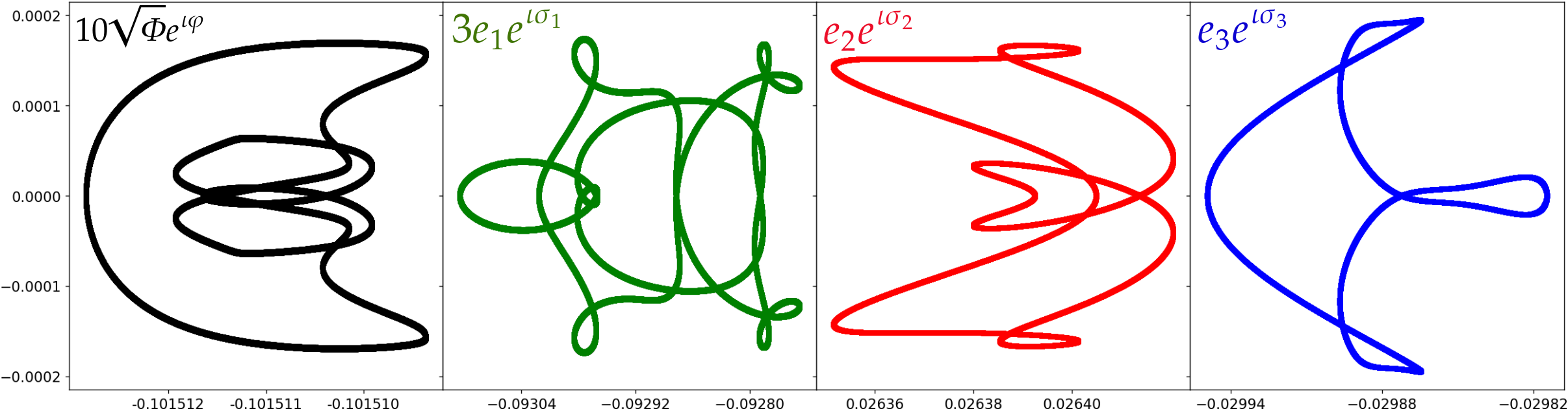}
	\caption{A libration center of the 3P0R $n_1-4n_2+3n_3=0$ in the complex plane with the masses of Fig. \ref{fig:stability_map}. The locus of the 3P0R is close to the 2P1R ($n_1/n_2=2.00096$ and $n_2/n_3=1.50078$). The trajectories are periodic with period $5.9976\times2\pi/n_1$. Quantities not invariant by rotation are quasiperiodic with periods $\left(5.9976,-2270.5\right)\times2\pi/n_1$. This initial condition is on the $x$-axis of Fig. \ref{fig:stability_map}.}\label{fig:LibCen123}
\end{figure*}
In order to describe Fig. \ref{fig:stability_map} further, it is useful to introduce libration centers. We consider that the system can be close to the 2P1R $pn_1-\left(p+1\right)n_2=0$ and $\left(q-1\right)n_2-qn_3=0$, and we use the following transformation \citep{Delisle2017}
\begin{equation}\label{eq:change_var_close_angle}
		\begin{pmatrix}\varphi\\\varphi_2\\\varphi_3\end{pmatrix}=
		\begin{pmatrix}
			p&-\left(p+q\right)&q\\
			0&1&-1\\
			0&1-q&q
		\end{pmatrix}
		\begin{pmatrix}\lambda_1\\\lambda_2\\\lambda_3\end{pmatrix},
\end{equation}
adapted to \smash{$\Delta\lesssim\OO(\varepsilon^{2/3})$}, along with $\sigma_j=\varphi_3-\varpi_j$. The action are transformed in a canonical way. A fixed point of the averaged Hamiltonian corresponds to \smash{$\left(\Phi;\varphi\right)$} and \smash{$\left(D_j;\sigma_j\right)$} being constant. The angles $\varphi_2$ and $\varphi_3$ are absent from the Hamiltonian (because there are averaged upon and not invariant by rotation, respectively) and are not constant at a fixed point.

However, the complete (unaveraged) Hamiltonian does not have fixed point ($N$ masses cannot stay at rest when $N\geq2$) and along the flow of the complete Hamiltonian, \smash{$\left(\Phi;\varphi\right)$} and \smash{$\left(D_j;\sigma_j\right)$} are not constant. Let us assume that in the vicinity of a fixed point of the averaged Hamiltonian, a variable $f(t)$ of the complete Hamiltonian behaves quasiperiodically as
\begin{equation}\label{eq:quasiperiodic}
	f(t)=\sum_{\vect{k}\in\mathbb{Z}^6}f_\vect{k}e^{\iota \vect{k}\cdot\vect{\omega}t},
\end{equation}
where \smash{$\vect{\omega}=\left(\omega_1,\omega_2,\omega_3,\omega_1',\omega_2',\omega_3'\right)$} are the fundamental frequencies of the complete Hamiltonian. $\omega_1$ is approximated by Eq. (\ref{eq:Imsampi}). $\omega_2$ and $\omega_3$ are the circulation frequencies of $\varphi_2$ and $\varphi_3$, while \smash{$\omega_j'$} is the libration frequency of $\sigma_j$.

We call "libration centers" of the complete Hamiltonian points in the phase space that are elliptic fixed points of the averaged Hamiltonian. Because the averaged Hamiltonian is obtained from the complete Hamiltonian by averaging over $\varphi_2$, quantities invariant by rotation, that is, quantities that are variables of the complete Hamiltonian, such as \smash{$\left(\Phi;\varphi\right)$}, \smash{$\left(\Gamma;\varphi_2\right)$} and \smash{$(D_j;\sigma_j)$} are periodic with frequency\footnote{More precisely, $\omega_2$ is an harmonic of the fast frequency $\omega_\text{fast}=\omega_2/D$, where \smash{$D=\lcm(p_j)/p_{N-1}-\lcm(p_j)/p_N$} and the chain is $p_1:\cdots:p_N$.} $\omega_2$ at a libration center. They verify $\vect{k}=\left(0,k,0,0,0,0\right)$ for $k\in\mathbb{Z}$ in Eq. (\ref{eq:quasiperiodic}). Quantities not invariant by rotation, such as $\varphi_3$, are quasiperiodic and verify $\vect{k}=\left(0,k_1,k_2,0,0,0\right)$ for $k_1, k_2\in\mathbb{Z}$ in Eq. (\ref{eq:quasiperiodic}), but these quantities are not variables of the complete Hamiltonian.

Therefore, elliptic fixed points of the averaged Hamiltonian are periodic orbits of the complete Hamiltonian. However, the fixed points of the averaged model Hamiltonian do not coincide with the libration centers of the complete Hamiltonian, because of truncated expansions (in eccentricities, masses and $\Lambda_j$) and because of the inherent change of variables associated with the averaging process (see Sect. \ref{ssec:Lie}).

A method to find them numerically from a fixed point is to first integrate the complete Hamiltonian. Then the Fourier expansion of Eq. (\ref{eq:quasiperiodic}) is computed for all variables of the Hamiltonian. Finally, all terms such that $\vect{k}\neq\left(0,k,0,0,0,0\right)$ are removed in order to generate a new initial conditions from which the process is iterated \citep{CoRoCo2022,CoLaCo2010}.

We implemented this iterative method in \href{https://github.com/Jeremycouturier/Aptidal}{\texttt{Aptidal}} in the general $N$-planet case. The implementation details of our libration center finder, which makes use of the NAFF algorithm of \citet{Laskar2003}, are given in Appendix \ref{ssec:NAFF}. As an example, we show in Fig. \ref{fig:LibCen123} a libration center of the complete Hamiltonian in the case $\left(p,q\right)=\left(1,3\right)$, corresponding to the chain $1:2:3$.

\subsection{About the stability map}\label{ssec:about_stability_map}

The stability map presented in Fig. \ref{fig:stability_map} is made of $2414\times1079$ systems for which chaos must be measured. Different techniques exist to evaluate how chaotic an orbit is. Here, we choose the method of diffusion of the fundamental frequencies \citep[][]{Laskar1988}. More precisely, for each pixel of Fig. \ref{fig:stability_map}, we integrated the complete Hamiltonian with \texttt{Aptidal} for a time $16\times\max\left(2\pi/\lambda,2\pi/\omega_3\right)$ ($13000$ to $77000$ inner periods, depending on the pixel), where $\lambda$ (Eq. \ref{eq:Imsampi}) is the libration frequency of $\varphi$ far from 2P1R and $\omega_3$ is the precession frequency of the periapses, obtained while searching the family of libration centers, that is, the $x$-axis of Fig. \ref{fig:stability_map} (see Sect. \ref{ssec:libration}).

For each half of the integration, we evaluated with high precision the fundamental frequencies $n_j$ associated with the mean longitudes $\lambda_j$, for $1\leq j\leq3$, using the NAFF procedure of Laskar. That is, we computed the frequency $n_j$ that extremizes \smash{$\left<\cos\lambda_j,e^{\iota n_j t}\right>$}, where the dot product \smash{$\left<\cdot,\cdot\right>$} is defined in Eq. (\ref{eq:dot_product}). To this aim, we applied a Newton-Raphson method on \smash{$d\left<\cos\lambda_j,e^{\iota n_j t}\right>/dn_j$} using as first guess for $n_j$ a linear fit of $\lambda_j$. A rather small timestep is needed so the dot product \smash{$\left<\cdot,\cdot\right>$}, which is an integral, can be computed properly, and we used a timestep \smash{$2^{-5}$}. With such a timestep, a $\mathcal{SABA}_2$ integrator is enough (see Sect. \ref{ssec:SABA}), and we made this choice to save computational time. We used a Hanning filter $\chi_4(t)$ of order 4 \citep[see][Proposition 2]{Laskar2003} to improve the determination of the fundamental frequencies. For each pixel, we ended up with $n_{j,1}$ for the first half of the trajectory and $n_{j,2}$ for the second half, and we defined \citep[][]{RobutelGabern2006,CoRoCo2022}
\begin{equation}
	\mathcal{D}=\frac{1}{3}\left(\frac{\left|n_{1,1}-n_{1,2}\right|}{n_{1,1}}+\frac{\left|n_{2,1}-n_{2,2}\right|}{n_{2,1}}+\frac{\left|n_{3,1}-n_{3,2}\right|}{n_{3,1}}\right).
\end{equation}
The diffusion $\mathcal{D}$ is the color-axis of Fig. \ref{fig:stability_map}, which took $45$ hours to produce on $36$ cores. A function \texttt{FundamentalFrequency} is now publicly available in \texttt{Aptidal} (see Sect. \ref{ssec:NAFF}) and allows similar figures to be easily produced. The masses of Fig. \ref{fig:stability_map} are $m_j/m_0=\left(1.64,2.32,2.01\right)\times10^{-5}$, as in simulation \texttt{pq13}.

\subsection{Tidal dissipation}\label{ssec:tides}

In order to implement tidal dissipation in our pendulum model of 3P0R, we consider tides raised by the star on the planets and interacted with by the star, as all other tidal contributions are negligible \citep[][Sect. 3.1]{CoRoCo2021}. The tidal potential due to the redistribution of mass within planet $j$ reads
\begin{equation}\label{perturbedS}
	V(\vect{r}_j)=-\frac{1}{2}\kappa_2^{(j)}\frac{\G m_0 R_j^5}{r_j^3r_j^{\star3}}\left(3\cos^2\left(w_j-w_j^\star-\theta_j+\theta_j^\star\right)-1\right),
\end{equation}
where $w_j$ is the true longitude, $\theta_j$ is the sideral angle, \smash{$\kappa_2^{(j)}$} is the second Love number, \smash{$f^\star=f\left(t-\Delta_t^{(j)}\right)$} for any quantity $f$, and \smash{$\Delta_t^{(j)}$} is the tidal timelag. Refer to \cite{Couturier2026}, Sect. 2.1.3 or to \href{https://jeremycouturier.com/img/PDF/PoincareTides.pdf}{this document} for a comprehensive tidal framework. Refer to \cite{BoueCorreiaLaskar2019} for a rigorous and modern mathematical description of tides.

Tides are included by adding the potential \smash{$H_j=m_0V(\vect{r}_j)$} to the Hamiltonian. The resulting scalar field is pseudo-Hamiltonian because it depends on past times (in that case on time \smash{$t-\Delta t^{(j)}$}). However, the equations of motions with tidal dissipation can still be obtained from the Hamilton equations if the starred variables are kept constant while differentiating (because \smash{$\vect{r}_j^\star$} is not a variable but a time-dependent forcing).

The literature lacks an expression of the equations of motion in the constant timelag model using Poincar\'e's rectangular coordinates \smash{$\left(\Lambda_j,x_j;\lambda_j,-\iota\bar{x}_j\right)$}, defined in Eq. (\ref{eq:Poincare}) and below, and we now fill this void. Specifically, we write the equations of motions in the non-canonical Poincar\'e variables \smash{$X_j=\sqrt{2/\Lambda_j}\,x_j$}. We assume a coplanar problem, where planets have zero obliquity, even though \cite{Millholland2019} showed that captures into Cassini states can lock the obliquities to high values. By conservation of the total angular momentum, the sideral rotation $\theta_j$ of planet $j$ is also a variable of the problem. Its conjugated action is \smash{$\Theta_j=\alpha_jm_jR_j^2\dot{\theta}_j$}, where \smash{$\alpha_jm_jR_j^2$} is the principal moment of inertia.
The kinetic energy \smash{$T_j=\Theta_j^2/\left(2\alpha_jm_jR_j^2\right)$} of planetary rotation is added to the Hamiltonian. In order to work with dimensionless variables, we define
\begin{equation}
	\Lam_j=\!\frac{\Lambda_j}{\Lambda_{j,0}}\!=\left(\frac{n_j}{n_{j,0}}\right)^{-1/3}\!\!\!\!\!,\tilde{\Theta}_j=\frac{\Theta_j}{\Lambda_{j,0}},X'_j=\!\sqrt{\frac{2}{\Lambda_{j,0}}}\,x_j=\Lam_j^{1/2}X_j,
\end{equation}
and we renormalize the tidal Hamiltonian with \smash{$\Ham_j=H_j/\Lambda_{j,0}$}. The tidal equations of motion are
\begin{equation*}
	\dot{\Lam}_j=-\drond{\Ham_j}{\lambda_j},\dot{\lambda}_j=\drond{\Ham_j}{\Lam_j},\dot{X}_j'=-2\iota\drond{\Ham_j}{\bar{X}_j'},\dot{\tilde{\Theta}}_j=-\drond{\Ham_j}{\theta_j},\dot{\theta}_j=\drond{\Ham_j}{\tilde{\Theta}_j},
\end{equation*}
all these partial derivatives being computed while keeping the starred quantities constant. The expression of \smash{$\Ham_j$} averaged over the mean motion of planet $j$ is
\begin{equation*}
	\Ham_j\!=\!-n_{j,0}q_j\frac{m_0}{m_j}\Lam_j^{-6}\Lam_j^{\star-6}\left(\frac{1}{4}\!+\!\frac{3}{4}\cos2(\lambda_j\!-\!\lambda_j^\star\!-\!\theta_j\!+\!\theta_j^\star)\!+\!\!\sum_{k=1}^{+\infty}\!\Xi_{2k}^{(j)}\right),
\end{equation*}
where
\smash{$q_j=\kappa_2^{(j)}R_j^5/a_{j,0}^5$}. The coefficients \smash{$\Xi_{2k}^{(j)}$} of the expansion in eccentricity are given by \citep[][Eq. (42) and Appendix B]{CoRoCo2021} up to $k=2$ (beware that $\mathcal{R}_j$ and $X_j$ are written in this work instead of $\Lam_j$ and $X_j'$).

\subsection{Equations of motion with tides}\label{ssec:eq_tides}

Once the equations of motions are obtained, the starred variables are given as a function of the non-starred variables. Because \smash{$\Delta_t^{(j)}$} is much smaller than the secular timescales, we make the substitution \smash{$\Lam_j^\star=\Lam_j$} and \smash{$X_j'^\star=X_j'$}. We define \smash{$Q_j^{-1}=n_{j,0}\Delta_t^{(j)}$} and \smash{$\omega_j=\dot{\theta}_j/n_{j,0}$} and a first-order expansion in \smash{$\Delta_t^{(j)}$} yields
\begin{equation}
	\begin{split}
		&\lambda_j-\lambda_j^\star=n_j\Delta_t^{(j)}=\frac{n_j}{n_{j,0}}n_{j,0}\Delta_t^{(j)}=\Lam_j^{-3}Q_j^{-1},\\
		&\theta_j-\theta_j^\star=\dot{\theta}_j\Delta_t^{(j)}=\omega_jQ_j^{-1}.
	\end{split}
\end{equation}
We obtain, for the regular Poincaré coordinates
\begin{equation}\label{differential}
	\begin{split}
		&\dot{X}_j=-3n_{j,0}\frac{q_j}{Q_j}\frac{m_0}{m_j}\Lam_j^{-13}X_j\left(\sum_{n=1}^{+\infty}\left(p_{2n}^{(j)}-p_{2n}\iota Q_j\right)X_j^{n-1}\bar{X}_j^{n-1}\right),\\
		&\dot{\lambda}_j=6n_{j,0}q_j\frac{m_0}{m_j}\Lam_j^{-13}\left(1+\sum_{n=1}^{+\infty}u_{2n}X_j^n\bar{X}_j^n\right),\\
		&\dot{\Lambda}_j=-3n_{j,0}\Lambda_{j,0}\frac{q_j}{Q_j}\frac{m_0}{m_j}\Lam_j^{-12}\left(\Lam_j^{-3}-\omega_j+\sum_{n=1}^{+\infty}k_{2n}^{(j)}X_j^n\bar{X}_j^n\right),\\
		&\ddot{\theta}_j=3n_{j,0}^2\alpha_j^{-1}\frac{q_j}{Q_j}\frac{a_{j,0}^2}{R_j^2}\frac{m_0}{m_j}\Lam_j^{-12}\left(\Lam_j^{-3}-\omega_j+\sum_{n=1}^{+\infty}h_{2n}^{(j)}X_j^n\bar{X}_j^n\right),
	\end{split}
\end{equation}
where we recall that \smash{$q_j=\kappa_2^{(j)}R_j^5/a_{j,0}^5$} and
\begin{equation*}
		p_{2n}^{(j)}\!=\!p_{2n}'\Lam_j^{-3}-p_{2n}''\omega_j,\;k_{2n}^{(j)}\!=\!k_{2n}'\Lam_j^{-3}-k_{2n}''\omega_j,\;h_{2n}^{(j)}\!=\!h_{2n}'\Lam_j^{-3}-h_{2n}''\omega_j,
\end{equation*}
Using \texttt{celeries}, we computed the equations of motions up to order $9$ in eccentricity and stored the corresponding coefficients in Table \ref{tab:coefficients}. Order $0$ can be directly read in Eqs. (\ref{differential}).
The conservation of the total angular momentum reads \smash{$h_{2n}^{(j)}-k_{2n}^{(j)}+p_{2n}^{(j)}=0$} \citep[][Appendix D]{CoRoCo2021},
and Eqs. (\ref{differential}) preserve the total angular momentum at all orders in eccentricity.
\begin{table}[h]
	\setlength{\tabcolsep}{3.6pt}
	\renewcommand{\arraystretch}{1.25}
	\caption{Coefficients of Eqs. (\ref{differential}) up to order $9$ in eccentricity}\label{tab:coefficients}
	\vspace{-2mm}
	\centering
	\begin{tabular}{ccccc|ccccc}
		\hline
		\hline
		$u_0$&$u_2$&$u_4$&$u_6$&$u_8$&$p_2$&$p_4$&$p_6$&$p_8$&$p_{10}$\\
		$1$&$\frac{65}{8}$&$\frac{455}{16}$&$\frac{525}{8}$&$\frac{5750747}{49152}$&$\frac{5}{2}$&$\frac{65}{4}$&$\frac{105}{2}$&$\frac{3745}{32}$&$\frac{52605}{256}$\\
		\hline
		$p_2'$&$p_4'$&$p_6'$&$p_8'$&$p_{10}'$&$p_2''$&$p_4''$&$p_6''$&$p_8''$&$p_{10}''$\\
		$\frac{19}{2}$&$106$&$\frac{4351}{8}$&$\frac{58395}{32}$&$\frac{1185825}{256}$&$6$&$\frac{351}{8}$&$\frac{637}{4}$&$\frac{12705}{32}$&$\frac{12411}{16}$\\
		\hline
		$k_0'$&$k_2'$&$k_4'$&$k_6'$&$k_8'$&$k_0''$&$k_2''$&$k_4''$&$k_6''$&$k_8''$\\
		$1$&$23$&$\frac{697}{4}$&$\frac{6045}{8}$&$\frac{170284187}{73728}$&$1$&$\frac{27}{2}$&$\frac{273}{4}$&$\frac{847}{4}$&$\frac{35742107}{73728}$\\
		\hline
		$h_0'$&$h_2'$&$h_4'$&$h_6'$&$h_8'$&$h_0''$&$h_2''$&$h_4''$&$h_6''$&$h_8''$\\
		$1$&$\frac{27}{2}$&$\frac{273}{4}$&$\frac{847}{4}$&$\frac{35742107}{73728}$&$1$&$\frac{15}{2}$&$\frac{195}{8}$&$\frac{105}{2}$&$\frac{6469787}{73728}$\\
		\hline
		\hline
	\end{tabular}
\end{table}

The coefficients $p_{2n}$ and $u_{2n}$ are not involved in the equation of conservation of the total angular momentum because they correspond to elastic tides\footnote{They are the terms of Eqs. (\ref{differential}) independent on $Q_j$, for which $\ddot{\vect{r}}_j\propto\vect{r}_j$} and they do not affect the angular momentum of the system. These coefficients are evaluated to zero in Eqs. (\ref{differential}) when only the dissipative contributions of tides matter (\textit{e.g.} when computing the real parts of the eigenvalues around the equilibria). The evaluation $\Lambda_j=\Lambda_{j,0}$, that is, $\Lam_j=1$, kills all the dissipation in the libration amplitude of the Laplace angle and is not a viable approximation. Sometimes, it is more useful to know the quantities $\dot{D}_j$ and $\dot{\varpi}_j$ instead of $\dot{X}_j$, where \smash{$X_j=\sqrt{2D_j/\Lambda_j}e^{\iota\varpi_j}$} and we recall that \smash{$X_j'=\sqrt{2D_j/\Lambda_{j,0}}\,e^{\iota\varpi_j}=\Lam_j^{1/2}X_j$}. Writing
\begin{equation}
	\dot{X}_j'=X_j'\left(\frac{1}{2}\frac{\dot{D}_j}{D_j}+\iota\dot{\varpi}_j\right)=\frac{1}{2}\dot{\Lam}_j\Lam_j^{-1/2}X_j+\Lam_j^{1/2}\dot{X}_j,
\end{equation}
we get\footnote{Hence an error on $\dot{D}_j$ in Eqs. (56) of \cite{CoRoCo2022}}
\begin{equation}
	\begin{split}
		&\dot{D}_j\!=\!-3n_{j,0}\frac{q_j}{Q_j}\frac{m_0}{m_j}\Lam_j^{-13}D_j\left(20\Lam_j^{-3}\!-\!13\omega_j\!+\!\sum_{n=1}^{+\infty}\left(k_{2n}^{(j)}\!+\!2p_{2n+2}^{(j)}\right)D_j^n\right),\!\!\!\!\!\!\!\!\!\!\!\!\!\!\!\!\\	
		&\dot{\varpi}_j=3n_{j,0}q_j\frac{m_0}{m_j}\Lam_j^{-13}\left(\frac{5}{2}+\sum_{n=1}^{+\infty}p_{2n+2}D_j^{n}\right).
	\end{split}
\end{equation}

\subsection{Eigenvalues and tidal timescales}\label{ssec:eigen}

With tides, the problem goes from the pair of variables $\left(\Delta\Phi, \varphi\right)$, to the seven variables $\left(\Delta\Phi, \varphi, \Delta\Gamma, \Delta\Upsilon, \vartheta_1, \vartheta_2, \vartheta_3\right)$, where we use the variable \smash{$\vartheta_j=1-\dot{\theta}_j/n_{j,0}$} for the rotations. The goal of this section is to obtain the eigenvalues of the differential system derived from Hamiltonian (\ref{eq:Ham_1dof}), to which we add the tidal contributions (\ref{differential}) at order zero in eccentricity. We only keep the dissipative contributions of tides and discard from Eqs. (\ref{differential}) terms independent on $Q_j$, since they do not contribute to the real part of the eigenvalues. The resulting differential system has two independent first integrals, that are
\begin{equation}\label{eq:first_integrals}
	\begin{split}
		&J_1=\Delta\Upsilon-\sum_{1\leq j\leq3}\alpha_j\Lambda_{j,0}\frac{R_j^2}{a_{j,0}^2}\vartheta_j,\\
		&J_2=\Delta\Gamma-\frac{p+q}{p}\alpha_1\Lambda_{1,0}\frac{R_1^2}{a_{1,0}^2}\vartheta_1-\alpha_2\Lambda_{2,0}\frac{R_2^2}{a_{2,0}^2}\vartheta_2.
	\end{split}
\end{equation}
These first integrals traduce the conservation of the total angular momentum and of the scaling parameter with rotations. They can be used to reduce the size of the differential system from 7 to 5 variables. We could get rid of $\Delta\Gamma$ and $\Delta\Upsilon$ and keep the $\vartheta_j$ only, but it turns out to be much more convenient analytically to get rid of $\vartheta_2$ and $\vartheta_3$. Therefore, we express $\vartheta_2$ and $\vartheta_3$ as a function of $J_1$, $J_2$, $\Delta\Gamma$, $\Delta\Upsilon$ and $\vartheta_1$ using Eqs. (\ref{eq:first_integrals}) and we are left with the five variables \smash{$\mathcal{X}=\,^t\!\left(\Delta\Phi, \varphi, \Delta\Gamma, \Delta\Upsilon, \vartheta_1\right)$}. The equations of motion can be written $d\mathcal{X}/dt=F(\mathcal{X})$ where $F$ denotes the vector field.
If $\mathcal{X}_0$ is a fixed point of $F$, then, with $\mathcal{X}=\mathcal{X}_0+\delta\mathcal{X}$, the differential system in the vicinity of $\mathcal{X}_0$ can be written
\begin{equation}
	\frac{d}{dt}\delta\mathcal{X}=F(\mathcal{X}_0+\delta\mathcal{X})=\nabla F(\mathcal{X}_0)\delta\mathcal{X}+\mathcal{O}\left(\left|\delta\mathcal{X}\right|^2\right).
\end{equation}
The eigenvalues $\lambda_i$ of the Jacobian matrix $\mathcal{M}=\nabla F(\mathcal{X}_0)$ characterize the dynamics of the system in the vicinity of the fixed point since the proper modes $P_i$ behave like $P_i(t)=P_{i,0}e^{\lambda_it}$. The three steps are
\begin{itemize}
	\item Find a fixed point \smash{$\mathcal{X}_0$} of the differential system $F$ with tides.
	\item Linearize $F$ in the vicinity of \smash{$\mathcal{X}_0$} in order to obtain the Jacobian matrix \smash{$\mathcal{M}=\nabla F(\mathcal{X}_0)$}.
	\item Compute the eigenvalues of the Jacobian.
\end{itemize}
These steps are performed analytically. For the first step, we look for the fixed point that is a perturbation of $\varphi=\pi$, since the fixed point that is a perturbation of $\varphi=0$ corresponds to the separatrix of the pendulum (we consider $\sign\Xi=\sign\delta=\sign\Delta=1$).

Let \smash{$\varepsilon_t=\varepsilon^{-1}q_j/Q_j$} be the characteristic size of tides. Looking at Eqs. (\ref{differential}), and especially the quantities \smash{$\mathcal{L}_j^{-12}$} in them, the Jacobian matrix \smash{$\nabla F(\mathcal{X}_0)$} has a tractable expression only if $J_1=J_2=0$. The eigenvalues of the system do not depend on $J_1$ and $J_2$ at first order in $\varepsilon_t$ as long as $J_j\ll1$, and so we choose $J_1=J_2=0$.
With this choice, $\mathcal{X}_0=\,^t\!\left(\Delta\Phi_0, \varphi_0, \Delta\Gamma_0, \Delta\Upsilon_0, \vartheta_{1,0}\right)=\,^t\!\left(0,\pi,0,0,0\right)$ is a fixed point of the differential system with tides. Using the algebraic manipulator \texttt{Maxima}, we linearized the equations of motions with tides in the vicinity of $\mathcal{X}_0$ and found for the Jacobian matrix $\mathcal{M}=\mathcal{M}_0+\varepsilon_t\mathcal{M}_1$, with
\begin{equation*}
	\small
	\mathcal{M}_0=\begin{pmatrix}
		0\!\!\!\!&\displaystyle{\frac{m_1m_2}{m_0^2}n_{3,0}\Lambda_{3,0}\Xi}\!\!\!\!\!\!\!\!\!\!\!\!&0\!\!\!\!&0\!\!\!\!&0\\
		-3K\!\!\!\!&0\!\!\!\!\!\!\!\!\!\!\!\!&\displaystyle{3\left(p\!+\!q\right)\frac{n_{2,0}}{\Lambda_{2,0}}\!+\!3q\frac{n_{3,0}}{\Lambda_{3,0}}}\!\!\!\!&\displaystyle{-3q\frac{n_{3,0}}{\Lambda_{3,0}}}\!\!\!\!&0\\
		0\!\!\!\!&0\!\!\!\!\!\!\!\!\!\!\!\!&0\!\!\!\!&0\!\!\!\!&0\\
		0\!\!\!\!&0\!\!\!\!\!\!\!\!\!\!\!\!&0\!\!\!\!&0\!\!\!\!&0\\
		0\!\!\!\!&0\!\!\!\!\!\!\!\!\!\!\!\!&0\!\!\!\!&0\!\!\!\!&0
	\end{pmatrix},
\end{equation*}
\begin{equation*}
	\small
	\varepsilon_t\mathcal{M}_1=
	\begin{pmatrix}
		\!\displaystyle{9n_{1,0}\frac{m_0}{m_1}\frac{q_1}{Q_1}}\!\!\!\!\!\!\!&0\!\!\!\!\!\!&0\!\!\!\!\!\!&0\!\!\!\!\!\!&\star\\
		\!0\!\!\!\!\!\!\!&0\!\!\!\!\!\!&0\!\!\!\!\!\!&0\!\!\!\!\!\!&0\\
		\!\displaystyle{9\left(p+q\right)\left(n_{1,0}\frac{m_0}{m_1}\frac{q_1}{Q_1}+n_{2,0}\frac{m_0}{m_2}\frac{q_2}{Q_2}\right)}\!\!\!\!\!\!\!&0\!\!\!\!\!\!&\star\!\!\!\!\!\!&0\!\!\!\!\!\!&\star\\
		\!\displaystyle{9pn_{1,0}\frac{m_0}{m_1}\!\frac{q_1}{Q_1}\!-\!9\!\left(p\!+\!q\right)n_{2,0}\frac{m_0}{m_2}\!\frac{q_2}{Q_2}\!+\!9qn_{3,0}\frac{m_0}{m_3}\!\frac{q_3}{Q_3}}\!\!\!\!\!\!\!&0\!\!\!\!\!\!&\star\!\!\!\!\!\!&\star\!\!\!\!\!\!&\star\\
		\!\displaystyle{9n_{1,0}p\frac{m_0}{m_1}\frac{q_1}{Q_1}\Lambda_{1,0}^{-1}\alpha_1^{-1}\frac{a_{1,0}^2}{R_1^2}}\!\!\!\!\!\!\!&0\!\!\!\!\!\!&0\!\!\!\!\!\!&0\!\!\!\!\!\!&\star
	\end{pmatrix},
\end{equation*}
where the stars denote unspecified non-zero coefficients. In order to analytically obtain the eigenvalues of $\mathcal{M}_0+\varepsilon_t\mathcal{M}_1$, we use the following result, whose demonstration is provided in Appendix \ref{append:eigen}. Let $\mathcal{Q}$ be a near-diagonal $n\times n$ complex matrix of the form $\mathcal{Q}=\mathcal{D}_0+\varepsilon_t\mathcal{Q}_1$, where $\mathcal{D}_0=\text{diag}(\lambda_{j,0})$ and \smash{$\mathcal{Q}_1=(q_{i,j})_{1\leq i,j \leq n}$}. Let $\lambda_j$ be the eigenvalues of $\mathcal{Q}$. Then, if $\lambda_{k,0}$ is a well-separated eigenvalue of $\mathcal{D}_0$, in the sense that \smash{$\forall i\in\left[|1,n|\right], \;i\neq k \Rightarrow \left|\lambda_{k,0}-\lambda_{i,0}\right|\gg\varepsilon_t$}, then, at first order in $\varepsilon_t$, the eigenvalue $\lambda_k$ can be read on the diagonal of $\mathcal{Q}$, in the sense that
\begin{equation}
	\lambda_k=\lambda_{k,0}+\varepsilon_t\,q_{k,k}+\mathcal{O}(\varepsilon_t^2).
\end{equation}
However, $\mathcal{M}_0$ is not diagonal, and we first need to write $\mathcal{M}$ in a basis that diagonalizes $\mathcal{M}_0$. A matrix of eigenvectors of $\mathcal{M}_0$ is
\begin{equation*}
	\small
	\mathcal{P}_0\!=\!\begin{pmatrix}
		\displaystyle{\frac{m_1m_2}{m_0^2}}\!\!\!\!\!\!&\displaystyle{\frac{m_1m_2}{m_0^2}}\!\!\!\!\!\!&\displaystyle{q\frac{n_{3,0}}{\Lambda_{3,0}}}\!\!\!\!\!\!&0\!\!\!\!\!\!&0\\
		\displaystyle{-\iota\left(\frac{3K}{n_{3,0}\Lambda_{3,0}\Xi}\right)^{1/2}}\!\!\!\!\!\!&\displaystyle{\iota\left(\frac{3K}{n_{3,0}\Lambda_{3,0}\Xi}\right)^{1/2}}\!\!\!\!\!\!&0\!\!\!\!\!\!&0\!\!\!\!\!\!&0\\
		0\!\!\!\!\!\!&0\!\!\!\!\!\!&0\!\!\!\!\!\!&\displaystyle{q\frac{n_{3,0}}{\Lambda_{3,0}}}\!\!\!\!\!\!&0\\
		0\!\!\!\!\!\!&0\!\!\!\!\!\!&-K\!\!\!\!\!\!&\displaystyle{\left(p\!+\!q\right)\frac{n_{2,0}}{\Lambda_{2,0}}\!+\!q\frac{n_{3,0}}{\Lambda_{3,0}}}\!\!\!\!\!\!&0\\
		0\!\!\!\!\!\!&0\!\!\!\!\!\!&0\!\!\!\!\!\!&0\!\!\!\!\!\!&1
	\end{pmatrix}.
\end{equation*}
Since a change of basis does not alter the eigenvalues, the matrix $\mathcal{Q}=\mathcal{D}_0+\varepsilon_t\mathcal{Q}_1=\mathcal{P}_0^{-1}\mathcal{M}\mathcal{P}_0$ has the same eigenvalues as $\mathcal{M}$. The principal part is $\mathcal{D}_0=\text{diag}(-\iota\lambda,\iota\lambda,0,0,0)$
where $\lambda$ is given in Eq. (\ref{eq:Imsampi}). The last three eigenvalues of $\mathcal{D}_0$ are not well-separated, and so we do not obtain them. These eigenvalues are related to the planetary rotations and are purely real and negative. However, the first two eigenvalues, which govern the behavior of the Laplace angle, are well-separated and we use the previous result to obtain them. We only need the first two diagonal coefficients of $\mathcal{Q}_1=\mathcal{P}_0^{-1}\mathcal{M}_1\mathcal{P}_0$. Since the first two columns of $\mathcal{P}_0$ end with three zeros, only the first two columns of $\mathcal{M}_1$ contribute to the first two diagonal coefficients of $\mathcal{Q}_1$. This is why we did not specify the non-zero coefficients of the last three columns of $\mathcal{M}_1$, even though we obtained them with \texttt{Maxima}. The first two diagonal coefficients of $\mathcal{Q}_1$ are equal and read $\left(\mathcal{Q}_1\right)_{1,1}=\left(\mathcal{Q}_1\right)_{2,2}=\mathfrak{R}\text{e}\,\sampi=$
\begin{equation}\label{eq:Resampi}
	\frac{9}{2}p^2\!\frac{n_{1,0}^2}{K\!\Lambda_{1,0}}\frac{m_0}{m_1}\frac{q_1}{Q_1}\!+\!\frac{9}{2}(p\!+\!q)^2\!\frac{n_{2,0}^2}{K\!\Lambda_{2,0}}\frac{m_0}{m_2}\frac{q_2}{Q_2}\!+\!\frac{9}{2}q^2\!\frac{n_{3,0}^2}{K\!\Lambda_{3,0}}\frac{m_0}{m_3}\frac{q_3}{Q_3},
\end{equation}
where $K$ was defined in Eq. (\ref{eq:K}) and we recall that \smash{$q_j=\kappa_2^{(j)}R_j^5/a_{j,0}^5$}. The two eigenvalues that control the pendulum of the 3P0R $pn_1-\left(p+q\right)n_2+qn_3=0$ with tidal dissipation are $\sampi=\mathfrak{R}\text{e}\,\sampi\pm\iota\lambda$. In particular, the libration amplitude $\mathcal{A}_{\varphi}$ of the 3-planet angle $\varphi=p\lambda_1-\left(p+q\right)\lambda_2+q\lambda_3$ is given by
\begin{equation}
	\mathcal{A}_{\varphi}=\mathcal{A}_{\varphi,0}\,\exp\left(\mathfrak{R}\text{e}\,\sampi t\right),
\end{equation}
and $1/\mathfrak{R}\text{e}\,\sampi$ is the e-folding time of increase of the 3-planet angle's libration amplitude. In order to evaluate how accurate our analytical computation of $\mathfrak{R}\text{e}\,\sampi$ is, we compare in Table \ref{tab:Resampi} the value of $1/\mathfrak{R}\text{e}\,\sampi$ evaluated using Eq. (\ref{eq:Resampi}) with a direct numerical computation of the eigenvalues of $\mathcal{M}=\mathcal{M}_0+\varepsilon_t\mathcal{M}_1$. For each computation in Table \ref{tab:Resampi}, we choose the locus of the 3P0R corresponding to $\Delta=0.02$ (Eq. \ref{eq:rho}). The masses and radii are the same as different triplets of our numerical simulations (Table \ref{tab:simulations}) and are indicated by giving the name of the simulation and the corresponding triplet. The tidal parameters $\kappa_2/Q$ are chosen arbitrarily and are not necessarily the same as in the numerical simulations.
\begin{table}[h]
	\setlength{\tabcolsep}{4.9pt}
	\renewcommand{\arraystretch}{0.5}
	\caption{Analytical and numerical value of $1/\mathfrak{R}\text{e}\,\sampi$}\label{tab:Resampi}
	\vspace{-2mm}
	\centering
	\begin{tabular}{ccccc}
		\hline
		Triplet&$10^4\,\kappa_2/Q$&Analytical&Numerical&Error\\
		\hline
		\hline
		\texttt{pq23}, 012&$3.1,3.1,7.6$&$56.578\;10^9$&$56.574\;10^9$&$0.006$\%\\
		\texttt{pq23}, 012&$\times\,100$&$\times\,1/100$&$0.5837\;10^9$&$3.08$\%\\
		\hline
		\texttt{pq23}, 012&$5.3,1.6,3.8$&$64.187\;10^9$&$64.186\;10^9$&$0.002$\%\\
		\texttt{pq23}, 012&$\times\,100$&$\times\,1/100$&$0.7011\;10^9$&$8.45$\%\\
		\hline
		\texttt{pq13}, 012&$4.6,7.2,3.9$&$48.855\;10^9$&$48.864\;10^9$&$0.019$\%\\
		\texttt{pq13}, 012&$\times\,10$&$\times\,1/10$&$4.985\;10^9$&$1.99$\%\\
		\texttt{pq13}, 012&$\times\,100$&$\times\,1/100$&$1.113\;10^9$&$56.1$\%\\
		\hline
		\texttt{pq13}, 012&$2.4,6.6,3.2$&$78.311\;10^9$&$78.311\;10^9$&$.0007$\%\\
		\texttt{pq13}, 012&$\times\,100$&$\times\,1/100$&$1.100\;10^9$&$28.8$\%\\
		\hline
		\texttt{703}, 345&$5.1,4.7,1.3$&$11.753\;10^9$&$11.752\;10^9$&$0.007$\%\\
		\texttt{703}, 345&$\times\,100$&$\times\,1/100$&$0.1497\;10^9$&$21.5$\%\\
		\hline
		\texttt{703}, 345&$2.6,3.7,3.5$&$14.748\;10^9$&$14.747\;10^9$&$0.008$\%\\
		\texttt{703}, 345&$\times\,100$&$\times\,1/100$&$0.1644\;10^9$&$10.3$\%\\
		\hline
		\texttt{33}, 123&$1.6,4.2,2.7$&$7.518\;10^9$&$7.520\;10^9$&$0.025$\%\\
		\texttt{33}, 123&$\times\,100$&$\times\,1/100$&$0.3496\;10^9$&$78.5$\%\\
		\hline
	\end{tabular}
	\tablefoot{The e-folding time $1/\mathfrak{R}\text{e}\,\sampi$ is given in units of $P_\text{in}$. The numerical values come from a direct numerical evaluation of the eigenvalues of $\mathcal{M}=\mathcal{M}_0+\varepsilon_t\mathcal{M}_1$. The analytical values come from Eq. (\ref{eq:Resampi}).}
\end{table}

Each computation of Table \ref{tab:Resampi} comes in pair. The first computation of the pair corresponds to physical $\kappa_2/Q$ (on the order of $10^{-4}$), whereas the second computation corresponds to the same $\kappa_2/Q$ multiplied by $100$, making them typical of a numerical simulation where $\kappa_2/Q$ has to be artificially increased to limit the running time (see Sect. \ref{ssec:simulations}).

In Table \ref{tab:Resampi}, we notice that with physical values for $\kappa_2/Q$, Eq. (\ref{eq:Resampi}) gives the e-folding time $1/\mathfrak{R}\text{e}\,\sampi$ to within $0.1$\%. However, with the large nonphysical $\kappa_2/Q$ used for numerical simulations, Eq. (\ref{eq:Resampi}) often estimates the e-folding time poorly. This is because with nonphysically large $\kappa_2/Q$, the tidal small parameter $\varepsilon_t$ is not small enough for the theorem of Appendix \ref{append:eigen} to apply.

Although Table \ref{tab:Resampi} shows the validity of our analytical expression of the e-folding time, it does not assess if the 1-dof pendulum model is a good approximation of the complete system. In order to verify that, we ran numerical integrations in \texttt{Aptidal} with a $\mathcal{SABA}_{10}$ integrator and a timestep $0.26$ of the inner orbital period of two triplets of Table \ref{tab:Resampi}. To limit the runtime, we use larger than physical $\kappa_2/Q$. The initial conditions are close to the libration centers at $\Delta=0.02$, that is $n_1/n_2\approx2.04$ in the left panel and $n_1/n_2\approx1.53$ in the right panel.

\begin{figure}[h]
	\includegraphics[width=\linewidth]{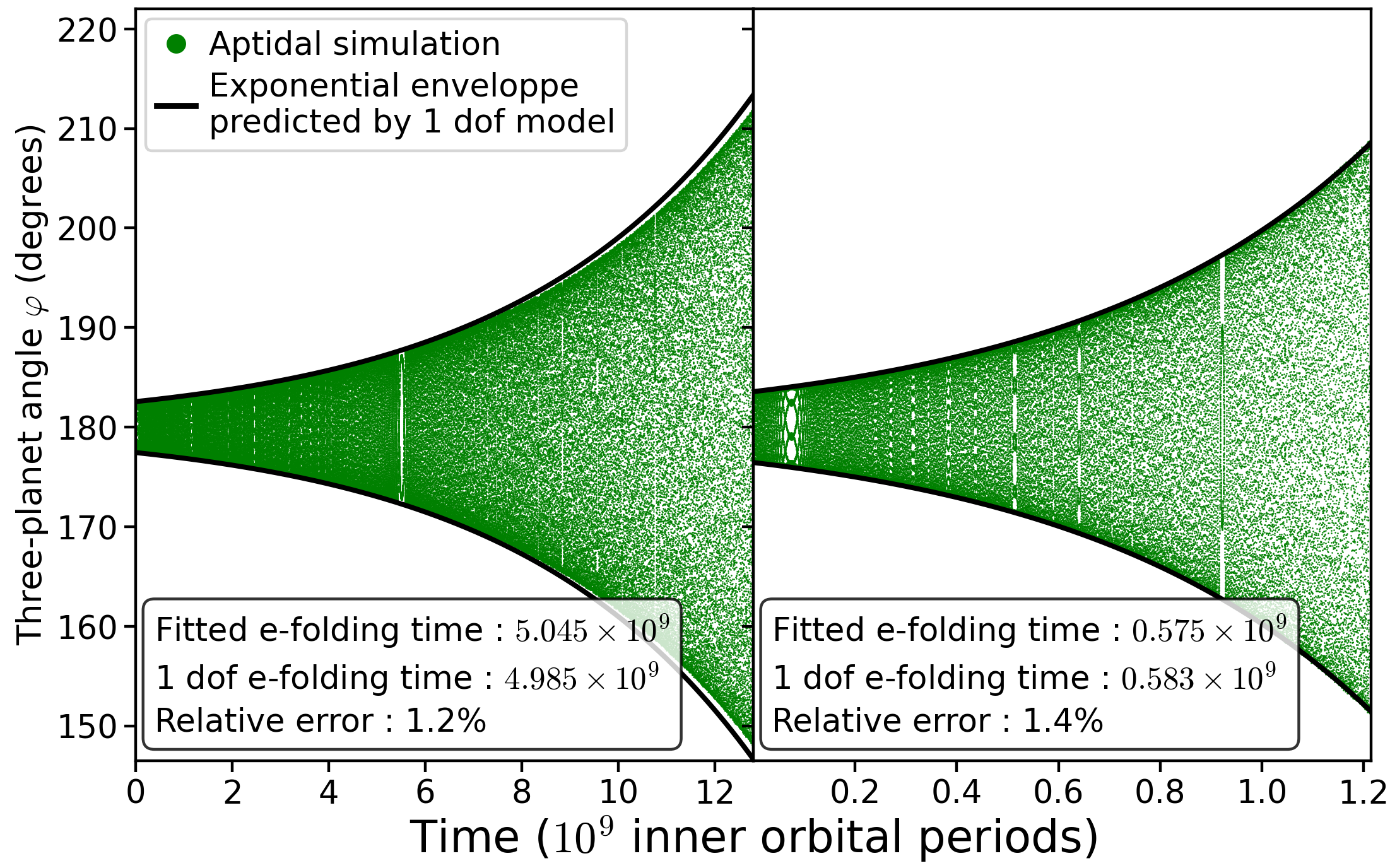}
	\caption{Tidal evolution of the 3-planet angle as a function of time for \texttt{pq13} (left) and \texttt{pq23} (right). See text and Table \ref{tab:Resampi}.}\label{fig:e-fold_3P0R}
\end{figure}

In Fig. \ref{fig:e-fold_3P0R}, we plot the 3-planet angle as a function of time, along with the exponential envelope predicted from $\mathfrak{R}\text{e}\,\sampi$. We use the exact numerical value of $\mathfrak{R}\text{e}\,\sampi$, since the analytical expression given in Eq. (\ref{eq:Resampi}) is not valid with large nonphysical $\kappa_2/Q$. We see that the 1-dof pendulum model is able to predict the e-folding time of exponential increase of the 3-planet angle within $\sim1$\% error. We expect that the agreement would have been even better with physical $\kappa_2/Q$, but verifying it is computationally too expensive. Combining Table \ref{tab:Resampi} with Fig. \ref{fig:e-fold_3P0R}, we conclude that for physical $\kappa_2/Q$, Eq. (\ref{eq:Resampi}) gives the true e-folding time with a precision better than $1$\%.

\section{Discussion and Conclusion}\label{sec:conclusion}

In this work, we elaborated a 1-dof model of 3P0R, with and without planetary tides. We validated this model with the new $N$-body software \texttt{Aptidal}, using a stability map for the conservative case (Fig. \ref{fig:stability_map}), and comparing the e-folding timescale between the analytical prediction (Eq. \ref{eq:Resampi}) and the $N$-body evolution for the tidal case (Fig. \ref{fig:e-fold_3P0R}).

We used the 3P0R conservative model to classify the exoplanet population (Fig. \ref{fig:classification}). When available, posteriors of dynamical analysis
(mostly TTVs) were used. Since our model only depends on the periods, mean longitudes, and masses, we also used it to classify systems detected in transits or radial velocities, using a mass-radius relation when the mass was unknown.

We showed that most triplets of the known chains Kepler-60, 80, 223, K2-138, TOI-178, TOI-1136, HD 110067, TRAPPIST-1 and GJ-876 \citep{Millholland2018} are in 3P0R, although for some of these triplets, the 3P0R is due to a proximity to 2P1R. We also confirmed the resonant state of Kepler-221 bce \citep{GoldbergBatygin2021} and YZ-Ceti bcd \citep{Stock2020} to the best of our knowledge of these systems. We found a new triplet in 3P0R, unlabeled in Fig. \ref{fig:classification}, that we will discuss in a subsequent study.

We studied the tidal evolution of chains of planets initially in pairs of 2P1R using the new software \texttt{Aptidal}, able to simulate multi-planetary systems under adiabatic tides. With a $\mathcal{SABA}_{10}$ integrator \citep{LaskarRobutel2001}, it can use timesteps as large as $1/3$ of the inner period.
Initially, the systems move away from 2P1R following the equilibria of the chain \citep{Delisle2017}.
Once the systems are far from 2P1R, 3P0R dominate the dynamics, triplets diverge from the $180^\circ$ equilibrium, and exit the 3P0R.
Using our analytical model, we showed that 3P0R are always unstable under tides, and we computed the tidal e-folding time. This complements results from the 2P0R \citep[that is, the coorbital resonance, ][]{CoRoCo2021}, and is consistent with the innermost planets of some chains being outside the resonant configuration, or having the largest libration amplitude of the chain \citep{Macdonald2016,Leleu2024TOI178,Leleu2026a}.

Another consequence of the instability of 3P0R under tides is that, while a triplet can be captured in 3P0R by a double capture in 2P1R, direct capture in 3P0R is inefficient. Indeed, in the case of direct capture, triplets exit the 3P0R as soon as they enter because of the instability, and they tend to simply cross 3P0R. This happened to the outer triplet of simulation \texttt{703}, in yellow in Fig. \ref{fig:classification}.
This qualitatively differs from previous works where direct capture in 3P0R could occur \citep[\textit{e.g.},][]{Papaloizou2018,Yi2025}.
To our understanding, this qualitative difference can be due to two approximations:
\begin{itemize}
	\item In their simulations, tides comes from eccentricity damping prescriptions, However, in a 3P0R at eccentricities $\lesssim10^{-3}$, the libration of the semi-major axes is the main motor of convergence or divergence of the Laplace angle away from the equilibrium, as we showed in Sect. \ref{ssec:eigen}. This is true for the 2P0R as well \citep{Leleu2019,CoRoCo2021}.
	\item A prescription of the form $\dot a = a \tau_a$ with constant $\tau_a$ does not capture the true dependency of tides in semi-major axis.
	Indeed, \cite{Leleu2019} showed that with a prescription $\dot a = a^K \tau_a$, the stability of the 2P0R strongly depends on $K$.
\end{itemize}
Our work models tides at the level of the planetary spin and alleviates both issues.

In our simulations, the triplets typically exit the 3P0R by the bottom in Fig. \ref{fig:Period_ratio}, which is expected in the case of stronger tidal dissipation in the innermost planet. Then, as they keep evolving in the period ratio plane due to residual eccentricity damping, they are often captured in 3P1R\footnote{By combining Eq. (28) of \cite{Petit2021} with $\mathcal{I}_0$ below their Eq. (40), one can show that 3P1R emanating from 2P1R (schematized in Fig. \ref{fig:Period_ratio}) are oriented such that the opening of 3P1R is facing the 3P0R, allowing capture in 3P1R after exit from 3P0R.}.
We note that a simulation cannot predict in which 3P1R the capture will occur, since thin 3P1R are crossed before thick 3P1R (see Fig. \ref{fig:stability_map}). In numerical simulations with $P_\text{in}\sim1$ days, tides are less adiabatic than in reality (to speed up the simulation), favoring thicker 3P1R.
We identified in Table \ref{tab:3P1R} triplets possibly in 3P1R after following the schematic evolution of Fig. \ref{fig:Period_ratio}, such as Kepler-1987 cde. The outer triplet of K2-138 is also suspected to be in 3P1R \citep{Cerioni2023, Leleu2026a} after following an evolution similar to the outer triplet of simulation \texttt{703} (in yellow in Fig. \ref{fig:classification}). The actual fraction of systems actually in 3P1R is hard to determine, as these depend on the eccentricities and longitude of periapses, and 3P1R typically take decades to affect TTV signatures \citep{Leleu2026b}.

Finally, we used our model to study the overall impact of 3P0R on the architecture of multi-planetary systems. When considering the entire population, there is a significant overabundance of triplets in 3P0R compared to a bootstrap population (see Fig. \ref{fig:cumulative}, \cite{GoldbergBatygin2021} and \cite{Cerioni2022}). 
We showed that this overabundance comes from the subpopulation close enough from their star to undergo tides (\smash{$p=2.36\,10^{-5}$}).
We also showed an overabundance of tidally affected systems below 3P0R, consistent with a fraction of close-in systems having followed the evolution presented in Fig. \ref{fig:Period_ratio}, although the overabundance is not significant ($p=0.198$), and this is probably not a major mechanism of evolution in the exoplanet population.

\begin{acknowledgements}
The authors acknowledge support of the Swiss National Science Foundation under grant number TMSGI2\_211697. The authors acknowledge the computer resources provided by the LTE lab of Paris Observatory, as well as fruitful discussions with Michaël Gastineau and Philippe Robutel. No generative AI was used in the work of this paper or in its writing.
\end{acknowledgements}

%
\bibliographystyle{aa} 
\bibliography{p_pq_q.bib} 

\onecolumn
\makeatletter
\@fleqnfalse
\makeatother
\begin{appendix}
	
\nolinenumbers

\section{Expansion to second order in mass in the planetary Hamiltonian}\label{sec:second_mass}

In this section, we perform a Lie serie expansion to second order in mass in the Hamiltonian in order to obtain analytical expressions of the terms in front of $\exp(\iota\vect{k}\cdot\vect{\lambda})$ with $\vect{k}\in\mathbb{Z}^3$ and $\vect{\lambda}\in\left(\mathbb{R}/2\pi\mathbb{R}\right)^3$. We extend the python package \href{https://gitlab.unige.ch/delisle/celeries}{\texttt{celeries}} with a class \href{https://www.astro.unige.ch/~delisle/celeries/doc/_autosummary/celeries.perham3pla.PerHam3pla.html#celeries.perham3pla.PerHam3pla}{\texttt{PerHam3pla}} outputting that term at any degree in eccentricity and inclination.

\subsection{First order in mass}\label{ssec:first_mass}

We consider a planetary system with $n$ planets of masses \smash{$\left(m_i\right)_{1\leq i\leq n}$} orbiting a star of mass $m_0$. In this section, $I$ is an inclination, while $i$ and $\iota$ are, as in Sect. \ref{sec:chaos}, an index and \smash{$\sqrt{-1}$}, respectively. With $\vect{r}_j$ the heliocentric positions and $\vect{v}_j$ the barycentric speeds, the Hamiltonian of the system can be written \citep[][]{LaskarRobutel1995}
\begin{equation}\label{eq:nbody}
	\Ham(\vect{\tilde{r}},\vect{r})=\sum_{i=1}^n\left(\frac{\vect{\tilde{r}}_i^2}{2\beta_i}-\frac{\mathcal{G}m_0m_i}{r_i}\right)+\sum_{i=1}^n\sum_{j=i+1}^n\left(\frac{\vect{\tilde{r}}_i\cdot\vect{\tilde{r}}_j}{m_0}-\frac{\mathcal{G}m_im_j}{\left|\vect{r}_i-\vect{r}_j\right|}\right),
\end{equation}
where the momenta $\vect{\tilde{r}}_j=m_j\vect{v}_j$ are conjugated to $\vect{r}_j$. The reduced masses are \smash{$\beta_j=m_0m_j/\left(m_0+m_j\right)$}. We work with the Poincar\'e coordinates defined as
\begin{equation}\label{eq:Poincare}
		\Lambda_j=\beta_j\sqrt{\mu_j a_j},\;\;\;\;D_j=\Lambda_j\left(1-\sqrt{1-e_j^2}\right),\;\;\;\;H_j=\Lambda_j\sqrt{1-e_j^2}\left(1-\cos I_j\right),
\end{equation}
conjugated to $\lambda_j=M_j+\varpi_j$, $-\varpi_j=-\omega_j-\Omega_j$ and $-\Omega_j$, where \smash{$\mu_j=\mathcal{G}\left(m_0+m_j\right)$} and $\mathcal{G}$ is the gravitational constant. We use the complex Poincar\'e variables defined by the canonical transformation \smash{$\left(D,H;-\varpi,-\Omega\right)\mapsto\left(x=\sqrt{D}\exp(\iota\varpi),y=\sqrt{H}\exp(\iota\Omega);-\iota\bar{x},-\iota\bar{y}\right)$}, as well as the non-canonical variables \smash{$X=x\sqrt{2/\Lambda}=e\exp(\iota\varpi)+\OO(e^3)$} and \smash{$Y=y/\sqrt{2\Lambda}=I\exp(\iota\Omega)/2+\OO(I^3)+\OO(Ie^2)$}.

\subsubsection{Expansion in Poincar\'e coordinates}\label{sssec:expansionPoincare}
The first term of Eq. (\ref{eq:nbody}) is the Keplerian part of the Hamiltonian and corresponds to star$-$planet interactions. It can be written as a function of $\vect{\Lambda}$ only, and reads
\begin{equation}\label{eq:Keplerian}
	\Ham_K(\vect{\Lambda})=-\sum_{j=1}^n\frac{\beta_j^3\mu_j^2}{2\Lambda_j^2}.
\end{equation}
The second term of Eq. (\ref{eq:nbody}) is the perturbative part of the Hamiltonian and corresponds to planet$-$planet interactions. It is of characteristic size $\varepsilon$ with respect to the Keplerian part, where $\varepsilon=\left(m_1+m_2+\cdots+m_n\right)/m_0$. To emphasize the small size of the perturbation, we will write it $\varepsilon\Ham_P$. Hereafter, a quantity of order $r$ in mass will be written with a factor $\varepsilon^r$. The perturbative part can be written
\begin{equation}\label{eq:ij}
	\varepsilon\Ham_P=\sum_{i=1}^n\sum_{j=i+1}^n\varepsilon\Ham_{i,j},
\end{equation}
which emphasizes the fact that it only contains interactions between two planets. In particular, at order $1$ in mass, combinations of three or more angles do not appear. The perturbative part due to pair $\left(i,j\right)$ can be expanded in $X_i$, $X_j$, $Y_i$ and $Y_j$ as \citep[][Eq. (73)]{LaskarRobutel1995}
\begin{equation}\label{eq:HP}
	\varepsilon\Ham_{i,j}=\sum_{\vect{k}\in\mathbb{Z}^2}\left(\sum_{\vect{q}\in\mathbb{N}^8}\Xi_{\vect{k},\vect{q}}(\Lambda_i,\Lambda_j) X_i^{q_1}X_j^{q_2}\bar{X}_i^{q_3}\bar{X}_j^{q_4}Y_i^{q_5}Y_j^{q_6}\bar{Y}_i^{q_7}\bar{Y}_j^{q_8}\right)e^{\iota\left(k_1\lambda_i+k_2\lambda_j\right)}.
\end{equation}
In order to prevent an infinite number of terms in front of a given \smash{$e^{\iota\left(k_1\lambda_i+k_2\lambda_j\right)}$}, this expression is generally truncated at some degree $r$ in eccentricity and inclination, and only terms such that $q_1+\cdots+q_8\leq r$ are retained. Additionally, the symmetries of the Hamiltonian reveal that most of the $\Xi_{\vect{k},\vect{q}}$ are zero.

\subsubsection{Extended d'Alembert rule}\label{ssec:dAlembert}
The first symmetry is the invariance around the $z$-axis. Rotating around the $z$-axis, that is, redefining the $x$ and $y$ axes, is equivalent to translate the angles of the problem by some constant angle $\vartheta$. Injecting the transformation $\left(\lambda,\varpi,\Omega\right)\to\left(\lambda+\vartheta,\varpi+\vartheta,\Omega+\vartheta\right)$ into Eq. (\ref{eq:HP}), shows that the Hamiltonian is invariant by rotation around the $z$-axis if, and only if, the tuples \smash{$\vect{q}=\left(q_1,\cdots,q_8\right)\in\mathbb{N}^8$} and \smash{$\vect{k}=\left(k_1,k_2\right)\in\mathbb{Z}^2$} verify the d'Alembert rule \citep[][Sect. 6.1]{ChencinerLaskar1989}
\begin{equation}\label{eq:dAlembert}
	k_1+k_2+q_1+q_2-q_3-q_4+q_5+q_6-q_7-q_8=0.
\end{equation}
Clearly, the dynamics of the system should not be affected by a redefinition of the $x$-axis, and so $\Xi_{\vect{k},\vect{q}}=0$ whenever the d'Alembert rule is not verified. Another symmetry to consider is the symmetry with respect to the plane $\left(x,y\right)$. Changing $z$ into $-z$, the ascending node becomes the descending node and vice-versa, and $\Omega$ is transformed into $\Omega+\pi$. Because $\omega$ is the angle between the ascending node and the periapsis, it is also transformed into $\omega+\pi$. The mean anomaly $M$ is unchanged as it is defined from the periapsis which is unaffected by that symmetry. The angles $\varpi=\omega+\Omega$ and $\lambda=M+\varpi$ are also unchanged. Injecting this transformation into Eq. (\ref{eq:HP}) yields an extension to the d'Alembert rule \citep[][Sect. 6.1]{ChencinerLaskar1989}
\begin{equation}
	q_5+q_6+q_7+q_8 = 0\mod2.
\end{equation}
Since the dynamics should also not be affected by a symmetry with respect to the plane $\left(x,y\right)$, $\Xi_{\vect{k},\vect{q}}=0$ whenever the extension to the d'Alembert rule is not verified. The last symmetry is the conservation of the Hamiltonian by a $180^\circ$ rotation around the $x$-axis. Such rotation is equivalent to multiplying the angles $\lambda,\;\varpi$ and $\Omega$ by $-1$. Injecting this into Eq. (\ref{eq:HP}) yields
\begin{equation}
	\Xi_{k_1, k_2, q_1, q_2, q_3, q_4, q_5, q_6, q_7, q_8}=\Xi_{-k_1, -k_2, q_3, q_4, q_1, q_2, q_7, q_8, q_5, q_6},
\end{equation}
ensuring that the Hamiltonian is real and contains only cosine terms. The d'Alembert rule and its extension, along with a truncation in eccentricity and inclination at degree $r$, allow only a modest number of terms in front of \smash{$e^{\iota\left(k_1\lambda_i+k_2\lambda_j\right)}$} at a small degree $r$.

However, Eq. (\ref{eq:HP}) overall still contains infinitely many terms. For example, each \smash{$e^{\iota\left(p\lambda_i-\left(p+1\right)\lambda_j\right)}$} contains terms of degree $1$ for $p\in\mathbb{Z}$. In practice, only the terms \smash{$e^{\iota\left(k_1\lambda_i+k_2\lambda_j\right)}$} where $k_1\lambda_i+k_2\lambda_j$ is a slow angle are retained, the others being discarded (averaged upon). Even when fast angles are retained, a finite number of terms is ensured by the fact that $\Xi$ goes to $0$ exponentially quickly when $\left(k_1,k_2\right)$ goes to infinity (see Table \ref{tab:decay}).
\begin{table}[h]
	\setlength{\tabcolsep}{1.8pt}
	\renewcommand{\arraystretch}{0.8}
	\sidecaption
	\begin{tabular}{cccccc}
		\hline
		$p$&$1$&$2$&$5$&$20$&$50$\\
		\hline
		\hline
		$\Xi_i$&$0.81133$&$0.69751$&$0.33960$&$0.0030273$&$0.000000715$\\
		$-\Xi_j$&$0.44033$&$0.93170$&$0.47044$&$0.0042894$&$0.000000954$\\
		\hline
	\end{tabular}
	\caption{Coefficients $\Xi_i$ and $\Xi_j$ in front of \smash{$X_ie^{\iota\left(p\lambda_i-\left(p+1\right)\lambda_j\right)}$} and \smash{$X_je^{\iota\left(p\lambda_i-\left(p+1\right)\lambda_j\right)}$} evaluated at $\alpha_{ij}=0.7$, for some $p$. Both coefficients go to $0$ when $p\to\infty$.}\label{tab:decay}
\end{table}

\subsubsection{Expression as a function of the Laplace coefficients}
The coefficients $\Xi_{\vect{k},\vect{q}}(\Lambda_i,\Lambda_j)$ appearing in Eq. (\ref{eq:HP}) can be written as a function of the Laplace coefficients \smash{$b_{s/2}^{(k)}(\alpha_{ij})$} where $s$ and $k$ are integers ($k$ being odd), $\alpha_{ij}=a_i/a_j$ and\footnote{It is assumed that $i$ is the inner planet of the pair and $j$ is the outer one. Hence $0<\alpha_{ij}<1$.} \smash{$b_{s/2}^{(k)}(\alpha)$} is defined as
\begin{equation}
	b_{s/2}^{(k)}(\alpha)=\frac{2}{\pi}\int_0^\pi\left(1-2\alpha\cos\theta+\alpha^2\right)^{-s/2}\cos k\theta\;d\theta.
\end{equation}
Using the recurrence relation \citep[][p. 501]{BrouwerClemence1961}
\begin{equation}\label{eq:laplace_rec}
	b_{s/2}^{(k)}(\alpha)=\frac{2k-2}{2k-s}\left(\alpha+\alpha^{-1}\right)b_{s/2}^{(k-1)}(\alpha)-\frac{2k+s-4}{2k-s}b_{s/2}^{(k-2)}(\alpha),
\end{equation}
and the fact that \smash{$b_{s/2}^{(-k)}(\alpha)=b_{s/2}^{(k)}(\alpha)$}, it can be shown that $\Xi_{\vect{k},\vect{q}}$ depends on \smash{$b_{s/2}^{(0)}(\alpha)$} and \smash{$b_{s/2}^{(1)}(\alpha)$} only for some $s$. Furthermore, if $q_1+\cdots+q_8=r$, then it can be shown that $\Xi$ depends on a unique value of $s$, given by \citep[][Eq. (75)]{LaskarRobutel1995}
\begin{equation}\label{eq:smax_order1}
	s=1+r+(r\mod2).
\end{equation}
Looking back at Eq. (\ref{eq:nbody}), $\Xi_{\vect{k},\vect{q}}$ has contributions from the term in \smash{$1/\left|\vect{r}_i-\vect{r}_j\right|$}, usually called direct contributions, and contributions from the term in $\vect{\tilde{r}}_i\cdot\vect{\tilde{r}}_j$, usually called indirect contributions. The methods \texttt{direct}([$k_1,k_2$]) and \texttt{indirect}([$k_1,k_2$]) from class \texttt{PerHam} of the python package \href{https://gitlab.unige.ch/delisle/celeries}{\texttt{celeries}} give the full coefficient in front of \smash{$e^{\iota\left(k_1\lambda_i+k_2\lambda_j\right)}$} in Eq. (\ref{eq:HP}). The method \texttt{direct} leaves out a factor $-\G m_im_j/a_j=-m_in_j\Lambda_j/m_0$. The method \texttt{indirect} leaves out a factor \smash{$\alpha_{ij}^{-1/2}m_in_j\Lambda_j/m_0+\OO(\varepsilon^2)$}.

\subsection{Perturbation theory : Lie serie expansion and second order in mass}\label{ssec:Lie}

\begin{figure}[h]
	\sidecaption
	\includegraphics[width=0.68\linewidth]{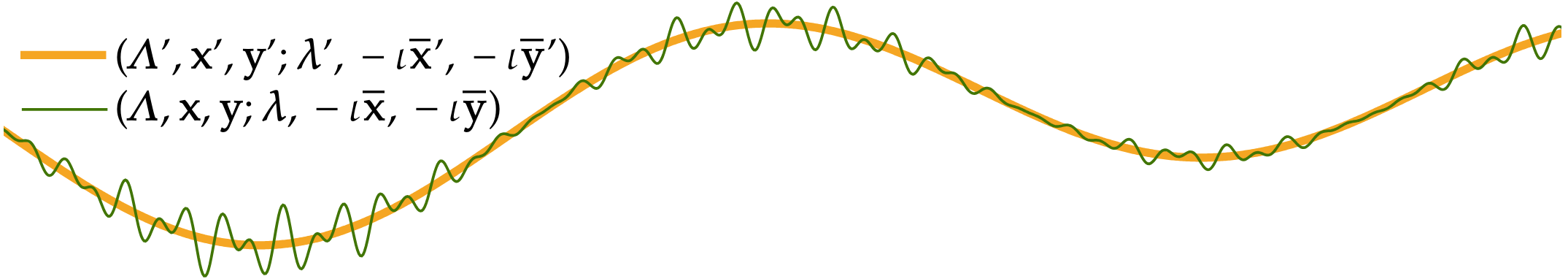}
	\caption{Schema of the transformation $\Psi$}\label{fig:LieTransformation}
\end{figure}

The Hamiltonian is $\Ham=\Ham_K+\varepsilon\Ham_P$. In particular, it contains only two-planet contributions and no terms of order $\varepsilon^2$. Terms of order $\varepsilon^2$ appear when averaging $\varepsilon\Ham_P$ over fast circulating angles. When the angle $k_1\lambda_i+k_2\lambda_j$ is fast-circulating (when the pair $\left(i,j\right)$ is far from resonance $k_1:-k_2$), then the term \smash{$e^{\iota\left(k_1\lambda_i+k_2\lambda_j\right)}$} in \smash{$\varepsilon\Ham_{i,j}$} averages to $0$ over large timescales and is generally removed from \smash{$\varepsilon\Ham_P$}. This procedure is justified mathematically using perturbation theory, and yields terms of order $\varepsilon^2$ in the Hamiltonian. The Hamiltonian of the planetary system is \smash{$\Ham(\vect{\Lambda},\vect{x},\vect{y};\vect{\lambda},-\iota\bar{\vect{x}}, -\iota\bar{\vect{y}})=\Ham_K(\vect{\Lambda})+\varepsilon\Ham_P(\vect{\Lambda},\vect{x},\vect{y};\vect{\lambda},-\iota\bar{\vect{x}}, -\iota\bar{\vect{y}})$}. We consider a transformation
\begin{equation}
	\begin{split}
		\Psi:\mathbb{R}^{6n}&\to\mathbb{R}^{6n}\\
		\mathcal{X}=\left(\vect{\Lambda},\vect{x},\vect{y};\vect{\lambda},-\iota\bar{\vect{x}}, -\iota\bar{\vect{y}}\right)&\mapsto\mathcal{X}'=\left(\vect{\Lambda}',\vect{x}',\vect{y}';\vect{\lambda}',-\iota\bar{\vect{x}}', -\iota\bar{\vect{y}}'\right),
	\end{split}
\end{equation}
close to the identity, such that
\begin{equation}\label{eq:lie_first_iter}
	\check{\Ham}(\mathcal{X}')=\Ham(\mathcal{X})=\check{\Ham}_0(\vect{\Lambda}')+\varepsilon\check{\Ham}_1+\varepsilon^2\check{\Ham}_2(\mathcal{X}'),
\end{equation}
where \smash{$\check{\Ham}=\Ham\circ\Psi^{-1}$} and \smash{$\check{\Ham}_1=\left<\Ham_P\right>$}. The operator $\left<\cdot\right>$ denotes the average over the fast combinations of 2-planet angles $k_1\lambda_i+k_2\lambda_j$ in Eq. (\ref{eq:HP}). That is to say, \smash{$\check{\Ham}_1$} depends only on slow combinations of angles. In order to ensure a canonical transformation, $\Psi$ is chosen as the flow at time $-1$ of a generator $\chi$
\begin{equation}\label{eq:transformation_lie}
	\Psi=\Phi_{\varepsilon\chi}(-1,\cdot),\quad\Leftrightarrow\quad\Psi^{-1}=\Phi_{\varepsilon\chi}(1,\cdot).
\end{equation}
For any choice of $\chi$, the transformation is canonical \citep[][Sect. 1.6]{Morbidelli2002}, and $\chi$ is constrained in such a way that only slow angles appear in \smash{$\varepsilon\check{\Ham}_1$}. We define the Lie derivative $L_\chi$ of $\chi$ as $L_\chi=\left\lbrace\chi,\cdot\right\rbrace$ where
\begin{equation}\label{eq:Lie_def}
	\left\lbrace\chi,\cdot\right\rbrace=\frac{d}{dt}\Bigr|_{\chi}=\drond{\chi}{\vect{\Lambda}'}\cdot\drond{}{\vect{\lambda}'}-\drond{\chi}{\vect{\lambda}'}\cdot\drond{}{\vect{\Lambda}'}+\iota\drond{\chi}{\vect{x}'}\cdot\drond{}{\bar{\vect{x}}'}-\iota\drond{\chi}{\bar{\vect{x}}'}\cdot\drond{}{\vect{x}'}+\iota\drond{\chi}{\vect{y}'}\cdot\drond{}{\bar{\vect{y}}'}-\iota\drond{\chi}{\bar{\vect{y}}'}\cdot\drond{}{\vect{y}'}.
\end{equation}
Denoting \smash{$L_{\varepsilon\chi}^{(j)}$} the $j^{\text{th}}$ iteration of the Lie derivative, \smash{$\check{\Ham}$} is Taylor expanded as
\begin{equation}\label{eq:exp_L}
	\check{\Ham}(\mathcal{X}')=\Ham(\mathcal{X})=\Ham\circ\Psi^{-1}(\mathcal{X}')=\Ham\circ\Phi_{\varepsilon\chi}(1,\mathcal{X}')=\sum_{j=0}^{+\infty}\frac{L_{\varepsilon\chi}^{(j)}}{j!}\Ham(\mathcal{X}')=\exp(\varepsilon L_\chi)\Ham(\mathcal{X}'),
\end{equation}
that is, \smash{$\check{\Ham}=\exp(\varepsilon L_\chi)\Ham$}. Expanding to second order in $\varepsilon$ yields
\begin{equation}\label{eq:Expansion_Lie}
	\check{\Ham}_0+\varepsilon\check{\Ham}_1+\varepsilon^2\check{\Ham}_2=\Ham_K+\varepsilon\left(\accolade\Ham_P+\left\lbrace\chi,\Ham_K\right\rbrace\right)+\varepsilon^2\left(\left\lbrace\chi,\Ham_P\right\rbrace+\frac{1}{2}\left\lbrace\chi,\left\lbrace\chi,\Ham_K\right\rbrace\right\rbrace\right),
\end{equation}
and in order to achieve the form (\ref{eq:lie_first_iter}), we constrain the generator with the cohomological equation
\begin{equation}\label{eq:cohomological}
	\Ham_P-\left<\Ham_P\right>=\left\lbrace\Ham_K,\chi\right\rbrace.
\end{equation}
Injecting Eq. (\ref{eq:cohomological}) into Eq. (\ref{eq:Expansion_Lie}), the expression of the Hamiltonian at second order in mass is
\begin{equation}\label{eq:H2bis}
	\check{\Ham}_2=\frac{1}{2}\left\lbrace\chi,\Ham_P+\left<\Ham_P\right>\right\rbrace.
\end{equation}
The Hamiltonian at second order in mass is entirely determined by Eq. (\ref{eq:H2bis}) provided that the generator $\chi$ can be extracted from Eq. (\ref{eq:cohomological}). To do so, we write, as in Eq. (\ref{eq:ij}), \smash{$\chi=\sum_{i=1}^n\sum_{j=i+1}^n\chi_{i,j}$} and \smash{$\Ham_P=\sum_{i=1}^n\sum_{j=i+1}^n\Ham_{i,j}$}, and then in the Fourier domain
\begin{equation}\label{eq:fourier}
	\chi_{i,j}=\sum_{\vect{k}\in\mathbb{Z}^2\backslash(0,0)}\chi_\vect{k}e^{\iota\left(k_1\lambda_i+k_2\lambda_j\right)},\;\;\;\;\Ham_{i,j}-\left<\Ham_{i,j}\right>=\sum_{\vect{k}\in\mathbb{Z}^2\backslash(0,0)}h_\vect{k}e^{\iota\left(k_1\lambda_i+k_2\lambda_j\right)}.
\end{equation}
Injecting into Eq. (\ref{eq:cohomological}) and using $\partial\Ham_K/\partial\vect{\Lambda}=\vect{n}$ yields
\begin{equation}\label{eq:sol_coho}
	\chi_\vect{k}=\frac{-\iota h_\vect{k}}{k_1n_i+k_2n_j}.
\end{equation}
When the 2-planet angle $k_1\lambda_i+k_2\lambda_j$ is among the combinations averaged by $\left<\cdot\right>$, the corresponding coefficient $h_\vect{k}$ is non-zero in Eq. (\ref{eq:sol_coho}). In this case, for the transformation to be $\varepsilon$-close to the identity, it is necessary for $k_1n_i+k_2n_j$ to be a quantity of order unity, that is, for $k_1\lambda_i+k_2\lambda_j$ to be a fast-circulating angle. As long as the operator $\left<\cdot\right>$ averages only over fast angles, $\Psi$ is a quasiperiodic transformation $\varepsilon$-close to the identity, schematically represented in Fig. \ref{fig:LieTransformation}. The primed variables contain the tendency of the trajectories, where the short-period variations have been removed. They are better for an analytical work than the unprimed variables. Hereafter, we work with \smash{$\check{\Ham}$} and the primed variables, and we omit both the prime and the $\check{\cdot}$ for clarity.

\subsection{Poincar\'e coordinates and d'Alembert rule}\label{ssec:d'Alembert}

Analyzing Eqs. (\ref{eq:H2bis}), it can be noticed that at second order in mass, the Hamiltonian does depend on 3-planet angles. For example, considering the pair $\left(i,j\right)=\left(1,2\right)$ on the left-side of the Poisson bracket in Eq. (\ref{eq:H2bis}), and the pair $\left(i,j\right)=\left(1,3\right)$ on the right-side yields terms of the form \smash{$e^{\iota\left(k_1\lambda_1+k_2\lambda_2+k_3\lambda_3\right)}$}. Likewise, dependency on 2-planet, 1-planet and 0-planet angles remain in $\Ham_2$. Clearly, $\Ham_2$ has no dependency on 5-planet angles or more. Dependency on 4-planet angles should come from brackets of the form \smash{$\left\lbrace e^{\iota\left(p\lambda_i+q\lambda_j\right)},e^{\iota\left(r\lambda_k+s\lambda_l\right)}\right\rbrace$} where $\left(i,j,k,l\right)$ are all different, but then the sets of variables on the left-side and right-side are disjoint and the bracket vanishes due to Eq. (\ref{eq:Lie_def}). Therefore, $\Ham_2$ has no dependency on 4-planet angles and can be written
\begin{equation}
	\varepsilon^2\Ham_2=\sum_{i=1}^n\sum_{j=i+1}^n\sum_{k=j+1}^n\varepsilon^2\Ham_{i,j,k},
\end{equation}
which emphasizes the fact that it only contains interactions between three planets. Using Eq. (\ref{eq:HP}), we have
\begin{equation}\label{eq:H2_expanded}
	\Ham_{i,j,k}=\sum_{\vect{k}\in\mathbb{Z}^3}\left(\sum_{\vect{q}\in\mathbb{N}^{12}}\Xi_{\vect{k},\vect{q}}X_i^{q_1}X_j^{q_2}X_k^{q_3}\bar{X}_i^{q_4}\bar{X}_j^{q_5}\bar{X}_k^{q_6}Y_i^{q_7}Y_j^{q_8}Y_k^{q_9}\bar{Y}_i^{q_{10}}\bar{Y}_j^{q_{11}}\bar{Y}_k^{q_{12}}\right)e^{\iota\left(k_1\lambda_i+k_2\lambda_j+k_3\lambda_k\right)}.
\end{equation}
The considerations of symmetry stated in Sect. \ref{ssec:dAlembert} are valid at any order in mass, and the d'Alembert rule and its extension read
\begin{equation}\label{eq:dAlembert3pla}
	\begin{split}
		&k_1+k_2+k_3+q_1+q_2+q_3-q_4-q_5-q_6+q_7+q_8+q_9-q_{10}-q_{11}-q_{12}=0,\\
		&q_7+q_8+q_9+q_{10}+q_{11}+q_{12} = 0\mod2,\\
		&\Xi_{k_1, k_2, k_3, q_1, q_2, q_3, q_4, q_5, q_6, q_7, q_8, q_9, q_{10}, q_{11}, q_{12}}=\Xi_{-k_1, -k_2, -k_3, q_4, q_5, q_6, q_1, q_2, q_3, q_{10}, q_{11}, q_{12}, q_7, q_8, q_9}.
	\end{split}
\end{equation}

\subsection{Practical considerations}\label{ssec:practical}

Hereafter we assume, without restriction to the generality and to lighten the notations, that the $3$-planet angle of interest is $k_1\lambda_1+k_2\lambda_2+k_3\lambda_3=\vect{k}\cdot\vect{\lambda}$. In order to obtain the term of $\varepsilon^2\Ham_2$ proportional to $e^{\iota\vect{k}\cdot\vect{\lambda}}$ at order $r$ in eccentricity and inclination, all Poisson brackets of the form
\begin{equation}\label{eq:bracket}
	\frac{1}{2}\left\lbrace\frac{-\iota h_\vect{p}}{p_1n_i+p_2n_j}e^{\iota\left(p_1\lambda_i+p_2\lambda_j\right)},h_{\vect{p}'}e^{\iota\left(p_1'\lambda_{i'}+p_2'\lambda_{j'}\right)}\right\rbrace,
\end{equation}
such that $p_1\lambda_i+p_2\lambda_j+p_1'\lambda_{i'}+p_2'\lambda_{j'}=\vect{k}\cdot\vect{\lambda}$, for $\left(i,j,i',j'\right)\in\left\lbrace1,2,3\right\rbrace^4$, have to be considered in Eq. (\ref{eq:H2bis}). If $k_1k_2k_3\neq0$, then $\left(i,j,i',j'\right)$ must belong to $\left\lbrace\left(1,2,1,3\right),\left(1,3,1,2\right),\left(1,3,2,3\right),\left(2,3,1,3\right),\left(1,2,2,3\right),\left(2,3,1,2\right)\right\rbrace$. If $k_1k_2k_3=0$, then there are more possibilities for the value of $\left(i,j,i',j'\right)$. In fact, that tuple does not even have to belong to $\left\lbrace1,2,3\right\rbrace^4$ in that case. Indeed, if $k_3=0$, even brackets of the form \smash{$\{e^{\iota\left(p_1\lambda_1+p_2\lambda_5\right)},e^{\iota\left(p_1'\lambda_2+p_2'\lambda_5\right)}\}$} can generate contributions of the form $e^{\iota\left(k_1\lambda_1+k_2\lambda_2\right)}$. However, when $k_1k_2k_3=0$, the angle $\vect{k}\cdot\vect{\lambda}$ also exists at order $1$ in $\varepsilon$ and its contribution of second order is often not important. Therefore, we restrict this work to the case $k_1k_2k_3\neq0$.

The Poisson bracket in Eq. (\ref{eq:bracket}) must be of degree $r$ or less in eccentricity and inclination. If $h_\vect{p}$ is of degree $s$ and $h_{\vect{p}'}$ is of degree $s'$, then the full bracket is of degree $s+s'$ if $ss'=0$ and of degree $s+s'-2$ if $ss'\neq0$ (by differentiating with respect to the eccentricities and inclinations in Eq. (\ref{eq:Lie_def})). The largest possible size for both $s$ and $s'$ is therefore $r+1$ and $\varepsilon\Ham_P$ needs to be expanded to degree $r+1$ to get $\varepsilon^2\Ham_2$ to degree $r$.

For the 3-planet angle $\vect{k}\cdot\vect{\lambda}$, there are infinitely many brackets to consider if $k_1k_2k_3=0$. Even if a truncation in the size of $\Xi$ can be set (see Table \ref{tab:decay}), far too many Poisson brackets must be computed and the final analytical expression is untractable. However, when $k_1k_2k_3\neq0$, it is straightforward to notice that there are only finitely many brackets to consider and to give them explicitly. This gives another argument in favor of restricting oneself to $k_1k_2k_3\neq0$. Consider for instance the case $\left(i,j,i',j'\right)=\left(1,2,1,3\right)$. The reasoning is essentially the same for the five other possible values of that tuple. Then the condition $p_1\lambda_i+p_2\lambda_j+p_1'\lambda_{i'}+p_2'\lambda_{j'}=k_1\lambda_1+k_2\lambda_2+k_3\lambda_3$ yields $p_1+p_1'=k_1$,$p_2=k_2$ and $p_2'=k_3$.
The condition that the degree of $h_\vect{p}$ and $h_{\vect{p}'}$ must be less than $r+1$ furthermore adds the bounds $|p_1+p_2|\leq r+1$ and $|p_1'+p_2'|\leq r+1$ because of the d'Alembert rule (see Eq. \ref{eq:dAlembert}). Since $p_2$ and $p_2'$ are forced to particular values, the equation $p_1+p_1'=k_1$ has only finitely many solutions that satisfy these two additional bounds. As an example, for $\left(k_1,k_2,k_3\right)=\left(2,-4,3\right)$ at degree $1$, assuming that all 2-planet angles are fast when defining $\left<\cdot\right>$, there are a total of $24$ Poisson brackets of the form (\ref{eq:bracket}) to compute, four for each value of the tuple $\left(i,j,i',j'\right)\in\left\lbrace\left(1,2,1,3\right),\left(1,3,1,2\right),\left(1,3,2,3\right),\left(2,3,1,3\right),\left(1,2,2,3\right),\left(2,3,1,2\right)\right\rbrace$.

\subsection{Expression as a function of the Laplace coefficients}\label{ssec:Laplace}

We implemented the method \texttt{angle} and the module \href{https://www.astro.unige.ch/~delisle/celeries/doc/_autosummary/celeries.perham3pla.PerHam3pla.html#celeries.perham3pla.PerHam3pla}{\texttt{PerHam3pla}} in the python algebraic manipulator package \href{https://gitlab.unige.ch/delisle/celeries}{\texttt{celeries}}. They give the full coefficient in front of $e^{\iota\left(k_1\lambda_1+k_2\lambda_2+k_3\lambda_3\right)}$ in $\varepsilon^2\Ham_2$ at any degree in eccentricity and inclination, when $k_1k_2k_3\neq0$. A factor $m_1m_2n_3\Lambda_3/m_0^2$ is left out. The output of this method depends on the Laplace coefficients \smash{$b_{s/2}^{(k)}(\alpha_{12})$}, \smash{$b_{s/2}^{(k)}(\alpha_{13})$} and \smash{$b_{s/2}^{(k)}(\alpha_{23})$} with $k\in\left\lbrace0,1\right\rbrace$ in virtue of Eq. (\ref{eq:laplace_rec}). Because the Laplace coefficients depend on $\vect{\Lambda}$, $\Xi$ should depend on \smash{$db_{s/2}^{(k)}/d\alpha$} as well. However, the relation
\begin{equation}\label{eq:dbsk}
	\frac{db_{s/2}^{(k)}}{d\alpha}=\frac{s^2}{s-2k}b_{\left(s+2\right)/2}^{(k-1)}(\alpha)+\left(\frac{s^2-ks}{2k-s}\alpha+\frac{ks}{2k-s}\alpha^{-1}\right)b_{\left(s+2\right)/2}^{(k)}(\alpha)
\end{equation}
is used to remove dependency on \smash{$db_{s/2}^{(k)}/d\alpha$}. From Eq. (\ref{eq:smax_order1}), the maximum value of $s$ in $\varepsilon\Ham_P$ at degree $r$ in eccentricity and inclination is $s=1+r+\left(r\mod2\right)$. Using Eq. (\ref{eq:dbsk}), the maximum value of $s$ in $\varepsilon^2\Ham_2$ is\footnote{No derivation with respect to $\vect{\Lambda}$ occurs when a term of the bracket in Eq. (\ref{eq:bracket}) is of degree $r+1$.}
\begin{equation}
	s\leq3+r+\left(r\mod2\right).
\end{equation}
The module \texttt{PerHam3pla} takes the following set of parameters as argument:
\begin{itemize}
	\item \texttt{degree}. The degree of expansion in eccentricity and inclination.  No default value.
	\item \texttt{ang2pla} $=\left(\left(i_1,j_1,p_1,q_1\right),\cdots,\left(i_m,j_m,p_m,q_m\right)\right)$. A set of 2-planet angles that are not averaged by the operator $\left<\cdot\right>$. Angles $p_1\lambda_{i_1}+q_1\lambda_{j_1}$ to $p_m\lambda_{i_m}+q_m\lambda_{j_m}$ are not averaged. $\left(i_k,j_k\right)\in\left\lbrace\left(1,2\right),\left(1,3\right),\left(2,3\right)\right\rbrace$. Only the fundamental of a 2-planet angle or its opposite must be mentioned. Harmonics must not be mentioned ($\gcd(p_k,q_k)=1$). By default, all $2$-planet angles are averaged.
	\item \texttt{n0} $=\left(n_1, n_2, n_3\right)$. The output can be evaluated at mean motions specified by this parameter. Defaults to ().
	\item \texttt{ev}. A boolean deciding if the output should be evaluated at the nominal mean motions \texttt{n0}. Defaults to False.
	\item \texttt{spatial}. A boolean deciding if the problem is planar or 3D. The output depends on $Y_i$ and $\bar{Y}_i$ when True. Defaults to False.
	\item \texttt{keplerian}. A boolean deciding if the equality $k_1n_1+k_2n_2+k_3n_3=0$ is used to simplify the output. Defaults to False.
	\item \texttt{disregard13}. A boolean parameter controlling if contributions from the pair $\left(1,3\right)$ are disregarded. Defaults to False.
	\item \texttt{disregardInd}. A boolean controlling if indirect terms in $\varepsilon\Ham_P$ (coming from $\vect{\tilde{r}}_i\cdot\vect{\tilde{r}}_j$) are disregarded. Defaults to False.
	\item \texttt{takeout\_kn}. In Eq. (\ref{eq:bracket}), the denominator \smash{$\left(p_1n_i+p_2n_j\right)^{-1}$} depends on $\vect{\Lambda}$ through $\partial n_j/\partial\Lambda_j=-3n_j/\Lambda_j$ and cannot be taken out of the Poisson bracket a priori as it disregards terms of order \smash{$\varepsilon^2$}. However, that denominator can be forcefully factored out of the bracket by setting this boolean to True. Defaults to False.
	\item \texttt{verbose}. A boolean deciding if the method \texttt{angle} prints what Poisson brackets it computes. Defaults to False.
\end{itemize}
The method \texttt{angle} takes as single argument the tuple $\vect{k}=\left(k_1, k_2, k_3\right)$, and the coefficient $R$ in front of $e^{\iota\vect{k}\cdot\vect{\lambda}}$ in $\varepsilon^2\Ham_2$ is obtained with the call \texttt{R = h3.PerHam3pla(parameters).angle(k)} where \texttt{h3 = celeries.perham3pla}.

\subsection{Case of zeroth-order resonances}\label{ssec:zeroth_order}

In this subsection, we consider 3-planet angles $k_1\lambda_1+k_2\lambda_2+k_3\lambda_3$ where $k_1+k_2+k_3=0$, and $k_1k_2k_3\neq0$. For these angles, there exist non-zero integers $\left(p,q\right)$ such that $k_1\lambda_1+k_2\lambda_2+k_3\lambda_3=p\lambda_1-\left(p+q\right)\lambda_2+q\lambda_3$. For these 3-planet angles, the d'Alembert rule (Eq. (\ref{eq:dAlembert3pla})) shows that terms of degree zero in eccentricity and inclination exist. We now compute analytically the coefficient $\Xi$ such that the Hamiltonian to second order in mass and to degree zero has contributions of the form
\begin{equation}\label{eq:0th}
	\varepsilon^2\Ham_2=\frac{m_1m_2}{m_0^2}n_3\Lambda_3\Xi\cos(p\lambda_1-\left(p+q\right)\lambda_2+q\lambda_3).
\end{equation}
Denoting $\varphi=p\lambda_1-\left(p+q\right)\lambda_2+q\lambda_3$, the term $\Xi\cos\varphi$ in Eq. (\ref{eq:0th}) comes from \smash{$\Xi/2\left(e^{\iota\varphi}+e^{-\iota\varphi}\right)$}, so we simply obtain and double the term in front of $e^{\iota\varphi}$. For the calculations to be doable by hand, we disregard contributions from pair $\left(1,3\right)$ and from indirect terms. However, we do not factor the denominator \smash{$1/\left(p_1n_i+p_2n_j\right)$} out of the Poisson brackets. We define the operator $\left<\cdot\right>$ (see Eq. (\ref{eq:lie_first_iter})) by averaging over all 2-planet angles. There are six brackets of the form of Eq. (\ref{eq:bracket}) that give a contribution in $e^{\iota\varphi}$. These are
\begin{equation}\label{eq:00}
	\begin{split}
		\frac{-\iota}{2}&\left\lbrace\frac{m_1}{m_0}n_2\Lambda_2\frac{W_{12}^{(p)}}{p\left(n_1-n_2\right)}e^{\iota p\left(\lambda_1-\lambda_2\right)},\frac{m_2}{m_0}n_3\Lambda_3W_{23}^{(q)}e^{-\iota q\left(\lambda_2-\lambda_3\right)}\right\rbrace,\\
		\frac{\iota}{2}&\left\lbrace\frac{m_2}{m_0}n_3\Lambda_3\frac{W_{23}^{(q)}}{q\left(n_2-n_3\right)}e^{-\iota q\left(\lambda_2-\lambda_3\right)},\frac{m_1}{m_0}n_2\Lambda_2W_{12}^{(p)}e^{\iota p\left(\lambda_1-\lambda_2\right)}\right\rbrace,
	\end{split}
\end{equation}
\begin{equation}\label{eq:11_q}
	\begin{split}
		\frac{-\iota}{2}&\left\lbrace\frac{m_1}{m_0}n_2\Lambda_2\frac{V_{12,>}^{(p)}}{pn_1-\left(p+1\right)n_2}x_2e^{\iota\left(p\lambda_1-\left(p+1\right)\lambda_2\right)},\frac{m_2}{m_0}n_3\Lambda_3V_{23,<}^{(q-1)}\bar{x}_2e^{-\iota\left(\left(q-1\right)\lambda_2-q\lambda_3\right)}\right\rbrace,\\
		\frac{\iota}{2}&\left\lbrace\frac{m_2}{m_0}n_3\Lambda_3\frac{V_{23,<}^{(q-1)}}{\left(q-1\right)n_2-qn_3}\bar{x}_2e^{-\iota\left(\left(q-1\right)\lambda_2-q\lambda_3\right)},\frac{m_1}{m_0}n_2\Lambda_2V_{12,>}^{(p)}x_2e^{\iota\left(p\lambda_1-\left(p+1\right)\lambda_2\right)}\right\rbrace,\\
	\end{split}
\end{equation}
\begin{equation}\label{eq:11_p}
	\begin{split}
		\frac{-\iota}{2}&\left\lbrace\frac{m_1}{m_0}n_2\Lambda_2\frac{V_{12,>}^{(-p)}}{pn_1-\left(p-1\right)n_2}\bar{x}_2e^{\iota\left(p\lambda_1-\left(p-1\right)\lambda_2\right)},\frac{m_2}{m_0}n_3\Lambda_3V_{23,<}^{(-q-1)}x_2e^{-\iota\left(\left(q+1\right)\lambda_2-q\lambda_3\right)}\right\rbrace,\\
		\frac{\iota}{2}&\left\lbrace\frac{m_2}{m_0}n_3\Lambda_3\frac{V_{23,<}^{(-q-1)}}{\left(q+1\right)n_2-qn_3}x_2e^{-\iota\left(\left(q+1\right)\lambda_2-q\lambda_3\right)},\frac{m_1}{m_0}n_2\Lambda_2V_{12,>}^{(-p)}\bar{x}_2e^{\iota\left(p\lambda_1-\left(p-1\right)\lambda_2\right)}\right\rbrace.\\
	\end{split}
\end{equation}
For the two brackets (\ref{eq:00}), the operator \smash{$\left\lbrace\cdot,\cdot\right\rbrace$} reduces to \smash{$\drond{}{\Lambda_2}\drond{}{\lambda_2}-\drond{}{\lambda_2}\drond{}{\Lambda_2}$} whereas \smash{$\left\lbrace\cdot,\cdot\right\rbrace=\iota\drond{}{x_2}\drond{}{\bar{x}_2}-\iota\drond{}{\bar{x}_2}\drond{}{x_2}$} for the last four brackets. When indirect terms are disregarded, \smash{$\varepsilon\Ham_P$} is expanded with \citep[][Eqs. 12 to 15]{Petit2021}
\begin{equation}\label{eq:Hp_expansion}
	W_{ij}^{(l)}=-\frac{1}{2}b_{1/2}^{(l)}(\alpha_{ij}),\;\;\;V_{ij,<}^{(l)}=\frac{1}{2}\sqrt{\frac{2}{\Lambda_i}}\left(l+1+\frac{\alpha_{ij}}{2}\drond{}{\alpha_{ij}}\right)b_{1/2}^{(l+1)}(\alpha_{ij}),\;\;\;V_{ij,>}^{(l)}=-\frac{1}{2}\sqrt{\frac{2}{\Lambda_j}}\left(l+\frac{1}{2}+\frac{\alpha_{ij}}{2}\drond{}{\alpha_{ij}}\right)b_{1/2}^{(l)}(\alpha_{ij}),
\end{equation}
and all six brackets are efficiently computed by hand using the expressions \smash{$\partial\alpha_{ij}/\partial\Lambda_i=2\alpha_{ij}/\Lambda_i$}, \smash{$\partial\alpha_{ij}/\partial\Lambda_j=-2\alpha_{ij}/\Lambda_j$} and \smash{$\partial n_j/\partial\Lambda_j=-3n_j/\Lambda_j$}. A factor \smash{$m_1m_2n_3\Lambda_3/m_0^2$} can be immediately factored out of the first two brackets. For the last four brackets, nearly everything can be factored out and the brackets reduce to \smash{$\left\lbrace x_2,\bar{x}_2\right\rbrace=-\left\lbrace \bar{x}_2,x_2\right\rbrace=\iota$}. In order to give the final result as a function of $n_2$ and $n_3$ only, we assume $pn_1-\left(p+q\right)n_2+qn_3=0$. We obtain
\begin{equation}\label{eq:Xi}
	\begin{split}
		\Xi=&\;\frac{n_2}{n_2-n_3}\left(b_{1/2}^{(p)}b_{1/2}^{(q)}+\alpha_{12}b_{1/2}^{(q)}\frac{db_{1/2}^{(p)}}{d\alpha_{12}}+\frac{p}{q}\alpha_{23}b_{1/2}^{(p)}\frac{db_{1/2}^{(q)}}{d\alpha_{23}}\right)+\frac{3p}{2q}\frac{n_2^2}{\left(n_2-n_3\right)^2}b_{1/2}^{(p)}b_{1/2}^{(q)}\\
		-\frac{1}{2}&\frac{n_2}{\left(q-1\right)n_2-qn_3}\left(q\left(2p+1\right)b_{1/2}^{(p)}b_{1/2}^{(q)}+q\alpha_{12}b_{1/2}^{(q)}\frac{db_{1/2}^{(p)}}{d\alpha_{12}}+\left(\frac{1}{2}+p\right)\alpha_{23}b_{1/2}^{(p)}\frac{db_{1/2}^{(q)}}{d\alpha_{23}}+\frac{\alpha_{12}\alpha_{23}}{2}\frac{db_{1/2}^{(p)}}{d\alpha_{12}}\frac{db_{1/2}^{(q)}}{d\alpha_{23}}\right)\\
		+\frac{1}{2}&\frac{n_2}{\left(q+1\right)n_2-qn_3}\left(q\left(2p-1\right)b_{1/2}^{(p)}b_{1/2}^{(q)}-q\alpha_{12}b_{1/2}^{(q)}\frac{db_{1/2}^{(p)}}{d\alpha_{12}}+\left(\frac{1}{2}-p\right)\alpha_{23}b_{1/2}^{(p)}\frac{db_{1/2}^{(q)}}{d\alpha_{23}}+\frac{\alpha_{12}\alpha_{23}}{2}\frac{db_{1/2}^{(p)}}{d\alpha_{12}}\frac{db_{1/2}^{(q)}}{d\alpha_{23}}\right),\\
	\end{split}
\end{equation}
where \smash{$b_{1/2}^{(p)}$} depends on $\alpha_{12}$ and \smash{$b_{1/2}^{(q)}$} depends on $\alpha_{23}$. The first, second and third lines of Eq. (\ref{eq:Xi}) correspond to brackets (\ref{eq:00}), (\ref{eq:11_q}) and (\ref{eq:11_p}), respectively.

The second term of the first line comes from deriving the denominators with respect to $\Lambda_2$ in Eq. (\ref{eq:00}) and would have been absent if we had factored the denominators out of the Poisson brackets.
If $\left<\cdot\right>$ is defined by not averaging over the 2-planet angles $p\lambda_1-\left(p+1\right)\lambda_2$ and $\left(q-1\right)\lambda_2-q\lambda_3$, then the second line of Eq. (\ref{eq:Xi}) should be removed. We checked the agreement between Eq. (\ref{eq:Xi}) and the output of \texttt{2*H3.angle((p,-p-q,q))} with \texttt{H3=h3.PerHam3pla(degree=0,keplerian=True,disregard13=True,disregardInd=True)}.

\subsection{Correctness of \texttt{PerHam3pla}}
An analytical computation of $\Xi$ was first attempted by \citet[][]{Quillen2011} and \citet[][]{Petit2020}. Petit's Eq. (42) and Quillen's Eq. (23) give $\Xi$ multiplied by \smash{$m_1m_2n_3\Lambda_3/m_0^2$}.
Hereafter, when referring to these equations, we will consider that this factor was removed.
\begin{figure}[t]
	\includegraphics[width=\linewidth]{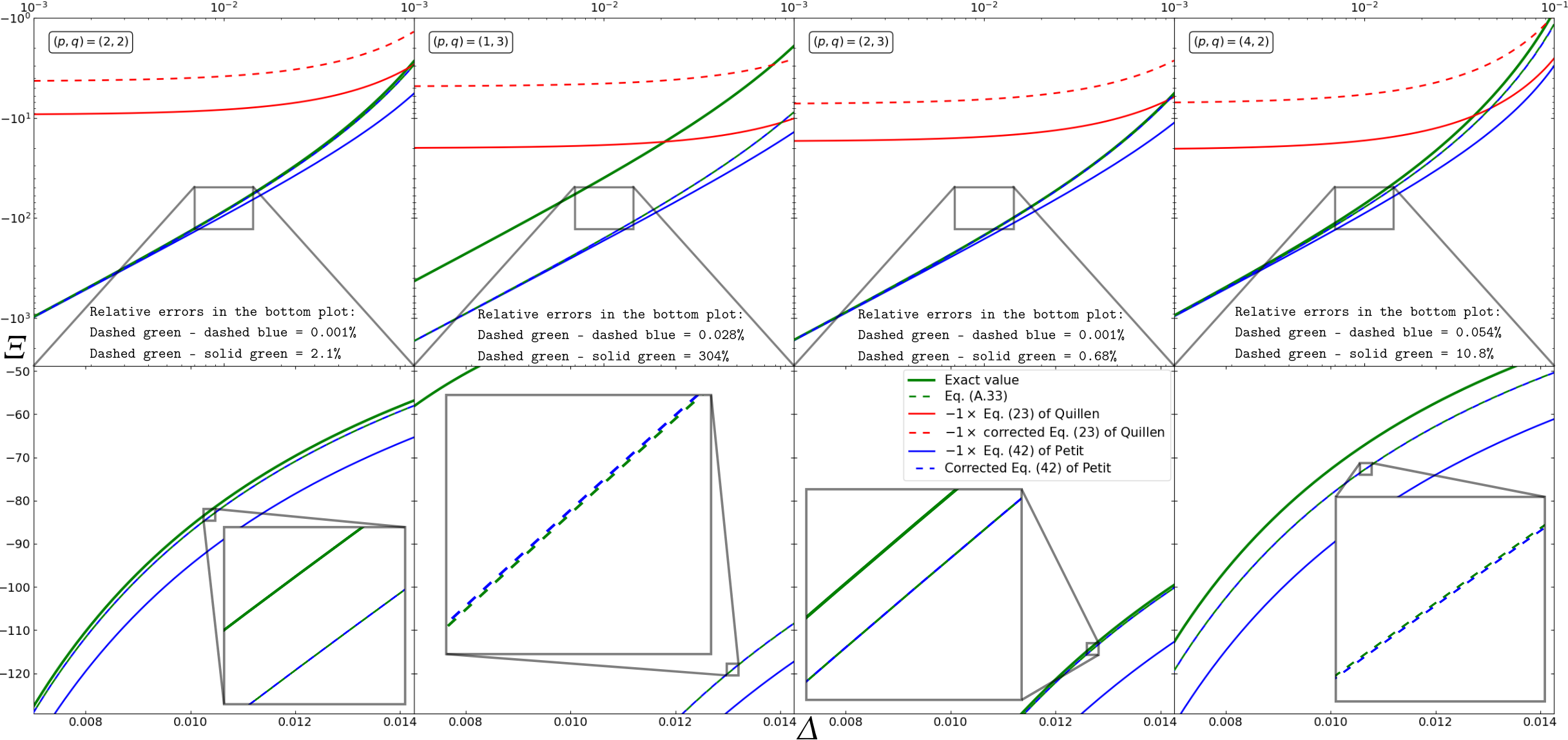}
	\caption{Exact value of $\Xi$ (solid green) on the loci of the 3P0R $pn_1-\left(p+q\right)n_2+qn_3=0$ parameterized by $\Delta$ (Eq. \ref{eq:rho}). The exact value is compared with Eq. (\ref{eq:Xi}) (dashed green), as well as with Petit's corrected (dashed blue) and original (solid blue) Eq. (42) and Quillen's corrected (dashed red) and original (solid red) Eq. (23). When needed, these curves are multiplied by $-1$ to fit on the figure. The bottom subplots are zoomed-in versions with linear axes of the top subplots in the region \smash{$\Delta\sim10^{-2}$}.}\label{fig:Xi}
\end{figure}

Quillen's and Petit's works neglect some Poisson brackets. Quillen disregards brackets (\ref{eq:11_q}) and (\ref{eq:11_p}), whereas Petit disregards brackets (\ref{eq:11_p}). We believe that they contain errors in their analytical expressions for $\Xi$. In Petit's Eq. (42), the $+$ sign in front of \smash{$1/\left(\nu-\left(p+q\right)^{-1}\right)$} should be replaced by a $-$ sign. With this correction, Eq. (42) of \cite{Petit2020} becomes identical to the first two lines of Eq. (\ref{eq:Xi}). In Quillen's Eq. (23), the parenthesis of the first term should be replaced by \smash{$1/(n_{ij}'n_{jk}')$}, and the last two terms should be divided by two. With this correction, Eq. (23) of \cite{Quillen2011} becomes identical to the first line of Eq. (\ref{eq:Xi}).

In Fig. \ref{fig:Xi}, we plot the exact value of $\Xi$ obtained with \texttt{Xi=2*h3.PerHam3pla(degree=0).angle((p,-p-q,q))}, for some values of $p$ and $q$ and for $10^{-3}\leq\Delta\leq10^{-1}$, where $\Delta$ is defined in Eq. (\ref{eq:rho}). We compare it with Eq. (\ref{eq:Xi}), Eq. (42) of Petit and Eq. (23) of Quillen. We include the corrected and original versions of these equations, as well as zooms in the region \smash{$\Delta\sim0.01$}. In this region, Eq. (\ref{eq:Xi}) matches Petit's corrected Eq. (42) to better than $0.1$\%, revealing that brackets (\ref{eq:11_p}), ignored by \cite{Petit2020}, are negligible. When $p\neq1$, Eq. (\ref{eq:Xi}) matches the exact value to about $1$\% to $10$\%, and contributions from the pair $\left(1,3\right)$ can be ignored in the region \smash{$\Delta\sim0.01$} if high precision on $\Xi$ is not needed, although pair $\left(1,3\right)$ can become important when $n_1/n_3$ is close to either $n_1/n_2$ or $n_2/n_3$. When $p=1$, Eq. (\ref{eq:Xi}) overestimates $\Xi$ by a factor $\sim4$, and indirect terms cannot be neglected in this case. Even after correction, Quillen's Eq. (23) is incorrect for $\Delta\leq0.1$, since it disregards brackets (\ref{eq:11_q}) which dominate $\Xi$ because of the denominators \smash{$\left(pn_1-\left(p+1\right)n_2\right)^{-1}$} and \smash{$\left(\left(q-1\right)n_2-qn_3\right)^{-1}$} that both diverge when $\Delta\to0$.

We do not have a proof that \texttt{2*h3.PerHam3pla(degree=0).angle((p,-p-q,q))} returns the exact value of $\Xi$, and our claim of correctness comes from two facts. On one hand, the separatrix in Fig. \ref{fig:stability_map} is located at $\Delta_{3P0R}=\pm1.97$, and matches the prediction of the pendulum model to about $1.5$\%. Since the resonance width scales with \smash{$\Xi^{1/2}$}, such a match between the 1-dof model and the actual 5-dof system would be surprising if $\Xi$ was incorrect\footnote{While the $1.5$\% error can be explained by several approximations (\textit{e.g.}, Table \ref{tab:Xi_gcd}), this fact alone only suggests correctness to about $3$\%.}. On the other hand, Eq. (\ref{eq:Xi}) and \texttt{PerHam3pla} were implemented independently. While Eq. (\ref{eq:Xi}) uses Eq. (\ref{eq:Hp_expansion}) for the expansion of $\varepsilon\Ham_P$, \texttt{PerHam3pla} uses the module \texttt{PerHam} of \texttt{celeries}, that returns different expressions of the Hamiltonian coefficients at first order in mass as a function of the Laplace coefficients. Despite that, Eq. (\ref{eq:Xi}) matches \texttt{2*h3.PerHam3pla(degree=0,disregardInd=True,disregard13=True).angle((p,-p-q,q))} to machine precision. Likewise, the first two lines of Eq. (\ref{eq:Xi}) match Eq. (42) of \cite{Petit2020} to machine precision after correction of a sign error, even though they obtained this equation independently from us.

However, \texttt{PerHam3pla} is able to output the coefficients at second order in mass of any 3PkR, at any degree in eccentricity and inclination, and these two facts do not suggest correctness when $k>0$, \texttt{degree}$\;>0$ or \texttt{spatial=True}. Our lighter claim of correctness in this general case comes from the standardized way of computing coefficients at second order in mass in \texttt{PerHam3pla}, although this does not guarantee the absence of a bug that would only trigger in the general case. An agreement between a model of 3P1R \citep[\textit{e.g.},][]{Petit2021} and a complete $N$-body system, using \texttt{PerHam3pla} for the computation of the coefficients at second order in mass, would strengthen the claim of correctness in the general case.

\twocolumn[{%
	\stepcounter{section}
	\centering
	\includegraphics[width=\linewidth]{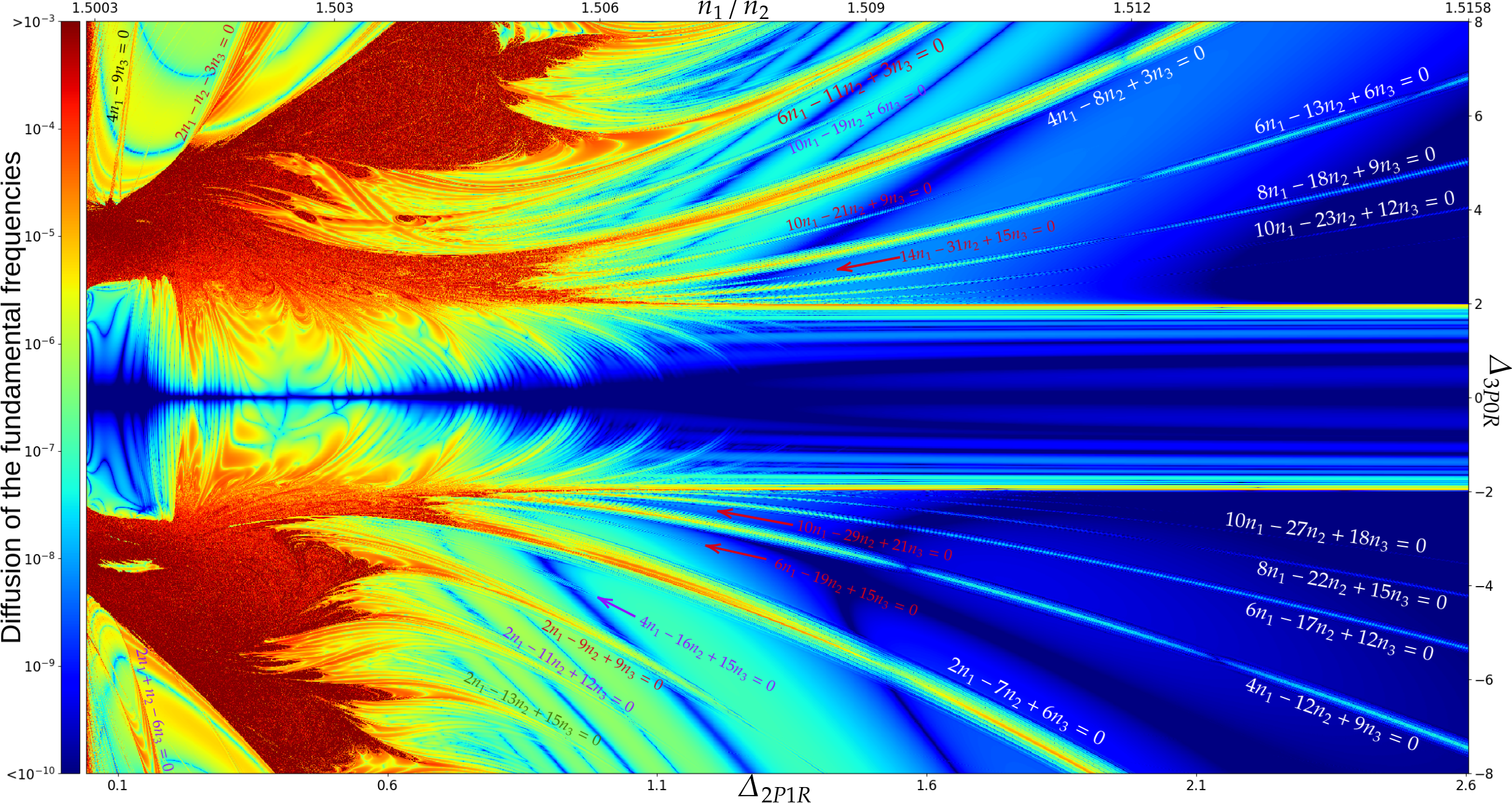}
	\captionof{figure}{Stability map of the 3P0R $2n_1-5n_2+3n_3=0$. On the $x$-axis are the libration centers of the 3P0R (Sect. \ref{ssec:libration}), parameterized by $n_1/n_2$ (top axis) or by $\Delta_\text{2P1R}$ (bottom axis). The separatrix of the 3P0R is at \smash{$\Delta_\text{3P0R}=\pm1.97$} close to the predicted \smash{$\Delta_\text{3P0R}=\pm2$} (Fig. \ref{fig:3P0RPhaseSpace}). The wide 3P0R $2n_1-5n_2+3n_3=0$ is visible. 3P1R, 3P2R, 3P3R and 3P4R are indicated in white, red, purple and green, respectively. See Sects. \ref{ssec:stability_map}, \ref{ssec:libration} and \ref{ssec:about_stability_map} for details. Similar figures for other $(p,q)$ and other masses are available \href{https://jeremycouturier.com/img/3P0R/}{here}\protect\footnotemark.}\label{fig:stability_map_23}
	\addtocounter{section}{-1}%
	\vspace{\dbltextfloatsep}%
}]
\makeatletter
\@fleqntrue
\makeatother
\footnotetext{\url{https://jeremycouturier.com/img/3P0R/}}

\section{Stability map of $2n_1-5n_2+3n_3=0$}\label{append:stability_map}

In this appendix, we present a stability map of the 3P0R $2n_1-5n_2+3n_3=0$. Similarly to Fig. \ref{fig:stability_map}, which is a stability map of the 3P0R $n_1-4n_2+3n_3=0$ with the masses of simulation \texttt{pq23} (Sect. \ref{ssec:simulations}), this stability map has $2414\times1079$ pixels. The masses are those of simulation \texttt{pq23} and are also the masses of the inner triplet of K2-138 according to the best posterior sample of \cite{Leleu2026a}. It is made in the same way as Fig. \ref{fig:stability_map}, and we refer the reader to Sects. \ref{ssec:stability_map} and \ref{ssec:about_stability_map} for details.

\section{\texttt{Aptidal}}\label{sec:Aptidal}

\texttt{Aptidal} is an open-source $N$-body integrator specifically designed for fast numerical integrations of planetary systems. It is implemented in \texttt{C} language and available \href{https://github.com/Jeremycouturier/Aptidal}{\texttt{here}}\footnote{\href{https://github.com/Jeremycouturier/Aptidal}{\texttt{https://github.com/Jeremycouturier/Aptidal}}}. \texttt{Aptidal} is able to
\begin{itemize}
	\item Integrate the complete Hamiltonian
	\item Integrate the averaged Hamiltonian at first order in mass
	\item Find libration centers of the complete Hamiltonian (Sect.$\!\,$\ref{ssec:libration})
	\item Find fixed points of the averaged Hamiltonian
	\item Evaluate the diffusion of the fundamental frequencies in order to make figures similar to Fig. \ref{fig:stability_map}.
\end{itemize}
There is no need to be familiar with \texttt{C} to use \texttt{Aptidal}. The user simply fills the file \texttt{parameters.h} with relevant parameters and initial conditions, and a numerical integration of the complete Hamiltonian (\textit{resp.} of the averaged Hamiltonian) with a $\mathcal{SABA}_n$ for a time $T$ with a timestep $dt$ is achieve by calling \texttt{X\_init(X)} and \texttt{SABAn(dt, T, outstep, X, n)} (\textit{resp.} \texttt{AveragedSABAn(dt, T, outstep, X, n)}) in the \texttt{main} function. Then, the numerical integration is launched with the command line \texttt{./aptidal.sh}. The software comes with several examples in order to help the user getting started. By default, \texttt{Aptidal} uses a system of unit where
\begin{itemize}
	\item The unit of mass is the star's mass,
	\item The unit of length is the inner planet's semi-major axis,
	\item The unit of time is the orbital period of a massless particle with semi-major axis the unit of length,
\end{itemize}
and the file of parameters should be filled in these units. They are often convenient and sensible because
\begin{itemize}
	\item The ratios $a_j/a_1$ are always much better constrained than $a_j$.
	\item The manipulated numerical quantities are $\OO(1)$.
	\item The inner orbital period is $P_1=1+\OO(\varepsilon)$, making it easier to choose the timestep.
	\item Gravity is scale-free and, in units of $P_1$, conservative physical timescales are unchanged when the system is rescaled.
	\item Even dissipative timescales scale more conveniently in units of $P_1$ rather than in an absolute unit like the year.
	\item Analytical works often use dimensionless units, which makes it easier to interpret simulation outputs.
\end{itemize}

\subsection{Convergence towards libration centers}\label{ssec:NAFF}

In this section, we describe the libration center finder of \texttt{Aptidal}. The algorithm is an iterative method outlined in Sect. \ref{ssec:libration}. It tries to converge to a libration center by computing the Fourier decomposition of Eq. (\ref{eq:quasiperiodic}) for each variable $f(t)$ of the Hamiltonian, and by only keeping the average and terms of the form $\vect{k}\cdot\vect{\omega}=k\omega_2$. While Sect. \ref{ssec:libration} assumes a 3-planet chain, we implemented our finder for chains of arbitrary length. We use the transformation proposed by \citet[][Appendix A]{Delisle2017}.

The difficulty lies in obtaining the decomposition of Eq. (\ref{eq:quasiperiodic}). Only the coefficients $f_\vect{k}$ in front of harmonics of the fast frequency $\omega_2$ are needed. We run a numerical integration of the complete Hamiltonian from $-T$ to $T$. At this point, both $\omega_2$ and $f_k$ are unknown. We first obtain the frequency $\omega_2$. According to \cite{Laskar1992}, if $T$ is significantly longer than the longest fundamental period, $\omega_2$ is obtained by looking for $\nu$ that extremizes the dot product \smash{$\left<\cos(\varphi_2),e^{\iota\nu t}\right>$}, where
\begin{equation}\label{eq:dot_product}
	\left<f(t),g(t)\right>=\frac{1}{2T}\int_{-T}^Tf(t)\bar{g}(t)\chi_p(t)dt,
\end{equation}
and $\chi_p$ is a weight function that improves greatly the quality of the determination for large enough $T$ \citep[][Proposition 2]{Laskar2003}. \texttt{Aptidal} first obtains an estimate of $\omega_2$ from a linear fit of $\varphi_2$, and the estimate can be refined by applying a Newton-Raphson method on \smash{$d\left<\cos(\varphi_2),e^{\iota\nu t}\right>/d\nu$}. When looking for fundamental frequencies to make stability maps like in Fig. \ref{fig:stability_map}, we refine the determination with a Newton-Raphson method (see Sect. \ref{ssec:about_stability_map}), but when looking for libration centers, this step is not needed. Once $\omega_2$ is known, the coefficients $f_k$ are given by the projection
\begin{equation}
	f_k=\left<f(t),e^{\iota k\omega_2t}\right>.
\end{equation}
\texttt{Aptidal} can converge to a libration center from a point $X$ with the call \texttt{LibrationCenterFind(X, dt, T, p, n, N, eps)} where $dt$ is the timestep, integrations are from $-T$ to $T$ with a $\mathcal{SABA}_n$, $p$ is the order of the weight function and $N$ the number of harmonics. The final point has the three-planet angles librating with less than \texttt{eps} degrees of amplitude due to slow terms. See \href{https://github.com/Jeremycouturier/Aptidal}{\texttt{Aptidal}}'s example section. In practice, the timestep must be small for the integrals to be computed properly (typically $dt\leq1/20$) and $n\leq2$ is enough.

\subsection{Implementation of $\mathcal{SABA}$ integrators}\label{ssec:SABA}

In this section, we present the $\mathcal{SABA}$ integrators implemented in \texttt{Aptidal}. We implemented two types of $\mathcal{SABA}$ integrators. The first type are the $\mathcal{SABA}_n$ for $1\leq n\leq10$ developed by \cite{LaskarRobutel2001}. The second type is the integrator $\mathcal{SABAH}_{10,6,4}$ described by \cite{Blanes2013}. All our $\mathcal{SABA}$ integrators integrate the Hamiltonian of Eq. (\ref{eq:nbody}) and work in heliocentric canonical coordinates \citep[heliocentric position and barycentric speed, see][]{LaskarRobutel1995}.

\subsubsection{General theory of $\mathcal{SABA}$ integrators}

A symmetric symplectic integrator takes the form
\begin{equation}\label{eq:shadow}
	e^{\tau L_\Kam}=e^{c_1\tau L_\Aam}e^{d_1\tau L_{\varepsilon\Bam}}e^{c_2\tau L_\Aam}\cdots e^{c_2\tau L_\Aam}e^{d_1\tau L_{\varepsilon\Bam}}e^{c_1\tau L_\Aam},
\end{equation}
where $\tau$ is the timestep and $\Aam$ and $\varepsilon\Bam$ are the Keplerian and perturbative part of Eq. (\ref{eq:nbody}), respectively. The notation \smash{$L_\chi=\left\lbrace\chi,\cdot\right\rbrace$} is the Lie derivative, defined in Sect. \ref{ssec:Lie}. By construction, the shadow Hamiltonian $\Kam$ is exactly integrated. As long as \smash{$\sum_j c_j=\sum_j d_j=1$}, $\Kam$ is close to the original Hamiltonian $\Ham$, and using \smash{$L_\Aam L_\Bam-L_\Bam L_\Aam=L_{\left\lbrace\Aam,\Bam\right\rbrace}$}, we have the Campbell-Baker-Hausdorff formula \citep[\textit{e.g.},][Eq. (7)]{LaskarRobutel2001}
\begin{equation}\label{eq:CBH}
	\Kam=\Ham+\sum_{p=1}^{+\infty}\sum_{q=1}^{p}k_{p,q}\,\OO\left(\tau^p\varepsilon^q\right),
\end{equation}
where $k_{p,q}$ depends on the integrator's coefficients $c_j$ and $d_j$. Refer to \citet[][Appendix E]{Couturier2025} for the calculation details in the case of the $\mathcal{SABA}_1$ integrator. When building a high-quality symplectic integrator, $c_j$ and $d_j$ are chosen as to cancel as many as possible of the lowest order terms in Eq. (\ref{eq:CBH}), while keeping at a minimum the number of steps in Eq. (\ref{eq:shadow}). \cite{LaskarRobutel2001} and \cite{Blanes2013} had different objectives. For their $\mathcal{SABA}_n$ integrators, \cite{LaskarRobutel2001} kept in Eq. (\ref{eq:CBH}) all terms $\OO(\tau^2\varepsilon^2)$ and smaller, in virtue of the small size of $\varepsilon$. On the other hand, they canceled out all terms $\OO(\tau^p\varepsilon)$ for $p<2n$. They obtained
\begin{equation}\label{eq:SABAn_CBH}
	\Kam_{\mathcal{SABA}_n}=\Ham+\OO\left(\tau^{2n}\varepsilon\right)+\OO\left(\tau^2\varepsilon^2\right).
\end{equation}
The strength of the $\mathcal{SABA}_n$ integrators is that there is a solution where $c_j>0$ and $d_j>0$. Therefore, as $n\to\infty$, $c_j\to0$ and $d_j\to0$, in virtue of \smash{$\sum_j c_j=\sum_j d_j=1$}. This makes the integrator stable at high timestep. For $10^{-4}<\varepsilon<10^{-5}$, typical of observed planetary systems, the $\mathcal{SABA}_{10}$ provides integrations of good quality with timesteps as large as $\tau=0.4$ (in units of the inner period). However, because of the \smash{$\OO\left(\tau^2\varepsilon^2\right)$} term, these integrators are formally second-order and are not the best at small timestep.

In order to design an integrator efficient at small timestep, \cite{Blanes2013} decided to forfeit the nice property $c_j>0$ and $d_j>0$ to be able to cancel all terms \smash{$\OO\left(\tau^2\varepsilon^q\right)$} and \smash{$\OO\left(\tau^3\varepsilon^q\right)$}. They obtained a fourth-order integrator
\begin{equation}\label{eq:SABAn_CBH}
	\Kam_{\mathcal{SABAH}_{10,6,4}}=\Ham+\OO\left(\tau^{10}\varepsilon\right)+\OO\left(\tau^6\varepsilon^2\right)+\OO\left(\tau^4\varepsilon^3\right),
\end{equation}
better than the $\mathcal{SABA}_n$ at small timesteps, but worse at large timesteps due to the larger size of the coefficients $c_j$ and $d_j$. The $\mathcal{SABA}_{10}$ is adapted for fast but moderately precise integrations (using $\tau>0.22$), whereas the $\mathcal{SABAH}_{10,6,4}$ is preferred for precise but slow integrations (using $\tau<0.12$). Refer to \citet[][Table 1]{LaskarRobutel2001} and \citet[][Table 4]{Blanes2013} for the numerical values of $c_j$ and $d_j$.

For the practical implementation of the integrators, the Hamiltonians $\Aam$ and $\Bam$ must be integrated. The step \smash{$e^{c_j\tau L_\Aam}$} consists in evolving the planets along Keplerian arcs for a time $c_j\tau$. This is done via the resolution of the Kepler equation and conversion from cartesian to elliptic coordinates. \texttt{Aptidal} relies on a \texttt{C} routine provided by Michaël Gastineau of the LTE lab of Paris Observatory (personal communication). The perturbative part $\Bam$ is not integrable as it mixes the momenta $\tilde{\vect{r}}$ and the coordinates $\vect{r}$. However, it can be written (Eq. \ref{eq:nbody}) $\Bam=\Bam_1(\tilde{\vect{r}})+\Bam_2(\vect{r})$ and can be integrated approximately but symplectically with a $\mathcal{SABA}_1$.

\subsubsection{Performances of $\mathcal{SABA}$ integrators}

Performances of symplectic integrators are often assessed by the conservation of the Hamiltonian. However, this can be misleading as the manifold of points with the same Hamiltonian value as the initial condition is of dimension $2n-1$ in a $2n$-dimensional phase space, while the true trajectory is $1$-dimensional. When the motion is not chaotic, a safer performance test is to compare the integrated trajectory with the true trajectory.

We evaluated our $\mathcal{SABA}$ integrators, both from the conservation of the Hamiltonian and the closeness to actual trajectory point of view. To ensure a non-chaotic initial condition, we used the system of Fig. \ref{fig:stability_map}, close to the 3P0R center at $\Delta=0.02$. That is, we used the initial condition\footnote{We introduced inclinations of about $2^\circ$ to run our tests in 3D.} of the left panel of Fig. \ref{fig:e-fold_3P0R} without tides. As the true trajectory is unknown, we took the output of $\mathcal{SABAH}_{10,6,4}$ with a timestep $\tau=0.008$ as the true trajectory.
\begin{figure}[h]
	\includegraphics[width=\linewidth]{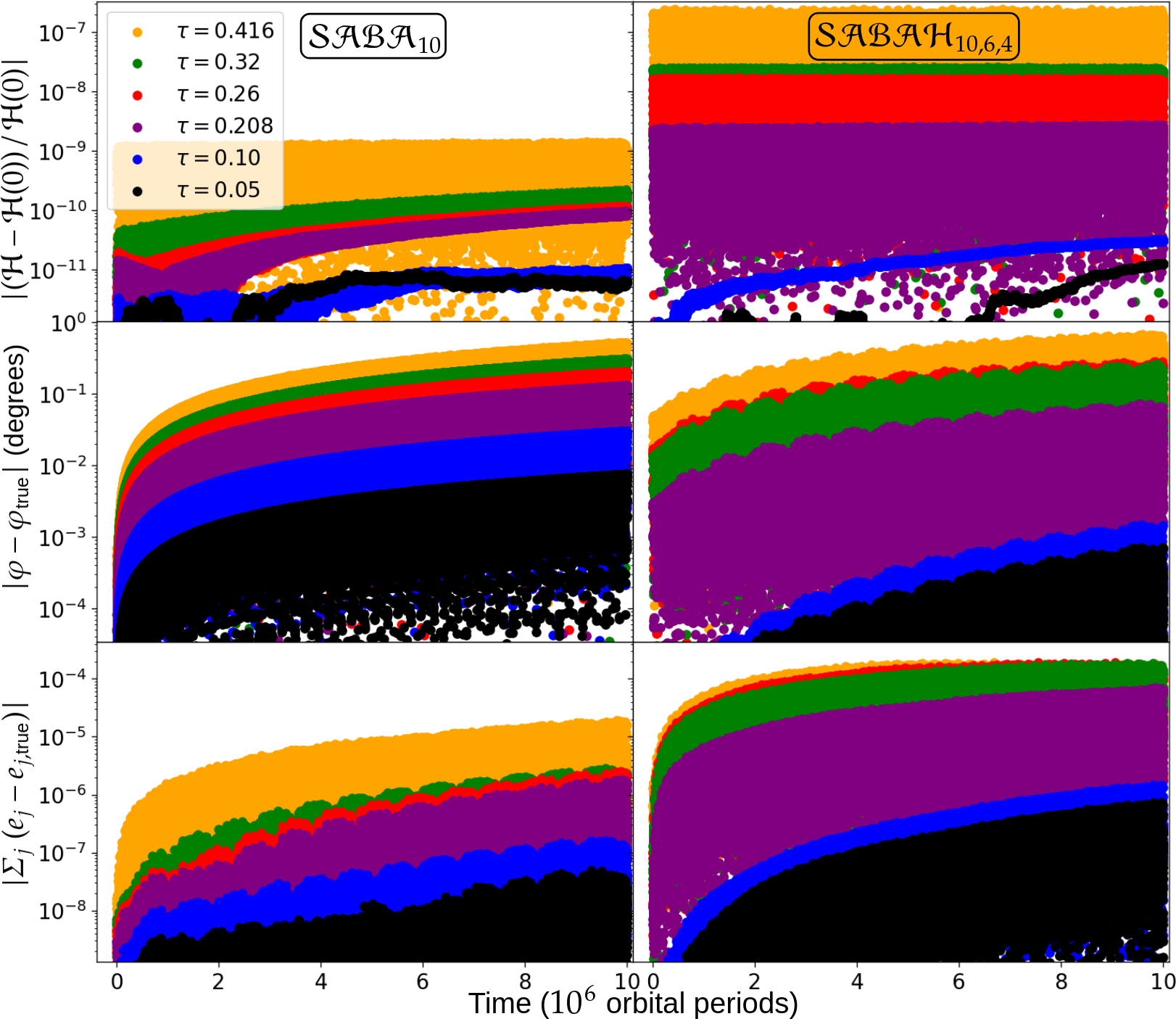}
	\caption{Conservation of the Hamiltonian of Eq. (\ref{eq:nbody}) (top), closeness of the 3-planet angle (middle) and closeness of the eccentricities to actual trajectory (bottom) for our $\mathcal{SABA}_{10}$ (left) and $\mathcal{SABAH}_{10,6,4}$ implementations (right).}\label{fig:SABA}
\end{figure}

In Fig. \ref{fig:SABA}, we show the quality of our integrators $\mathcal{SABA}_{10}$ (left) and $\mathcal{SABAH}_{10,6,4}$ (right) for the conservation of the Hamiltonian (top) and closeness to actual trajectory (middle and bottom). Overall, $\mathcal{SABAH}_{10,6,4}$ is better than $\mathcal{SABA}_{10}$ for small timesteps ($\tau\leq1/10$), whereas $\mathcal{SABA}_{10}$ is better than $\mathcal{SABAH}_{10,6,4}$ at large timesteps ($\tau\geq1/5$).

\subsubsection{Tides in $\mathcal{SABA}$ integrators}

In the constant$-\Delta t$ model, tides raised by the star on planet $j$ read \citep[][Eq. (11)]{Mignard1979, Couturier2026}
\begin{equation*}\label{eq:tidal_acceleration}
	\ddot{\vect{r}}_j\!=\!-3\kappa_2^{(j)}\G m_0\frac{R_j^5}{r_j^{10}}\left[r_j^2\vect{r}_j+\Delta t^{(j)}\left(2\left(\vect{r}_j\cdot\vect{v}_j\right)\vect{r}_j+r_j^2\left(\vect{v}_j+\vect{r}_j\times\vect{\Omega}_j\right)\right)\right],
\end{equation*}
where $\kappa_2^{(j)}$, $\Delta t^{(j)}$ and $\vect{\Omega}_j$ are the second Love number, the tidal timelag and the sideral rotation of planet $j$, respectively. The planetary rotations evolve in such a way that the pairwise angular momentum is conserved
\begin{equation}\label{eq:pairwise_Am_conservation}
	\frac{d\vect{\Omega}_j}{dt}=-m_j\mathcal{I}_j^{-1}\vect{r}_j\times\ddot{\vect{r}}_j.
\end{equation}
\texttt{Aptidal} assumes that planets remain spherical despite the tides and their rotations, and we take \smash{$\mathcal{I}_j=\alpha_jm_jR_j^2\mathbb{I}_3$}, where $\alpha_j$ is a dimensionless structure constant. Adiabatic tides can be incorporated in a $\mathcal{SABA}$ by adding a step $\tau\mathcal{S}_t=e^{\tau L_{H_t}}$ anywhere in Eq. (\ref{eq:shadow}), where $\mathcal{S}_t$ is the operator defined by the last two equations and $H_t$ is the pseudo-Hamiltonian introduced in Sect. \ref{ssec:tides}. In order to maintain the time-symmetry and to improve the precision, it is better to decompose the tidal operator in two steps and to put both steps in Eq. (\ref{eq:shadow}) in such a way that about half of the conservative timestep is performed between two tidal steps. As an example, for the $\mathcal{SABA}_{10}$ with tides, we write
\begin{equation*}
	e^{\tau L_\Kam}\!=e^{c_1\tau L_\Aam}\cdots e^{d_4\tau L_{\varepsilon\Bam}}e^{\tau L_{H_t}/2}\cdots e^{c_6\tau L_\Aam}\cdots e^{\tau L_{H_t}/2}e^{d_4\tau L_{\varepsilon\Bam}} \cdots e^{c_1\tau L_\Aam}.
\end{equation*}
However, the tidal operator $\mathcal{S}_t$ is not integrable and must be approximated. Since tides dissipate energy, a symplectic scheme in not necessary. Furthermore, since \smash{$e^{L_{H_t}}$} is much smaller than both \smash{$e^{L_\Aam}$} and \smash{$e^{L_{\varepsilon\Bam}}$} (this is the adiabaticity condition), a low-order scheme is enough. Therefore, we integrate tides with an explicit Euler method (Runge-Kutta 1) in \texttt{Aptidal}.

Since we integrate the planetary rotations in such a way that the pairwise angular momentum is conserved (Eq. \ref{eq:pairwise_Am_conservation}), we cannot use the conservation of the angular momentum or of the energy to evaluate the quality of our integrations. Instead, we rely on Fig. \ref{fig:e-fold_3P0R} that compares numerical integrations in \texttt{Aptidal} with what is expected from the theory. While we do not expect ground truth from the theory, Fig. \ref{fig:e-fold_3P0R} agrees with Eq. (\ref{eq:Resampi}) within 1\%, and such an agreement would be surprising if either the analytical theory or the numerical implementation were incorrect. Therefore, we conclude that both our numerical implementation of tides in \texttt{Aptidal}'s $\mathcal{SABA}$ integrators, and Eq. (\ref{eq:Resampi}) are correct, at least to a few percent.

\begin{table*}[h]
	\setlength{\tabcolsep}{5.5pt}
	\renewcommand{\arraystretch}{0.65}
	\caption{The numerical simulations run by \texttt{Aptidal} visible in Fig. \ref{fig:classification}}\label{tab:simulations}
	\centering
	\begin{tabular}{ccccccccc}
		\hline
		Name&Color&Type&Chain&$10^5m_j/m_0$&$10^3R_j/a_1$&$P_1$&$1/\mathfrak{R}\text{e}\,\sampi$&Length\\
		\hline
		\hline
		\href{https://github.com/Jeremycouturier/Aptidal/blob/main/example/NumericalIntegrations/Couturier_et_al_2027/33/parameters.h}{\texttt{33}}&\href{https://jeremycouturier.com/img/3P0R/logs/33.txt}{Blue}&Bern&$1:2:4:8$&$4.68,2.27,3.61,1.75$&$12.9, 6.35, 9.99, 4.91$&$1.21$&$4.64\;10^8$&$4.00\;10^{10}$\\
		\href{https://github.com/Jeremycouturier/Aptidal/blob/main/example/NumericalIntegrations/Couturier_et_al_2027/446/parameters.h}{\texttt{446}}&\href{https://jeremycouturier.com/img/3P0R/logs/446.txt}{Orange}&Bern&$1:2:4:8$&$3.11,3.80,2.76,2.46$&$8.61, 10.5, 7.66, 6.86$&$1.21$&$4.97\;10^8$&$4.00\;10^{10}$\\
		\href{https://github.com/Jeremycouturier/Aptidal/blob/main/example/NumericalIntegrations/Couturier_et_al_2027/703/parameters.h}{\texttt{703}}&\href{https://jeremycouturier.com/img/3P0R/logs/703.txt}{Yellow}&Bern&\begin{tabular}{@{}c@{}}$3:5,2:3,2:3$\\$2:3,3:4,3:5$\end{tabular}&\begin{tabular}{@{}c@{}}$0.344,0.501,0.510,$\\$1.21,1.81,1.02,1.12$\end{tabular}&\begin{tabular}{@{}c@{}}$1.85, 2.67, 2.72,$\\$6.34, 9.41, 5.34, 5.89$\end{tabular}&$0.48$&$5.04\;10^8$&$4.00\;10^{10}$\\
		\href{https://github.com/Jeremycouturier/Aptidal/blob/main/example/NumericalIntegrations/Couturier_et_al_2027/746/parameters.h}{\texttt{746}}&\href{https://jeremycouturier.com/img/3P0R/logs/746.txt}{Purple}&Bern&$1:2:4$&$3.92,3.17,2.50$&$10.3, 8.36, 6.65$&$1.30$&$4.99\;10^8$&$4.00\;10^{10}$\\
		\href{https://github.com/Jeremycouturier/Aptidal/blob/main/example/NumericalIntegrations/Couturier_et_al_2027/pq=13/parameters.h}{\texttt{pq13}}&\href{https://jeremycouturier.com/img/3P0R/logs/pq13.txt}{Black}&Fig. \ref{fig:stability_map}&$1:2:3$&$1.64,2.32,2.01$&$2.66, 2.90, 2.28$&$\cdots$&$4.87\;10^8$&$4.00\;10^{10}$\\
		\href{https://github.com/Jeremycouturier/Aptidal/blob/main/example/NumericalIntegrations/Couturier_et_al_2027/pq=23/parameters.h}{\texttt{pq23}}&\href{https://jeremycouturier.com/img/3P0R/logs/pq23.txt}{Green}&Fig. \ref{fig:stability_map_23}&$2:3,2:3$&$1.34,1.43,2.17$&$2.08, 2.81, 3.18$&$\cdots$&$5.34\;10^8$&$4.00\;10^{10}$\\
		\href{https://github.com/Jeremycouturier/Aptidal/blob/main/example/NumericalIntegrations/Couturier_et_al_2027/K1987/parameters.h}{\texttt{K1987}}&\href{https://jeremycouturier.com/img/3P0R/logs/K1987.txt}{Red}&Kepler-1987&$2:3,2:3,2:3$&$1.10,1.19,2.42,1.74$&$1.67, 1.80, 3.61, 2.60$&$2.35$&$1.48\;10^8$&$2.68\;10^{10}$\\
		\hline
	\end{tabular}
	\tablefoot{Clicking on the simulation's name (\textit{resp}. color) redirects to the parameter file (\textit{resp}. the log) of the simulation. The exact initial conditions are given in the parameter file. The inner period $P_1$ is in days and $1/\mathfrak{R}\text{e}\,\sampi$ and the simulation length are in units of $P_1$. The masses of \texttt{pq23} are those of the inner triplet of K2-138 according to the best posterior sample of \cite{Leleu2026a}.}
\end{table*}
\subsection{Numerical simulations}\label{ssec:simulations}

In this appendix, we present the numerical simulations that we ran with \texttt{Aptidal}, visible in Fig. \ref{fig:classification}. All simulations were performed with a $\mathcal{SABA}_{10}$ integrator, in a dimensionless system of units, as described in the introduction of Appendix \ref{sec:Aptidal}. Some of our simulations use initial conditions from simulations of planetary system formation in the Bern model and were extracted from the \href{https://dace.unige.ch/resonantStateObservationsCatalog/}{database} of the Geneva resonant state workshop \citep{Leleu2026b}. Some other simulations (\texttt{pq13} and \texttt{pq23}) are simulations of 3-planet systems with $\left(p,q\right)=\left(1,3\right)$ and $\left(p,q\right)=\left(2,3\right)$ with arbitrary masses (\texttt{pq13} uses the same masses as in Fig. \ref{fig:LibCen123}) and initial conditions close to the libration centers found by \texttt{Aptidal} (few degrees of amplitude for the Laplace angle) and near the 2P1R (close to the $x$-axis and close to the left edge of Fig. \ref{fig:classification}). A last simulation is named \texttt{K1987} and is an attempt at reproducing the capture in 3P1R experienced by Kepler-1987 (see Sect. \ref{ssec:3P0R23P1R}). It uses masses obtained from the mass-radius relation of \citep{Leleu2026a} for resonant systems and initial conditions close to the libration centers and the 2P1R. We present these simulations in Table \ref{tab:simulations}.

For all simulations, we used a timestep of about a quarter of the inner orbital period. Such a large timestep is acceptable for a $\mathcal{SABA}_{10}$ integrator (see Appendix \ref{ssec:SABA}). Over the simulations length, the relative variation of the Hamiltonian (on a witness simulation without tides) is less than $10^{-7}$. By design, our tidal implementation requires tides to be adiabatic, which we checked for each integration. For simulations \texttt{pq13} and \texttt{pq23}, the inner period in days is unspecified because gravity is scale-free, and these simulations were never in SI units. However, for the radii given in the parameter file to be physical, $P_1$ should be on the order of $1$ or $2$ days. The column $1/\mathfrak{R}\text{e}\,\sampi$ gives the e-folding time of exponential divergence of the Laplace angle's amplitude, given by Eq. (\ref{eq:Resampi}), in units of the inner planet's orbital period. It is computed for the inner triplet of the chain, except for \texttt{703}, where it is computed for the second triplet.

For each simulation, the duration of integration (last column) is about a factor $100$ times smaller than the characteristic age of the system, but the planetary tidal parameters \smash{$\kappa_2^{(j)}/Q_j$} (see line 66 of the parameters file by clicking on the simulation's name) are about a factor $100$ times larger than what is expected for super-Earth planets. Therefore, it can be considered that the simulations are run for a time similar, in order of magnitude, to the age of the system. Choosing larger \smash{$\kappa_2^{(j)}/Q_j$} can affect the capture in weak 3P1R, but it was a necessary choice to keep the running time reasonable ($32$ days for \texttt{703} and $5$ to $10$ days for the other simulations).

\section{Eigenvalues of a near diagonal matrix}\label{append:eigen}

We prove here that for a $n\times n$ near diagonal matrix $\mathcal{Q}=\mathcal{D}_0+\varepsilon_t\mathcal{Q}_1$ with \smash{$\mathcal{D}_0=\text{diag}(\lambda_{j,0})$} and \smash{$\mathcal{Q}_1=\left(q_{i,j}\right)_{1\leq i,j \leq n}$}, the well-separation
\begin{equation}
	\forall i\in\left[|1,n|\right], \;i\neq k \Rightarrow \left|\lambda_{k,0}-\lambda_{i,0}\right|\gg\varepsilon_t,
\end{equation}
of $\lambda_{k,0}$ implies that the eigenvalue $\lambda_k$ of $\mathcal{Q}$ is given by
\begin{equation}
	\lambda_k=\lambda_{k,0}+\varepsilon_t\,q_{k,k}+\mathcal{O}(\varepsilon_t^2),
\end{equation}
that is, it can be read on the diagonal. In order to prove this result, we use the definition of the eigenvalues as roots of the characteristic polynomial. The characteristic polynomial of $\mathcal{Q}$ is \smash{$\chi_{\mathcal{Q}}(\lambda)=\det(\lambda\mathbb{I}_n-\mathcal{Q})=\det(\lambda\mathbb{I}_n-\mathcal{D}_0-\varepsilon_t\mathcal{Q}_1)$}.
Using Leibniz formula for the determinant, this reads
\begin{equation}
	\chi_{\mathcal{Q}}(\lambda)=\sum_{\sigma\in\mathcal{S}_n}\text{sgn}(\sigma)\prod_{1\leq i\leq n}\left(\delta_{i,\sigma(i)}\left(\lambda-\lambda_{i,0}\right)-\varepsilon_tq_{i,\sigma(i)}\right),
\end{equation}
where $\mathcal{S}_n$ is the set of size $n$ permutations, $\text{sgn}(\sigma)$ is the signature of $\sigma$ and $\delta_{i,\sigma(i)}$ is $1$ if $i$ is a fixed point of $\sigma$ and zero otherwise. In the above sum, a permutation with at least two non-fixed points gives rise to terms of order $\varepsilon_t^2$ or smaller. Since the identity is the only permutation that does not have at least two non-fixed points, $\chi_{\mathcal{Q}}(\lambda)$ can be written
\begin{equation}
	\chi_{\mathcal{Q}}(\lambda)=\prod_{1\leq i\leq n}\left(\lambda-\lambda_{i,0}-\varepsilon_tq_{i,i}\right)+\mathcal{O}(\varepsilon_t^2).
\end{equation}
At this point, it is tempting to say that the eigenvalues of $\mathcal{Q}$ are $\lambda_{i,0}+\varepsilon_tq_{i,i}$ regardless of the non-separability condition, but this does not work because the term $\mathcal{O}(\varepsilon_t^2)$ cannot be neglected when $\chi_{\mathcal{Q}}$ is linearized in the vicinity of a non-well-separated eigenvalue. The characteristic polynomial can be expanded as
\begin{equation}
	\chi_{\mathcal{Q}}(\lambda)=\prod_{1\leq i\leq n}\left(\lambda-\lambda_{i,0}\right)-\varepsilon_t\sum_{1\leq i\leq n}q_{i,i}\prod_{j\neq i}\left(\lambda-\lambda_{j,0}\right)+\mathcal{O}(\varepsilon_t^2).
\end{equation}
We linearize $\chi_{\mathcal{Q}}$ in the vicinity of the well-separated eigenvalue $\lambda_{k,0}$. We write $\lambda=\lambda_{k,0}+d\lambda$ and neglect terms of size $\mathcal{O}(\varepsilon_td\lambda)$. We obtain $\chi_{\mathcal{Q}}(d\lambda)=$
\begin{equation*}
	\begin{split}
		&\prod_{1\leq i\leq n}\!\left(\lambda_{k,0}\!-\!\lambda_{i,0}\!+\!d\lambda\right)-\varepsilon_t\!\sum_{1\leq i\leq n}\!q_{i,i}\prod_{j\neq i}\left(\lambda_{k,0}-\lambda_{j,0}+d\lambda\right)+\mathcal{O}(\varepsilon_t^2)=\\
		&d\lambda\prod_{i\neq k}\left(\lambda_{k,0}-\lambda_{i,0}\right)-\varepsilon_tq_{k,k}\prod_{j\neq k}\left(\lambda_{k,0}-\lambda_{j,0}\right)+\mathcal{O}(\varepsilon_t^2)+\mathcal{O}(\varepsilon_td\lambda)=\\
		&\left(d\lambda-\varepsilon_tq_{k,k}\right)\prod_{i\neq k}\left(\lambda_{k,0}-\lambda_{i,0}\right)+\mathcal{O}(\varepsilon_t^2)+\mathcal{O}(\varepsilon_td\lambda),
	\end{split}
\end{equation*}
the root of which is \smash{$d\lambda=\varepsilon_tq_{k,k}+\mathcal{O}(\varepsilon_t^2)$}, ending the demonstration. The well-separability condition is necessary so that the quantity \smash{$\prod_{i\neq k}\left(\lambda_{k,0}-\lambda_{i,0}\right)$} is of order unity, making the term \smash{$\mathcal{O}(\varepsilon_t^2)$} negligible.






\end{appendix}

\end{document}